\documentclass[
reprint, 
superscriptaddress, 
amsmath,amssymb, 
aps, 
prl, 
]{revtex4-2}

\usepackage[T1]{fontenc}
\usepackage{graphicx} 
\usepackage{dcolumn} 
\usepackage{bm} 
\usepackage{makecell,multirow,array}
\usepackage{amsmath}
\usepackage{amsthm,amssymb, physics}
\usepackage{xcolor}
\usepackage{enumitem}
\usepackage{appendix}
\usepackage{float}
\usepackage{mathtools}
\usepackage{booktabs}
\usepackage{quantikz}
\usepackage{siunitx}
\usepackage{tikz}
\usepackage{makecell}
\usepackage{xurl}
\usepackage{dsfont}
\usepackage{lipsum}
\usepackage{hyperref}
\usepackage[normalem]{ulem}
\usepackage[version=4]{mhchem}
\begin{document}

\title{Hardware-Tailored Resource Estimation for Magic-State Distillation on Silicon Spin Qubits
}

\author{Songqinghao~Yang}
\affiliation{Cavendish Lab., Department of Physics, Univ. of Cambridge, Cambridge CB3 0HE, UK}
\affiliation{Hitachi Cambridge Lab., J.J. Thomson Avenue, Cambridge CB3 0HE, UK}

\author{Christopher~K.~Long}
\affiliation{Cavendish Lab., Department of Physics, Univ. of Cambridge, Cambridge CB3 0HE, UK}
\affiliation{Hitachi Cambridge Lab., J.J. Thomson Avenue, Cambridge CB3 0HE, UK}

\author{Rubén~M.~Otxoa}
\affiliation{Hitachi Cambridge Lab., J.J. Thomson Avenue, Cambridge CB3 0HE, UK}

\author{Prakash~Murali}
\affiliation{Department of Computer Science and Technology, Univ.~of~Cambridge}

\author{Crispin~H.~W.~Barnes}
\affiliation{Cavendish Lab., Department of Physics, Univ. of Cambridge, Cambridge CB3 0HE, UK}

\author{David~R.~M.~Arvidsson-Shukur}
\affiliation{Hitachi Cambridge Lab., J.J. Thomson Avenue, Cambridge CB3 0HE, UK}

\date{\today} 

\begin{abstract}

We present a resource analysis for generating high-fidelity logical magic states on silicon spin-qubit platforms. We consider a range of architectures, including a shuttling-based SpinBus design, a dense nearest-neighbor layout, and a hybrid scheme with shuttling-connected patches. We compare surface, color, and biased error-correcting codes, and analyze the $5\to1$ and $15\to1$ magic-state distillation protocols.
Our approach combines bottom-up and top-down methodologies. We construct a hardware-level noise model based on a silicon-processor Hamiltonian with realistic parameters and $1/f$ non-Markovian noise, enabling estimation of physical resources required to reach target logical error rates. These results are propagated to system-level overheads for applications including spin dynamics, integer factorization, and quantum chemistry. Conversely, we fix target logical fidelities and derive corresponding constraints on hardware performance.
Our framework enables systematic evaluation of resource-reduction strategies. We find that optimized control pulses reduce magic-state distillation overhead by 42\% compared to standard gate implementations. In addition, silicon-tailored biased error-correcting codes achieve an approximately threefold reduction in physical footprint relative to the surface code, even without physical-bias-preserving operations.

\end{abstract}

\maketitle

\section{Introduction}
\label{intro}
Quantum computers process information in ways that are fundamentally different from classical computers, offering the prospect of advances in cryptography~\cite{shor_factoring, gidney2021factor}, Hamiltonian simulation~\cite{lloyd1996universal, childs2018toward, berry2007efficient}, computational chemistry~\cite{babbush2018encoding, dong2022ground, lee2023evaluating}, and optimization~\cite{gilyen2019optimizing, farhi2014quantum}. Rapid experimental progress across multiple platforms has led to steady improvements in qubit number, wall-clock time, and control fidelity, strengthening the case for scalable quantum computation~\cite{zhou2025opportunities, gutierrez2019transversality, beverland2022assessing}. However, current quantum hardware remains intrinsically noisy: decoherence and control imperfections prevent reliable execution of complex algorithms~\cite{Fontana2021,StilckFranca2021,DePalma2023,Dalton2024, preskill2018quantum, bouland2019complexity, mcclean2018barren}. 

Scalable quantum computation, therefore, requires fault-tolerant architectures based on quantum error correction (QEC)~\cite{steane1996error, campbell2017roads, knill1998resilient, gottesman2010introduction, terhal2015quantum, fowler2012surface}. Recent years have seen several platforms take steps towards fault-tolerant architectures~\cite{bermudez2017assessing, google2023suppressing, google2025quantum, tripier2026fault}. In such architectures, logical qubits are encoded in multiple physical qubits. As long as the input states are restricted to stabilizer states and the operations to Clifford gates and Pauli measurements, standard QEC protocols actively detect and correct errors during computation. While QEC enables the detection and correction of noise, it does not provide a complete solution for universal quantum computation. In particular, Clifford gates and Pauli measurements are not universal~\cite{aaronson2004improved}. To harness the full power of quantum computation, one needs access to  non-Clifford operations.

Magic-state distillation (MSD) provides one path to promote error-corrected quantum computers to universal ones~\cite{reichardt2004improved, bravyi2005universal, reichardt2005quantum,reichardt2006quantum,meier2012magic,jones2012multilevel,bravyi2012magic, litinski2019magic}. 
In MSD, noisy (non-stabilizer) magic states are prepared at the physical-qubit level. These states are then injected into the hardware's logical space. In the logical space, the magic states are distilled to improve their fidelity. After the fidelity reaches some target value, the magic states can be injected into the error-corrected computer to implement high-fidelity non-Clifford operations, thus enabling universal quantum computing~\cite{bravyi2012magic, li2015magic}. Recent proof-of-principle experimental work has demonstrated high-fidelity preparation and distillation of logical magic states on error-corrected qubits~\cite{daguerre2025experimental,sales2025experimental}. 

In the last years, several studies have estimated the resources needed to utilize fault-tolerant hardware. These studies consider the total number of physical qubits and the wall-clock time needed to implement a specific quantum algorithm~\cite{BenchQ, harrigan2024qualtran, beverland2022assessing, van2023using, litinski2019game, mohseni2024build, litinski2023compute, filippov2025architecting, campbell2026resource, sunami2025transversal}.  Typically, the QEC and MSD protocols utilized in a specific implementation are identified as the major contributors to the computational overhead. The overheads incurred by a certain MSD protocol can only be quantified in light of a specific QEC protocol, and \textit{vice versa}~\cite{wan2024constant,litinski2019magic}. Moreover, noise profiles and qubit-connectivity architectures vary widely across such platforms, with direct implications on the suitability of various QEC and MSD protocols. Thus, to attain the highest accuracy in resource estimation, one must consider the specific physical platform in which the quantum hardware is implemented. Previous works have estimated quantum-computational resources for quantum computation with neutral atoms~\cite{ismail2025transversal,zhou2025resource}, photons~\cite{litinski2019game, litinski2025blocklet}, and superconducting qubits~\cite{gidney2025factor,huggins2025fluid}. However, to the best of our knowledge, no thorough resources estimation has been conducted for quantum computation with electron-spin qubits in silicon. 

Quantum-computing hardware with electron-spin qubits in silicon was originally proposed in the seminal work of Loss and DiVincenzo \cite{loss1998quantum}. Currently, such hardware is experiencing a surge in research interest~\cite{vandersypen2017interfacing, cai2023looped, sun2026folded, chadwick2025manufacturable, pataki2025compiling, boter2022spiderweb, siegel2024towards, otxoa2025spinhex, li2018crossbar, siegel2025snakes, micciche2025optimizing, gonzalez2021scaling, ladd2026digitally, moncy2026surface}. In Loss-DiVincenzo processors, single electrons are trapped in arrays of gate-based quantum dots within a silicon substrate embedded in a semiconductor heterostructure~\cite{burkard2023semiconductor, zwanenburg2013silicon}. The spin of these electrons encode the physical qubits. A main selling point is the small spatial footprint of the physical qubits (tens of nanometers~\cite{veldhorst2017silicon, neyens2024probing, zwerver2022qubits, elsayed2024low}). Moreover, the processors are compatible with current industrial production lines for CMOS and advanced semiconductor technologies, as well as with classical control electronics~\nocite{stano2022review,Stano2021}\cite{stano2022review,maurand2016cmos,zwerver2022qubits,clarke2025spin,swift2025large, swift2025superinductor}.  Spin qubits in silicon can have long $T_1$ times (around 1~\unit{\milli\second} to 1~\unit{\second}~\nocite{stano2022review,Stano2021}\cite{hendrickx2021four, stano2022review, otxoa2025spinhex}) and $T_2^*$ times (around 10 to $100~\unit{\micro \second}$~\cite{stano2022review}\nocite{Stano2021}). Current gate fidelities are competitive~\cite{crawford2023compilation, ginzel2024scalable, langrock2023blueprint, wang2024operating, zhang2025demonstration} with 99.9\%~\cite{mkadzik2022precision,noiri2022fast,mills2022two,steinacker2025industry} for two-qubit operations and up to 99.999\%~\cite{yoneda2018quantum, wu2025simultaneous} for single-qubit operations. Earlier concerns about silicon spin-qubit connectivity have been partially alleviated by the demonstration of coherent electron shuttling over micrometer distances~\cite{seidler2022conveyor,volmer2025reduction,ciroth2025numerical} at fidelities surpassing 99.5\% over $10~\unit{\micro \meter}$~\cite{de2025high}. To summarize, electron spin-qubits in silicon provide a competitive platform for future quantum computation. 

Motivated by these considerations, we present a hardware-tailored resource analysis of logical magic-state production in silicon spin-qubit architectures. We evaluate different physical-qubit-connectivity models, QEC codes and distillation protocols. With hardware-tailored noise modeling, we estimate the resulting space-time volumes for MSD. This allows us to estimate also the overheads for the implementation of specific quantum algorithms. Going beyond resource estimation, we also invert the analysis: for a target logical magic-state fidelity, we identify the corresponding hardware requirements on the physical noise and operational parameters. Thus, we can study the effect of hardware or software improvements on the resource budget for a specific algorithm. In particular, we gauge the potential resource benefits of using silicon-tailored QEC codes compared to standard QEC protocols. Together, our results clarify the tradeoffs between connectivity models, QEC codes , and MSD protocols in silicon spin-qubit systems. Further, our work provides a package for the analysis of experimental bottlenecks and a testbed for the construction and evaluation of new QEC and MSD protocols.

The remainder of this Article is structured as follows. In Sec. \ref{sec: main}, we summarize our study's computational pipeline and main results. In Sec. \ref{sec: background} we provide background material for the silicon hardware (physical level) as well as the MSD and QEC protocols (logical level) studied in this work. In Sec. \ref{sec: methodology}, we detail our methodology for pulse design and optimization. In Sec. \ref{noise} we present our noise analysis. And in Sec. \ref{sec:st_modeling} we provide MSD resource overheads. Finally, in Secs.  \ref{sec: discussion}, we discuss future directions and conclude our work.

\section{Main Result}
\label{sec: main}

\begin{figure*}
    \centering
    \includegraphics[width=\linewidth]{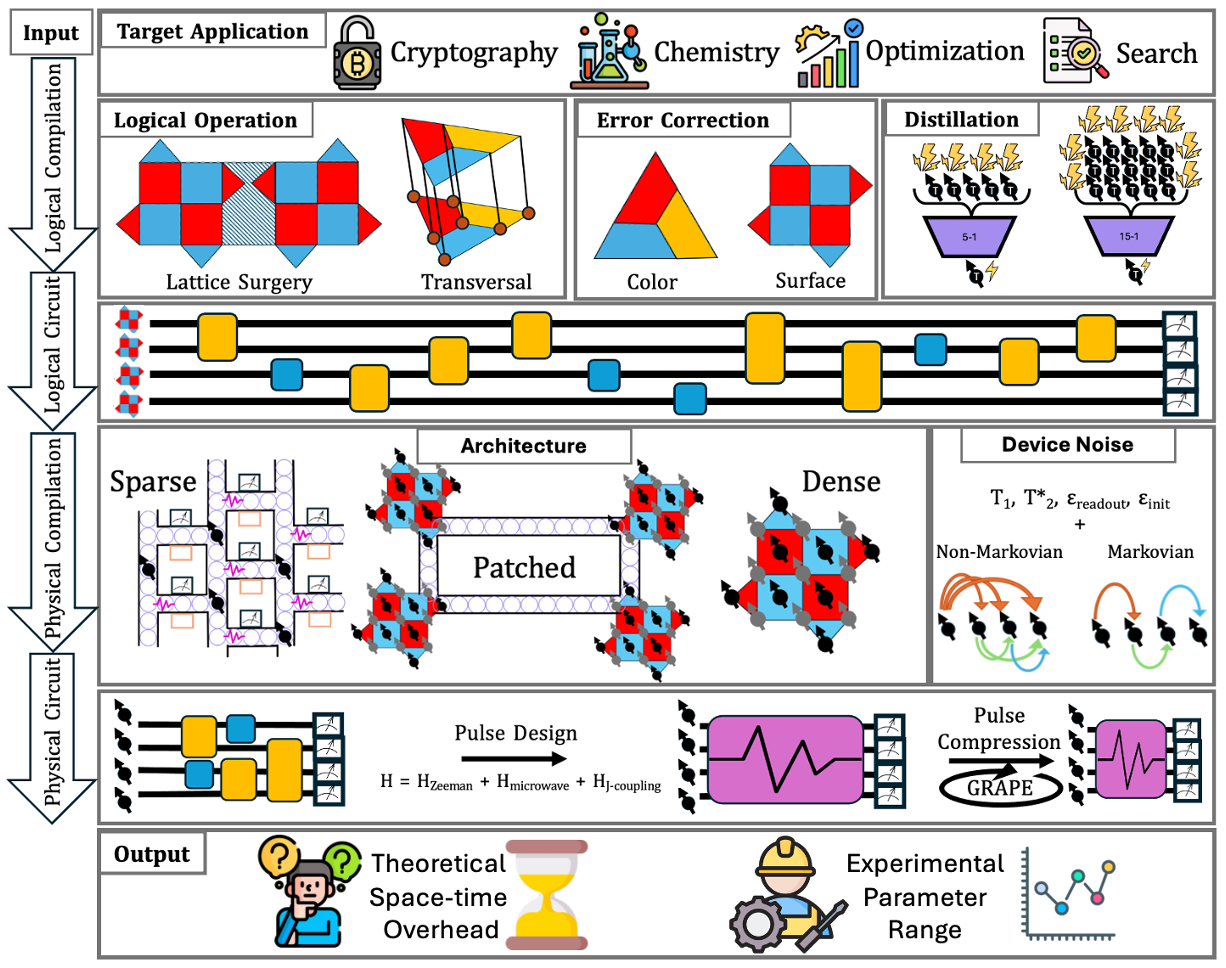}
    \caption{\textbf{Resource-estimation pipeline.} From top to bottom: A target quantum application is compiled into a logical circuit based on the choice of (1) logical operations, (2) QEC code, and (3) MSD protocol. The logical circuits are compiled into physical circuits by considering specific hardware architectures. The relevant QEC and MSD overheads depend on a hardware-specific noise model. The physical circuits are compiled into a sequence of hardware-tailored control pulses using pulse design. We use GRAPE to optimize the controls, thereby compressing the pulse sequence and potentially reducing errors and runtime. Finally, we produce theoretical resource metrics---space-time volume, physical-qubit overhead, runtime---and experimental  bounds and target ranges for device parameters---\textit{e.g.}, required coherence times, gate fidelities, and operation timing thresholds.}
    \label{fig:workflow}
\end{figure*}

Here, we give a brief summary of our work. We start by describing the hardware and software settings which we analyze. Then, we outline our main findings. The goal of our work is to address the question:
\begin{center}
\textit{Assuming that current experimental results on small-scale devices can be scaled up, what experimental resources will be needed to run quantum algorithms on silicon hardware?}
\end{center}
To answer this question, we construct the following pipeline, illustrated in Fig.~\ref{fig:workflow}:
\begin{enumerate}
\item First, we identify the quantum algorithm that we aim to execute on a silicon spi--qubit quantum computer. We investigate three prototypical algorithms for quantum dynamics, integer factoring and quantum chemistry; details are given in Sec.~\ref{RE}. The target algorithm is cast in terms of a logical circuit. 
\item Next, we determine the target algorithm's logical implementation by fixing (1) the form of logical operations (lattice surgery or transversal gates), (2) the QEC code (color, standard surface, or XZZX surface), and (3) the MSD protocol ($5\to1$ or $15\to1$). Further details are given in the Sec. \ref{sec: background}.
\item After completing the logical compilation, we select a silicon spin–qubit architecture. We consider three architectures: a dense 2D-grid layout, a shuttling-mediated SpinBus architecture, and a hybrid `patched' layout (Sec.~\ref{arch}).
\item We then convert the logical circuits into physical circuits. We do so by optimizing hardware control pulses for the required circuit components (Sec. \ref{sec:pulse_optimization}). Our pulse optimizer considers a realistic silicon-device Hamiltonian and compresses pulses using the GRAPE algorithm. In conjunction with this analysis, we construct a noise and timing model that accounts for the device-specific decoherence times, with either Markovian noise or the $1/f$-type temporal noise commonly observed in silicon devices~\cite{cywinski2008enhance, cerfontaine2021filter}, shuttling-induced errors, and operational latency (Sec.~\ref{noise}).
\item Finally, having mapped out a pipeline from the target algorithm to the physical-level implementation, we simulate the combined resource cost of the required MSD and QEC protocols. This allows us to estimate the total resource budget needed for the target application.
\end{enumerate}

\begin{figure*}
    \centering
    \includegraphics[width=\linewidth]{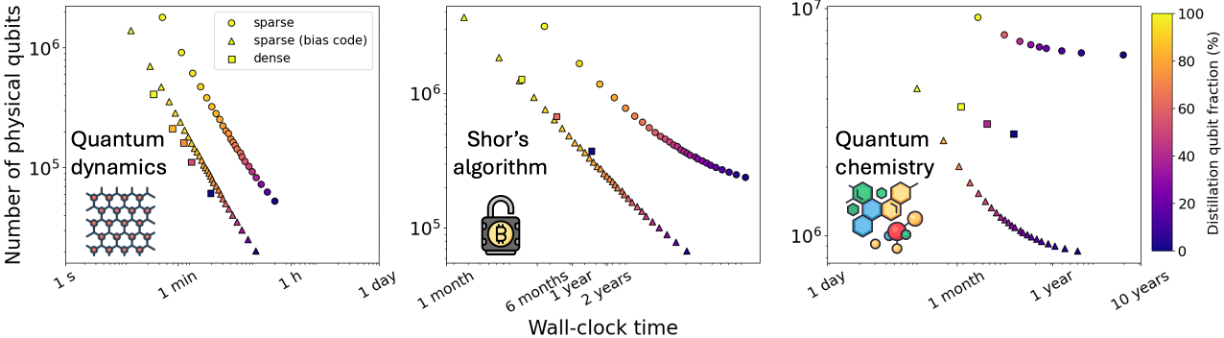}
    \caption{\textbf{Algorithmic resource requirements.} We plot the wall-clock time (horizontal axis) and the number of physical qubits (vertical axis) needed to run three different quantum algorithms. The color map indicates the fraction of the qubits used for MSD. The circles label the sparse SpinBus architecture with the standard surface code; the triangles label the SpinBus architecture with the XZZX surface code; and  the squares label the dense architecture with the standard surface code. The logical operations are implemented with lattice surgery and the `grow-and-distill' MSD protocol is chosen to minimize the space-time overhead.}
    \label{fig:tradeoff}
\end{figure*}

Figure~\ref{fig:tradeoff} shows the physical-qubit footprint of three representative quantum algorithms---quantum dynamics, Shor's factoring (where we follow the logical construction from~\cite{gidney2021factor}), and quantum chemistry---as a function of computation time. We restrict the analysis to the dense architecture operated with the standard surface code and the sparse architecture operated with either the standard or the XZZX surface code.  We demonstrate our framework on these popular and well-studied QEC schemes. However, our techniques are readily adaptable to other QEC schemes, which may reduce overheads. The noise model is described in Sec. \ref{nonmark}. Moreover, we plot data only for lattice-surgery operations and the $15\to1$ MSD protocol. A  detailed comparison between alternative protocols is deferred  Secs. \ref{MSD-in-silicon}. The resource overheads reported in Fig.~\ref{fig:tradeoff} can be considered as estimates of upper bounds: The data are produced using present-day physical parameters (see Tab.~\ref{tab:parameter_summary}) and can be improved by advances in algorithms and optimization techniques as well as by hardware and architecture advances. However, the resource estimates are also somewhat optimistic in that we assume that the demonstration of individual high-fidelity gates and shuttling can be scaled up efficiently to cover the entire hardware. For example, we assume that state-of-the-art shuttling fidelities will improve by an order of magnitude. Such improvements are necessary to enable the implementation of standard scalable QEC codes on shuttling-based processors. Moreover, we consider only first-order errors in the MSD process and ignore the effect of higher-order errors. 

\begin{table}[h]
\centering
\begin{tabular}{@{}lrrlll@{}}
\toprule
Parameter  & Range & \makecell{Default\\ Value} & Units  & Refs \\
\midrule
$T_1$ & 0.01--1 & 0.1 & s & \nocite{stano2022review,Stano2021}\cite{stano2022review,otxoa2025spinhex}\\
$T^*_2$ & 10--1000 & 100 & \unit{\micro\second} & \nocite{stano2022review,Stano2021}\cite{stano2022review,otxoa2025spinhex}\\
$t_{\text{I}}$ & 1--100 & 50 & ns & \cite{wu2025simultaneous,noiri2022fast,mills2022two}\\
$t_{\text{II}}$ & 10--1000 & 225 & ns &  \cite{wu2025simultaneous,noiri2022fast,mills2022two}\\
$t_{\text{readout}}$ & 0.1--10 & 1 & \unit{\micro\second} & \cite{blumoff2022fast, oakes2023fast, connors2020rapid}\\
$t_{\text{init}}$ & 0.01--1 & 0.1 & \unit{\micro\second} & \cite{keith2019single}\\
$\epsilon_{\text{defect}}$ & 0.01--1 & 0.1 & \% & \multicolumn{1}{c}{---}\\
$v_{\text{shuttle}}$ & \multicolumn{1}{c}{---} & 8 & m/s &\cite{kunne2024spinbus}\\
$N_{\text{hops}}$ & 10--100 & 10 & dots &\cite{kunne2024spinbus}\\
$d_{\text{dot}}$ & \multicolumn{1}{c}{---} & 100 & nm &\cite{kunne2024spinbus}\\
$\epsilon_{\text{readout}}$ & \multicolumn{1}{c}{---} & 0.01 & \% & \cite{oakes2023fast,keith2019single}\\
$\epsilon_{\text{shuttle}}$ & \multicolumn{1}{c}{---} & 0.001 & \% & \cite{de2025high} \\

\bottomrule
\end{tabular}
\caption{\textbf{Summary of parameters for silicon hardware.} The ``Range'' column denotes the parameters' typical range observed across experiments; the ``Default Value'' column specifies the values we used to produce the data in Fig.~\ref{fig:tradeoff}. We used the lower and upper values in the range for our optimistic and pessimistic analyses, respectively. $T_1$ and $T_2^*$ denote the relaxation and inhomogeneous dephasing times, respectively. $t_{\text{I}}$ and $t_{\text{II}}$ denote the implementation times for high-fidelity single- and two-qubit gates. $t_{\text{readout}}$ and $t_{\text{init}}$ denote the readout and initialization times. $\epsilon_{\text{defect}}$ denotes the percentage of faulty quantum dots that incure re-routing overheads.  $v_{\text{shuttle}}$ denotes the average electron-shuttling speed and $N_{\text{hops}}$ denotes the average number (plus one) of empty dots  between separated physical qubits in the sparse and patched architectures. $d_{\text{dot}}$ denotes the average quantum-dot separation. $\epsilon_{\text{readout}}$ and $\epsilon_{\text{shuttle}}$ denote the error probabilities of a single-spin readout and a shuttling operation over one quantum dot, respectively. To enable QEC simulations, we used a default value of $\epsilon_{\text{shuttle}}$ ten times smaller than the current state-of-the-art realisations~\cite{de2025high}.}
\label{tab:parameter_summary}
\end{table}

From Fig.~\ref{fig:tradeoff}, several trends emerge. The architecture with dense connectivity and fully local operations exhibit significantly lower space-time overheads than shuttling-based (sparse) layouts. This is intuitive as sparse architectures allocate  additional runtime to qubit shuttling and shuttling-induced errors increase the resources required for error correction. Moreover, across all three applications, the execution time can be reduced by parallelizing MSD at the expense of a larger physical-qubit footprint. Conversely, the spatial overheads can be reduced by limiting the fraction of physical qubits allocated to MSD. These findings are consistent with the observation that MSD dominates the resource cost in the fault-tolerant architectures that we study.
Whilst the dense architecture clearly outperforms the sparse one, practical wiring issues, etc., place the actual production of a dense platform well beyond current capabilities. It is, therefore, re-assuring to see that significant resource reductions can be obtained by means of improved and hardware-tailored software. We demonstrate this with our simulations of the sparse architecture with the XZZX code. Unlike the standard surface code, the XZZX code's performance benefits from the biased nature of noise, inherit to Silicon spin qubits. The use of the XZZX code significantly improves the space-time overheads of all three algorithms. 

In Sec. \ref{MSD-in-silicon}, we analyze the aforementioned trends in more detail.  We demonstrate how variations in hardware parameters directly affect the MSD resource overhead. We also introduce several techniques to reduce this overhead, including pulse-level optimization and the use of biased QEC codes.  We benchmark these approaches across different hardware architectures and consistently observe that dense layouts outperform sparse architectures in terms of resource overhead. In addition, we compare the performance of different QEC codes and MSD protocols. For example, we demonstrate the resource advantage of the $15\to1$ distillation protocol over the $5\to1$ protocol.

\section{Background}
\label{sec: background}

In this section, we first introduce our architectural models. Then we briefly outline the QEC codes we investigate. Finally, we summarize the MSD protocols used for benchmarking.
\subsection{Architectural Motivation}
\label{arch}
To structure our resource analysis, we focus on three architectural regimes that span reasonable designs of silicon spin-qubit processors. The three architectures are not intended as an exhaustive list of possible blueprints, but as representative designs aimed at addressing the key hardware tradeoffs of connectivity, shuttle overhead, wiring complexity, and fidelity optimization. The SpinBus architecture~\cite{kunne2024spinbus, escofet2025quantum, struck2024spin, langrock2023blueprint, losert2024strategies, seidler2022conveyor, xue2024si, volmer2024mapping}  captures a sparse, shuttle-dominated regime closest to near-term experimental demonstrations and highlights the cost of long-range transport. The patched architecture represents an intermediate design in which local dense regions are embedded within a globally sparse design, enabling efficient local error correction while requiring shuttling for logical operations. This layout is inspired by Ref.\cite{vandersypen2017interfacing} where shuttling connectivity is advocated to reduce the hardware burden of readout and power dissipation, \textit{etc}. Finally, the dense architecture serves as an optimistic upper bound on future silicon-hardware architectures. In the dense architecture, all spin qubits are coupled via nearest-neighbor connectivity, and the need for shuttling is eliminated. The performance of the dense and patched architectures may be further enhanced by reducing operation times with pulse-level optimization (Sec. \ref{sec:pulse_optimization}). By producing resource estimates for these three designs, we can systematically assess the impact of the hardware architecture on silicon quantum computing. 

\subsubsection{SpinBus architecture: sparse, shuttling-mediated}

\begin{figure*}[htb!]
    \includegraphics[width=\linewidth]{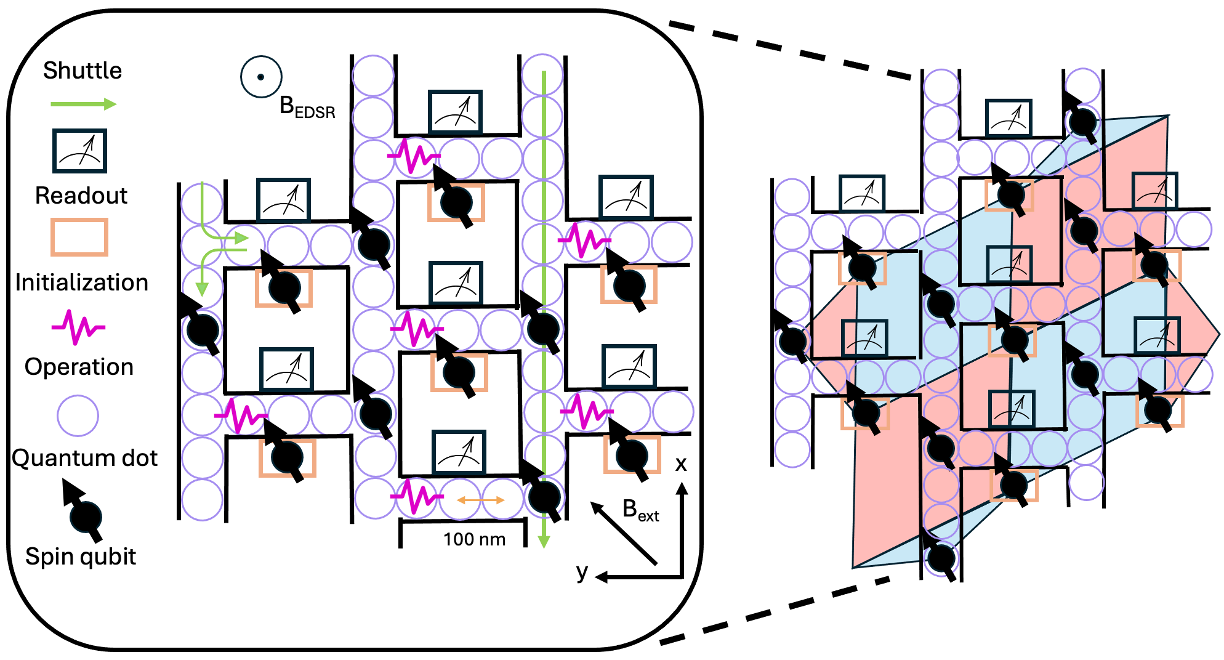}
    \caption{\textbf{Sparse SpinBus architecture.} (Left) Hollow circles represent quantum dots and dark dots represent physical spin qubits. The parallel lines represent shuttling channels.  An external magnetic field is applied in-plane in the diagonal direction to avoid bias in the $x$- and $y$-directions. Single-qubit operations are enabled by the  magnetic field perpendicular to the plane. Qubit initializations, operations, and readout are conducted in their designated manipulation zones. (Right) Mapping of a rotated surface code onto the sparse architecture. }
    \label{fig:spinbus}
\end{figure*}

The small physical footprint of silicon spin qubits is both a blessing and a curse. In principle, a small qubit size allows for a large number of qubits to be integrated on the same chip and fitted into a reasonably sized dilution refrigerator. However, if electron-spin qubits are packed densely on a chip, there are major implications for the possibility of adequate wiring of the control electronics. 
The \textbf{SpinBus} architecture~\cite{kunne2024spinbus, escofet2025quantum, struck2024spin, langrock2023blueprint, losert2024strategies, seidler2022conveyor, xue2024si, volmer2024mapping} was designed to overcome these difficulties. The architecture relies on a sparse connectivity regime in which several empty quantum dots sit between quantum dots occupied by electron-spin qubits. To facilitate computation, the electrons are coherently transported along shuttling lanes that connect regions with initialization and readout zones, as in Fig.~\ref{fig:spinbus}. Typical distances between dots are on the order of \mbox{$\sim 10$--$100~\unit{\nano\meter}$}, while straight shuttling lanes span lengths of $L\sim1~\unit{\micro\meter}$. This corresponds to approximately $N_{\rm hops}\approx 10$ discrete movements over quantum dots per connectivity lane. Assuming a coherent shuttling velocity of $v\approx 8~\unit{\meter\per\second}$, the transit time for a single lane is $t_{\rm lane}=L/v\approx 125~\unit{\nano\second}$, or $t_{\rm step}\approx 12.5~\unit{\nano\second}$ per hop~\cite{kunne2024spinbus}.

Experimental results  suggest that coherent spin-qubit transfer over one hop can be achieved with an overall infidelity on the order of $10^{-4}$ \cite{de2025high}. Optimistically, we set the per-dot shuttling error to be $\epsilon_{\rm shuttle} \sim 10^{-5}$. When shuttling around corners or T-junctions, the incurred infidelity rises by roughly a factor of 4~\cite{kunne2024spinbus}. In the SpinBus architecture, physical single-qubit operations are performed locally in manipulation zones using electric-dipole-spin-resonance (EDSR) pulses \cite{kunne2024spinbus, escofet2025quantum}. Physical two-qubit gates are typically implemented via exchange interactions between physical qubits that have been shuttled to neighboring quantum dots. A typical stabilizer cycle (see Sec. \ref{syndrom}), therefore, consists of ancillae initializations, multiple shuttling segments interleaved with local interactions, and a return shuttle that enables measurements in the readout zones.

\subsubsection{Patched architecture: semi-dense layout}

\begin{figure}
    \centering
    \includegraphics[width=\linewidth]{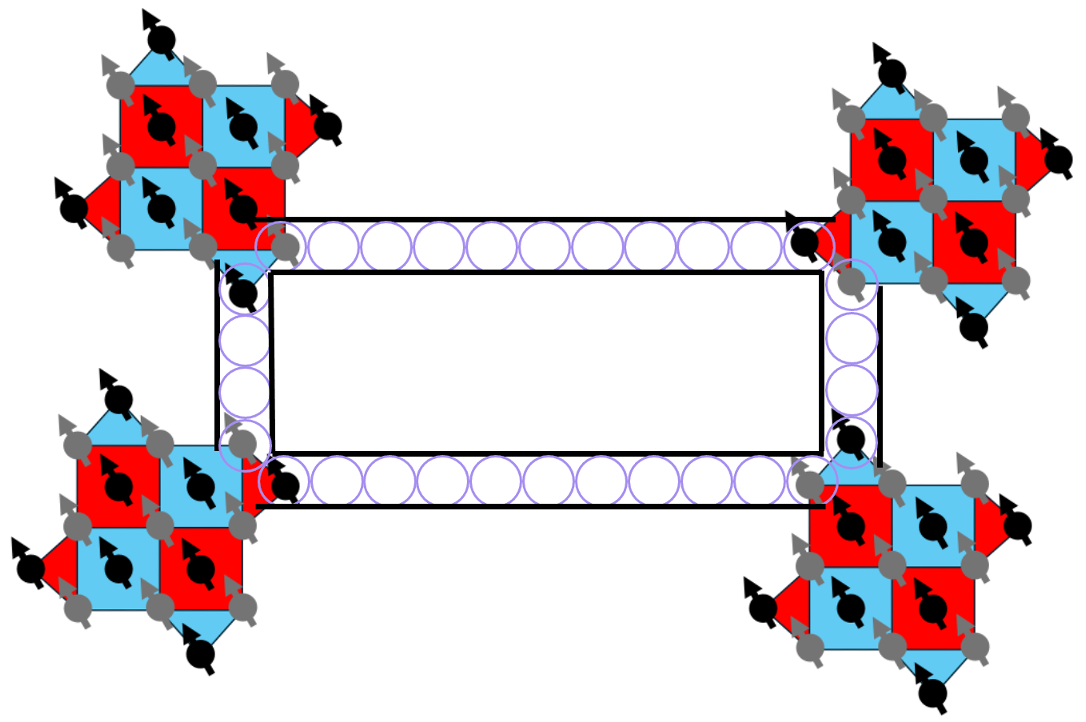}
    \caption{\textbf{Patched architecture.} $m$ logical qubits (here, we show $m=1$ logical qubits at distance 3) are encoded onto the densely connected physical qubits within each patch. The logical qubits are interconnected via shuttling channels.}
    \label{fig:patched}
\end{figure}

The \textbf{patched} architecture in Fig.~\ref{fig:patched} interpolates between sparse and dense regimes by partitioning the processor into locally dense patches connected by a reduced number of shuttling lanes~\cite{vandersypen2017interfacing}. Within each patch, physical single- and two-qubit operations are executed locally using EDSR and exchange coupling, respectively. The operations implemented on a specific patch can be compressed in time using pulse-level optimization (Sec. \ref{sec:pulse_optimization}). In this work, we consider the encoding of one or two logical qubits per patch. The number of physical qubits we consider per patch thus ranges from 17 (in a distance 3 rotated surface code) to 841 (in a distance 21 unrotated surface code). During the MSD process, the patch size is fixed at the number of physical qubits needed to implement the logical qubit with the largest code distance.

\begin{figure*}[htb!]
    \centering
    \includegraphics[width=\linewidth]{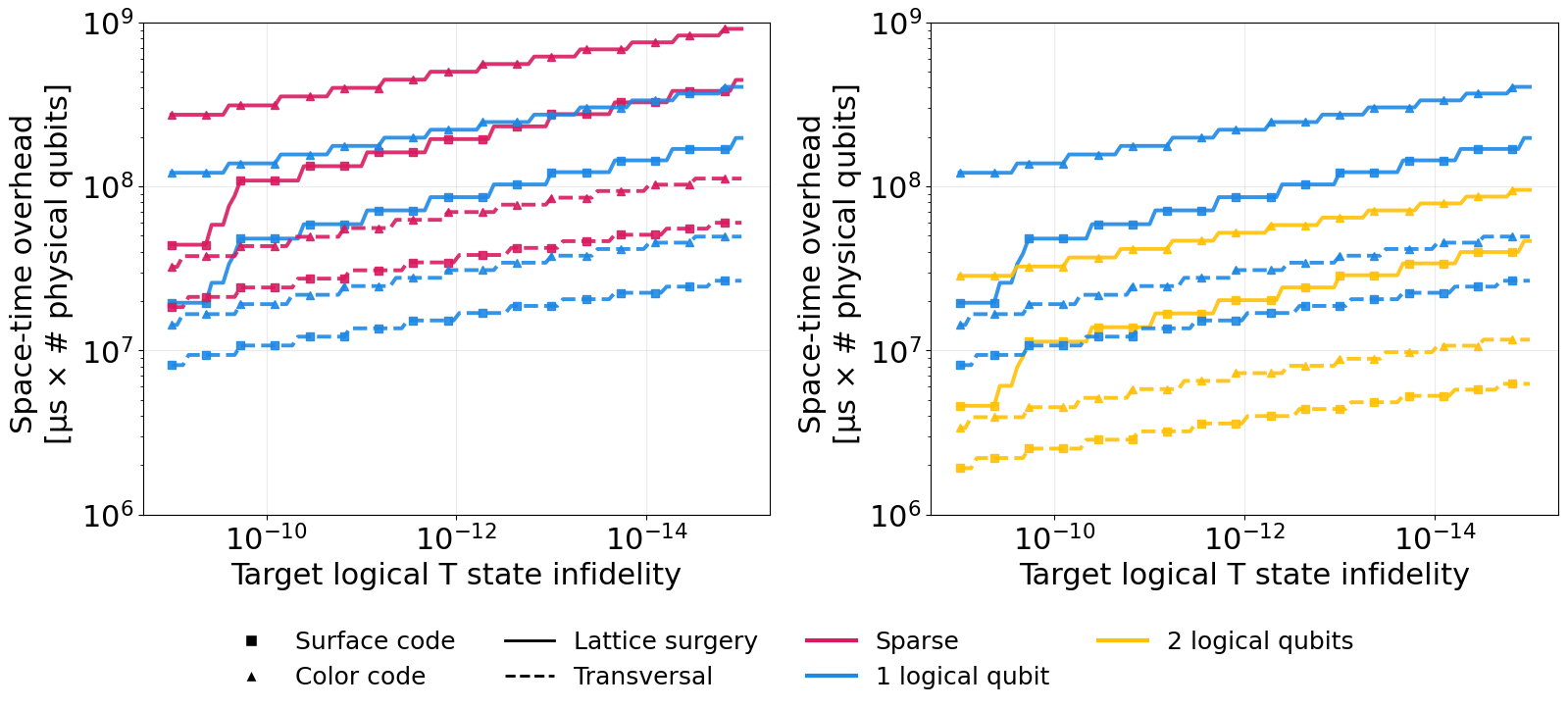}
    \caption{\textbf{Space-time volume required for one MSD factory.} Space-time overhead reduction relative to baseline (sparse layout). We study sparse (separated) physical qubits and patched architectures where 1 or 2 logical qubits are one the same dense patch and direct pulse operations generate the desired evolutions. On the vertical axis we have the space-time overhead in unit of microseconds times the number of physical qubits.}
    \label{fig:ost_reduction_pf}
\end{figure*}

Inter-patch operations, including long-range logical interactions and MSD-factory routing, rely on SpinBus shuttling. Moreover, the separation of local and global operations simplifies modeling: local gates are treated as directly-connected operations subject to pulse compression, while inter-patch transfers incur the same shuttling penalties as in the sparse architecture. 
Thus, the patched architecture, significantly reduces the average shuttling overheads while retaining many of the wiring and integration advantages of sparse layouts. For example, stabilizer cycles require substantially fewer shuttling operations compared to the sparse architecture. A comparison between the space-time volume needed for one MSD factory in the sparse and patched architectures is shown in Fig.~\ref{fig:ost_reduction_pf}. Clearly, the resource benefits of the patched architecture increases as the fraction of physical qubits with a dense connectivity (or the number of logical qubits per patch) increases.

\subsubsection{Dense architecture: good connectivity, bad wiring}

\begin{figure}[htb!]
    \centering
    \includegraphics[width=\linewidth]{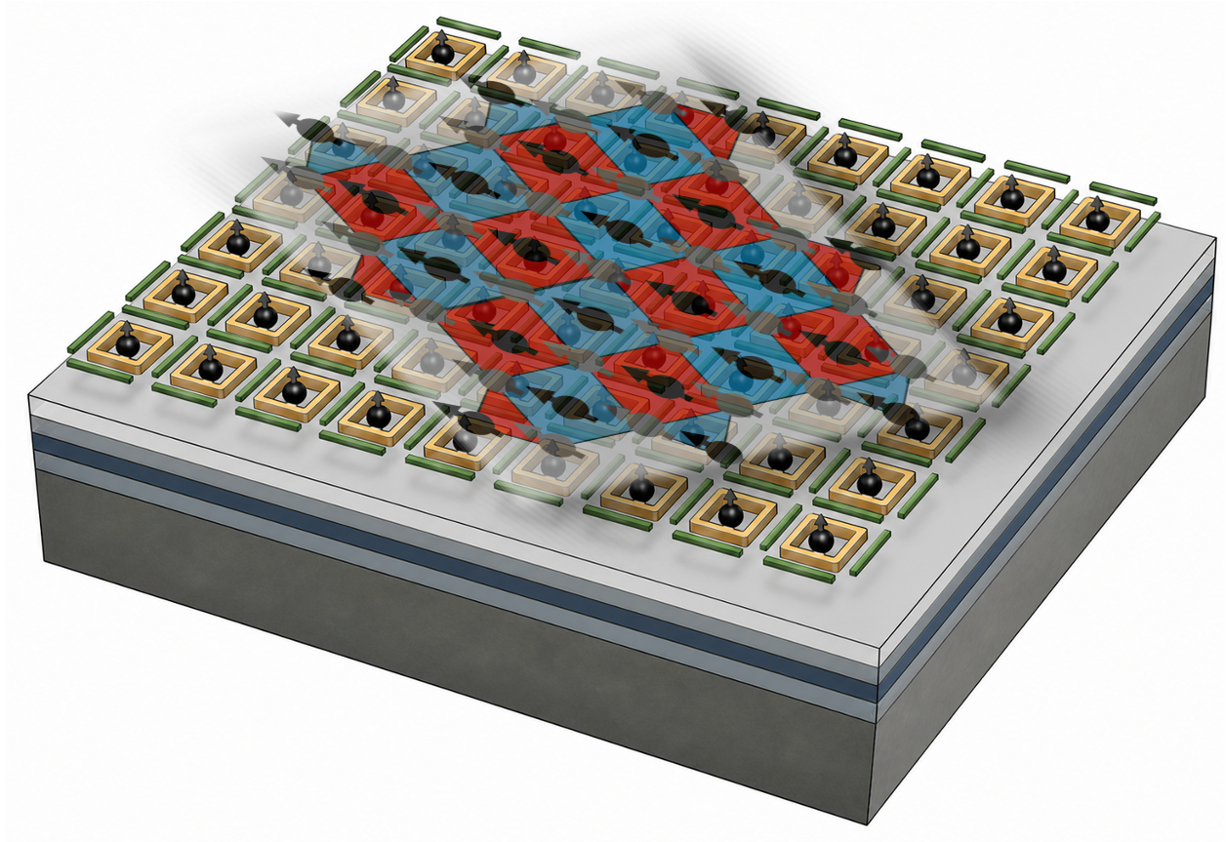}
    \caption{\textbf{Dense architecture.} A schematic of the silicon spin-qubit quantum processor. A global external magnetic field, which generates a Zeeman splitting, lies along the 2D array of the qubits. Plunger (brown square plates) and barrier (green outlines) gates create an electronic potential which traps individual electrons in the potential wells of gate-defined quantum dots.  The Hamiltonian description of the device is given in Eq. \eqref{eq:H_full}. For illustrative purposes, a small portion of surface code is projected on top of the array.}
    \label{fig:dense}
\end{figure}

The \textbf{dense} architecture represents a hardware-optimistic operating regime in which silicon spin qubits are arranged in a two-dimensional nearest-neighbor lattice with native exchange coupling available between all adjacent sites~\cite{veldhorst2017silicon}. See Fig.~\ref{fig:dense}. In the dense architecture, all logical operations---syndrome extraction, lattice surgery, and MSD---can be executed locally without explicit shuttling. Pulse-level optimization (Sec. \ref{sec:pulse_optimization}) can be applied directly to the underlying Hamiltonian controls used for all physical operations. Consequently, all physical operations can be implemented in times near the minimal evolution time in silicon structures~\cite{long2025minimal}. This substantially suppresses errors, making the dense layout the theoretically ideal setting for high-quality-output MSD factories.

However, from an experimental perspective, the dense architecture is often regarded as hypothetical. This is predominantly related to the fact that silicon spin qubits are far smaller than the spatial footprint of control and readout electronics. If the number of control lines needed per physical qubits is constant with the size of the processor, then the dense architecture is not scalable to the size needed for fault-tolerant computing. The number of interconnects and cryogenic I/O channels~\cite{boter2022spiderweb} grows too rapidly and quickly exceeds practical limits set by chip area and packing constraints~\cite{franke2019rent}. To implement a dense silicon-qubit architecture one must address the wiring issues that currently limit the scalability of solid-state quantum processors. 

Proposed solutions to these issues rely on multiplexing of control lines and sparse readout strategies that trade spatial density for wiring efficiency~\cite{franke2019rent}. While such approaches can dramatically reduce the effective wiring overhead, there are yet no experimental realizations at the scales needed for fault-tolerant quantum computing. Moreover, the proposed solutions introduce additional layout complexity, power dissipation, and control serialization. Consequently, we view the dense architecture as a forward-looking benchmark rather than as a near-term design target.

In our analysis, the dense layout serves two purposes. First, it provides an architectural best-case reference point against which the costs imposed by shuttling, routing, and operations in more practical architectures can be quantified. Second, it allows us to isolate and study the benefits of pulse-level compilation in the absence of shuttling overhead.

\subsection{Quantum error correction and thresholds}
\label{qec}
Quantum error correction (QEC) codes protect quantum information by encoding a number of logical qubits onto a larger number of  physical qubits and repeatedly detecting and correcting errors through measurements~\cite{devitt2013quantum}. A common class of QEC codes is the stabilizer codes.

Typically, a stabilizer code is defined by a set of commuting (Pauli) operators acting on subsets of the \textit{data qubits} (which hold the logical information. The protected code space corresponds to the simultaneous $+1$ eigenspace of these stabilizers. Experimentally, the stabilizer observables are measured indirectly using \textit{ancilla} qubits that couple to the data qubits through controlled operations. Measurements of the ancilla qubits produce outcomes corresponding to the stabilizer eigenvalues, from which an error syndrome can be extracted. Each round of syndrome extraction produces an error syndrome, which is decoded on classical hardware. The decoded syndrome specifies how to correct the physical errors. Repetition of this process suppresses the accumulation of errors and stabilizes the logical qubits over times exceeding the coherence of the underlying hardware.

Two central concepts for any QEC code are the \textit{distance} $d$ and the \textit{threshold error rate} $p^*$~\cite{steane2003overhead}. The code distance is defined as the smallest number of physical qubits that must be corrupted to cause an undetectable logical error. Increasing the code distance at the expense of more physical qubits suppresses the logical-error probability exponentially. 
However, a QEC code suppresses errors only when the physical error rate $p$ is below the threshold $p^*$. Then, the logical error rate $p_L$ is approximated by
\begin{equation}
    p_L \approx A \left( \frac{p}{p^*} \right)^{(d+1)/2} .
    \label{eq:pL}
\end{equation}
Here, $A$ is a code- and decoder-dependent constant. While Eq. \eqref{eq:pL} provides only an approximation which may break under certain noise profiles, it does highlight the hardware-parameter cutoff ($p<p^*$) for which fault-tolerant quantum computation is possible.

Among two-dimensional stabilizer codes, the surface code~\cite{fowler2012surface} and the color code~\cite{lacroix2025scaling, thomsen2024low, chamberland2020triangular} are leading candidates for solid-state platforms. In the unrotated surface code, a distance-$d$ logical qubit is encoded using $d^2$ physical data qubits arranged on a square lattice. Surface codes combine strictly local connectivity, high thresholds, and efficient decoding, which makes them well suited for current hardware. Triangular color codes encode one logical qubit using $(3d^2-1)/4$ physical data qubits, achieving a high encoding rate at the cost of larger stabilizer weights and typically small values of $p^*$. (The stabilizer weight is the number of qubits on which a stabilizer operator acts nontrivially.) Higher-weight stabilizers typically enforce stronger parity constraints across larger regions of the QEC code's lattice. However, high-weight stabilizers require ancilla qubits to interact with more data qubits during the syndrome extraction. This leads to more opportunities for gate errors to propagate to multiple qubits (\textit{e.g.,} hook errors~\cite{fujiu2026dense}).

\subsection{Biased codes}
\label{bias qec}
The XZZX surface code is a variant of the conventional surface code. It offers remarkable performance under  biased noise where one type of Pauli error occurs more frequently than others~\cite{bonilla2021xzzx, roffe2023bias, tuckett2018ultrahigh, tuckett2019tailoring, san2023cellular, tuckett2020fault}. The noise profile of semiconductor spin qubits is precisely of such a form, with dephasing noise being the dominant error mechanism.
The XZZX code is locally equivalent to the standard surface code~\cite{bonilla2021xzzx} but differs by a Hadamard rotation on alternate qubits, resulting in stabilizer operators that are the product XZZX of Pauli operators around each face of a square lattice of physical qubits. 

To summarize the performance of the XZZX code, we recall the general single-qubit Pauli noise channel:
\begin{equation}
\mathcal{E}(\rho) = (1 - p)\rho + p\left(r_X X\rho X + r_Y Y\rho Y + r_Z Z\rho Z\right),
\end{equation}
where $p\in [0,1]$ is the error probability and $\mathbf{r} = (r_X,r_Y,r_Z)$ is a positive vector, with unit $L_1$ norm, that characterizes the noise profile. For (Z-biased) biased noise, the bias parameter  is defined by
\begin{equation}
\eta = \frac{r_Z}{r_X + r_Y},
\end{equation}
with $r_X = r_Y$, such that $\eta = 1/2$ corresponds to depolarizing noise and $\eta \to \infty$ corresponds to completely Z-dominated noise. Following the result from~\cite{bonilla2021xzzx}, where the logical error rate is damped by a factor of $\sim \eta^{-d/4}$, we assume the following model:
\begin{equation}
p_L \approx A_z \left(\frac{p}{p^*_z}\right)^{(d_Z+1)/2}+ A_x \left(\frac{p}{p^*_x}\right)^{(d_X+1)/2}
\end{equation}
with $d_X$ and $d_Z$ being the code distance in the logical X and Z directions, respectively. In silicon electron-spin processors, the Z errors dominate the X errors. 
By splitting up the logical error rate into the X direction and the Z direction, one can allocate fewer resources (a smaller code distance and thus fewer physical qubits) to the X direction. This amounts to constructing a rectangular surface code where the logical X direction has a shorter length. 

Spin qubits exhibit strongly asymmetric error channels where dephasing dominates over bit-flip processes~\cite{messinger2025fault, otxoa2025spinhex}. As a result, Pauli-Z errors occur at a higher rate than Pauli-X or Pauli-$Y$ errors. This leads to a physical noise bias parameter reaching values $\eta \sim 10^2$--$10^3$ in experimentally relevant regimes~\cite{messinger2025fault, otxoa2025spinhex}. The advantage of the XZZX code relies critically on maintaining the physical noise bias throughout the full fault-tolerant circuit. That is to say, if the implemented gate set is \textit{bias-preserving}, meaning that it maps dominant Z errors to Z errors without symmetrizing the noise channel, then the effective bias $\eta_{\mathrm{eff}}$ remains large and the threshold continues to increase with $\eta$. By contrast, when operations are not bias-preserving, such as when Hadamard gates or CNOTs mix X and Z error channels, the effective noise seen by the code becomes more isotropic. In this case, even if the physical qubits exhibit large $\eta$, the circuit-level noise bias saturates around a factor of 5~\cite{etxezarreta2026leveraging}. Consequently, the fault-tolerant threshold does not increase indefinitely with physical bias. 

\begin{figure*}
    \centering
    \includegraphics[width=\linewidth]{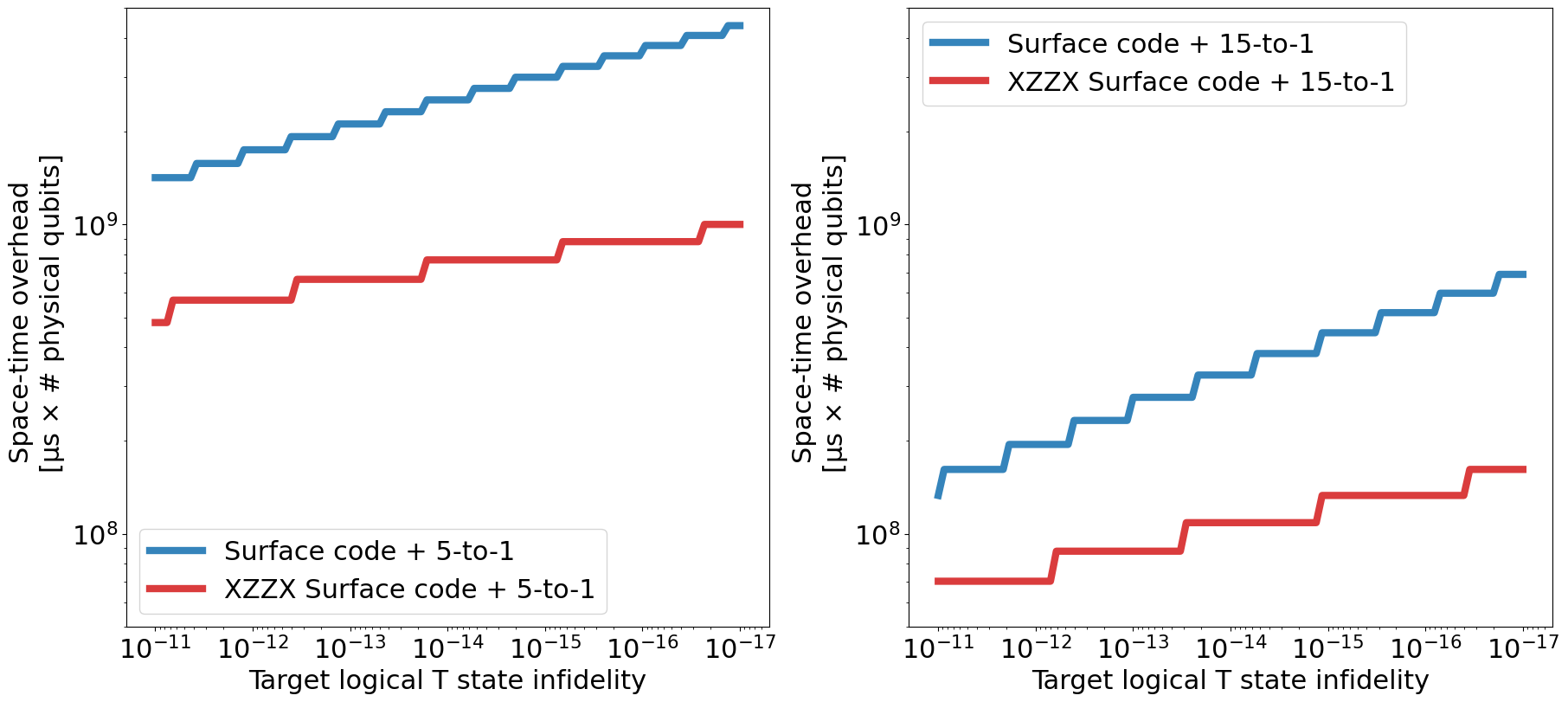}
    \caption{\textbf{Resource cost of one magical state with and without biased surface codes.} We plot the resource cost, in terms of the space-time volume (vertical axis, in unit of microseconds times the number of physical qubits) needed to generate one logical magic state at the targeted infidelity (horizontal axis). The standard and XZZX surface codes are plotted in red and blue, respectively. The left panel analyzes the $5\to1$ protocol and the right one the $15\to1$ protocol.}
    \label{fig:bias_overhead}
\end{figure*}

To illustrate the potential benefits of utilizing bias-preserving QEC codes on silicon hardware, we simulate and compare the space-time volume incurred by MSD factories operating with the unrotated surface code and with the XZZX-surface code. We limit our analysis to the sparse architecture. (Similar analyses for the patched and the dense architectures would require the investigation of bias-preservation through pulse optimization.)  Our results are presented in Fig.~\ref{fig:bias_overhead}. Our analysis shows a clear advantage for the XZZX code for both  $5\to1$ and $15\to1$ MSD factories (as defined below). For high-fidelity magic-state production, the improvement is roughly by a factor of 3 to 5.

\subsection{Syndrome checks}
\label{syndrom}

\begin{figure}[htb!]
    \centering
    \includegraphics[width=\linewidth]{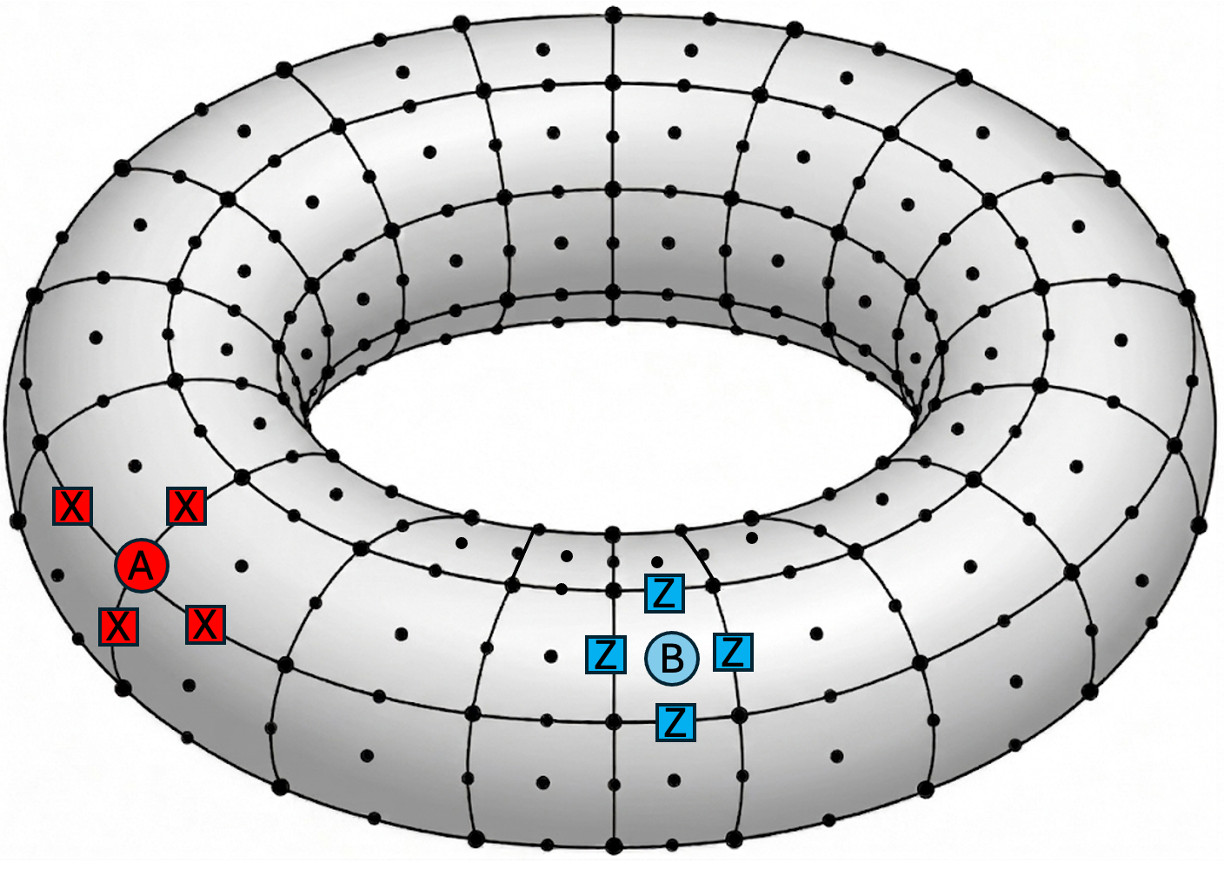}
    \caption{\textbf{Illustration of surface code on a torus.} An illustration showing the implementation of a 2D surface code on a toroidal geometry. Two stabilizer components are highlighted: a red X-type vertex operator ($A_v$) at a specific vertex A, and a blue Z-type plaquette operator ($B_p$) on a specific plaquette B.}
    \label{fig:torus}
\end{figure}

Next, we summarize the syndrome-check component of QECs, illustrating this with the standard surface code. The surface code is a two-dimensional stabilizer code defined on a square lattice of data qubits. Each edge of the lattice hosts a physical qubit, while the lattice's nodes (``vertices'') and faces (``plaquettes'') provide the locations at which stabilizer checks are applied, as shown in Fig.~\ref{fig:torus}. The code performs X-type checks on vertices and Z-type checks on plaquettes. The logical states are the simultaneous +1 eigenstates of a set of local stabilizer operators. The stabilizer group is generated by mutually commuting operators of the form 
\begin{equation}
    A_v = \prod_{j \in \mathrm{adjacent}(v)} X_j, 
    \qquad 
    B_p = \prod_{j \in \mathrm{plaquette}(p)} Z_j .
\end{equation}
$\mathrm{adjacent}(v)$ denotes the set of data qubits adjacent to vertex $v$ and $\mathrm{plaquette}(p)$ denotes the qubits surrounding plaquette $p$. For the unrotated surface code, these stabilizers typically have weight four inside the bulk of the device: every qubit is connected to four other qubits.

\begin{figure}[htb!]
\centering
\begin{quantikz}[row sep=0.3cm, column sep=0.35cm]
\lstick{$\ket{\psi_1}$} & \ctrl{4} & \qw  & \qw  & \qw  & \qw \\
\lstick{$\ket{\psi_2}$} & \qw & \ctrl{3}  & \qw  & \qw  & \qw \\
\lstick{$\ket{\psi_3}$}  & \qw  & \qw & \ctrl{2} & \qw  & \qw\\
\lstick{$\ket{\psi_4}$}  & \qw   & \qw  & \qw  & \ctrl{1}& \qw\\
\lstick{$\ket{0}_{\text{ancilla}}$}    & \targ{} & \targ{}  & \targ{}  & \targ{}  & \meter{}
\end{quantikz}
\caption{\textbf{Syndrome extraction.} A circuit for QEC syndrome extraction using a Z-type plaquette stabilizer $B_p = Z_1 Z_2 Z_3 Z_4$ in the surface code.}
\label{fig:surface_z_check}
\end{figure}
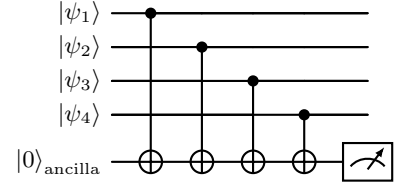

Syndrome extraction is performed by coupling each stabilizer generator to a dedicated ancilla qubit and then measuring that ancilla. Consider a Z-type plaquette stabilizer $B_p = Z_1 Z_2 Z_3 Z_4$. The corresponding ancilla is initialized in the state $\ket{0}_a$ and sequentially interacts with the data qubits via CNOT gates. The data qubits act as controls and the ancilla as the target. A computational-basis measurement of the ancilla yields a syndrome bit $s \in \{0,1\}$ such that
\begin{equation}
    (-1)^s = \bra{\psi} B_p \ket{\psi}.
\end{equation}
This procedure projects the data qubits onto an eigenstate of $B_p$ without revealing or altering any information about the encoded logical state.
The corresponding quantum circuit for a single plaquette measurement is shown in Fig.~\ref{fig:surface_z_check}. An X-type star stabilizer $A_v$ is measured analogously but with the ancilla initialized as $\ket{+}_a$.

The execution of these stabilizer checks across the lattice yields a syndrome vector $\bm{s}$ and defines a single syndrome-extraction round.  This process can be parallelized. Repeating the process over time yields a space-time syndrome history $\{\mathbf{s}(t)\}$. This history is decoded to infer the most likely configuration of physical errors with regard to spatial and temporal correlations. The decoded information is then used to construct error-correcting operations. 

\subsection{Logical operations: lattice surgery and transversal gates}\label{operation}

Logical operations between encoded qubits can be implemented using two broadly applicable methods: lattice surgery and transversal gates. These approaches differ in their connectivity requirements, time overheads, and suitability for different hardware platforms~\cite{gutierrez2019transversality}.

Lattice surgery is a measurement-based technique with only local interactions~\cite{fowler2012surface, landahl2014quantum}. Logical operations are performed by temporarily merging and splitting code patches and measuring joint stabilizers across their boundaries. For surface codes, a logical two-qubit gate implemented via lattice surgery typically requires $\mathcal{O}(d)$ rounds of syndrome extraction, making its space-time cost proportional to $\mathcal{O}(d^3)$. While lattice surgery is flexible and hardware-efficient, its sequential execution can limit the processor's clock speed, especially in resource-intensive protocols such as magic-state distillation.

Transversal gates, by contrast, apply logical operations through parallel physical gates between corresponding qubits in two code blocks~\cite{sahay2025error}. When available, transversal implementations can directly realize logical Clifford operations (and, in some codes, non-Clifford gates) with a constant number of syndrome extraction rounds independent of $d$~\cite{bravyi2005universal}. The primary advantage of transversal operations is their low temporal overhead and high compatibility with parallelization. On the negative side, the implementation of transversal gates generally requires either long-range connectivity or the ability to dynamically reposition qubits.

\subsection{Magic-state distillation}\label{msd}

Here, we review the workings of MSD protocol. We summarize a surface code's performances with $[[N,k,d]]$, where $N$ represents the total number of physical qubits on a QEC block, $k$ represents the number of logical qubits that are protected within this block, and $d$ represents the code distance. This triplet of parameters  characterizes the essential resources and power of the code, with larger $N$ indicating more physical overhead, larger $k$ indicating more encoded information, and larger $d$ indicating stronger error protection.

The canonical example of a MSD protocol is the $15\to1$ protocol, originally introduced by Bravyi and Kitaev~\cite{bravyi2005universal}. This protocol is based on the $[[15,1,3]]$ Reed-Muller code and takes 15 noisy $T$-type  magic states (\textit{i.e.,} non-stabilizer states $\ket{T}=(\ket{0} + e^{i \pi /4} \ket{1})/\sqrt{2}$) as input to produce a single output magic state of higher fidelity. The protocol suppresses the output error rate from $p$ to $\mathcal{O}(p^3)$. On the one hand, the protocol is resource intensive: it requires large qubit and circuit overheads. On the other hand, the protocol quickly improves the fidelity of the magic states.

An alternative approach is the $5\to1$ MSD protocol, which is based on the $[[5,1,3]]$ perfect stabilizer code~\cite{ryan2022implementing, laflamme1996perfect, gong2022experimental, chamberland2022circuit,litinski2018lattice}. This protocol consumes five noisy magic states and outputs one improved state. The protocol detects single-qubit errors and implements a quadratic error suppression: $p \mapsto \mathcal{O}(p^2)$. Compared to the $15\to1$ protocol, the $5\to1$ protocol requires fewer physical qubits and uses a simpler circuit, making it an attractive candidate for early distillation demonstration~\cite{sales2025experimental}. Alternative codes and optimized protocols can substantially reduce the associated overhead for MSD~\cite{zhang2025constant, yoshida2025concatenate, ruiz2025unfolded, wu2025bias, goto2014step}. In recent years, a growing body of work has demonstrated improved approaches~\cite{erew2025pre,fazio2025low,choi2023fault,gavriel2022transversal}, including code-specific optimizations~\cite{butt2024fault,daguerre2025code,zhang2024facilitating,yoshida2025low, lee2025low} and magic-state cultivation techniques~\cite{hirano2025efficient,chen2025efficient,vaknin2025magic,rosenfeld2025magic,gidney2024magic,itogawa2024even}. These techniques mitigate or circumvent conventional MSD costs. Nevertheless, we leave the study of such protocols for future work. 
 
An important practical consideration in MSD is a protocol's \textit{rejection rate}. This rate is the probability that a distillation attempt is discarded due to the detection of an error during stabilizer measurements. For small input error rates ($p \ll 1$), the rejection probability scales linearly with $p$ to leading order. However, the constant of proportionality is larger for protocols with a higher rates of error suppression. A detailed quantitative analysis of acceptance probabilities, rejection-induced overheads, and their dependence on architecture-specific noise models is deferred to later sections.

\section{Methodology}
\label{sec: methodology}

Our resource-estimation analysis is based on a theory-to-hardware pipeline that transforms a target algorithmic application into a silicon-processor-specific instruction set. By calculating the total overhead of that instruction set, we estimate the experimental resources required for the target application's implementation. Further, we produce quantitative resource metrics  for specific QEC codes and MSD protocols. In particular, we study the space-time volume, physical-qubit overhead and wall-clock runtime of quantum-computing subroutines. We also invert the analysis to calculate the experimental device performances needed to implement a subroutine within a fixed resource budget. In particular, we provide bounds on the single- and two-qubit gate fidelities, the coherence times, and the gate-implementation times needed to realize a given logical magic-state fidelity within a prescribed space-time budget.

The pipeline of the resource estimation is described in Sec.~\ref{intro}. In the remainder of this section, we describe our device modeling and pulse-optimization techniques (Sec.~\ref{sec:pulse_optimization}) and detail our results (Sec.~\ref{sec:memory_expt}). In the next section, we describe our noise model and our techniques for evaluating MSD overheads.

\subsection{Pulse optimization: setup}
\label{sec:pulse_optimization}

To transform an algorithm's logical implementation into an experimental instruction set, one must model the underlying hardware. 
One way to produce a clock-time estimate of an algorithm is to simply add up single- and two-qubit gate times according to experimental tables. However, real hardware implements operations using control pulses. If the physical qubits are directly connected (as in dense and patched architectures), these pulses can be optimized (in time and fidelity) to implement a number of gates in one go. 
Thus, pulse optimization can dramatically lower the temporal resources needed to implement quantum circuits. Pulse-level optimization allows circuits, like syndrome checks, to approach their minimal durations by exploiting the full native Hamiltonian rather than a discrete gate set. This can reduce idling errors in QEC cycles and the space-time volume of distillation factories. Moreover, pulse-optimized implementations naturally incorporate hardware constraints making the modeling  more realistic compared to hardware-agnostic studies~\cite{bolsmann2025fast,old2025fault, tasler2025optimizing}.
We incorporate pulse optimization in our analysis of the dense architecture and inside the patched architecture's dense regions. For the sparse layout, electron shuttling is used for qubit transport, which inhibits acceleration via pulse optimization.

There are several ways to simulate the implementation of quantum gates on semiconductor spin qubits \cite{burkard2023semiconductor, ArvidssonShukur17, Lepage20}.
We build our analysis around an effective Heisenberg model of a spin lattice \cite{burkard2023semiconductor}.  In particular, we describe the dynamics of an $N$-qubit array using a driven Heisenberg Hamiltonian \cite{burkard2023semiconductor}:
\begin{align}
H(t) &= -\frac{1}{2}\sum_{i=1}^N B_i\sigma^{(i)}_z 
-\frac{1}{2} g(t)\sum_{i=1}^N \sigma^{(i)}_x\nonumber
\\
&+ \sum_{1\le i<j\le N}\frac{J_{ij}(t)}{4}\sigma^{(i)}\cdot\sigma^{(j)},
\label{eq:H_full}
\end{align}
where $i,j$ label qubits and $\sigma^{(i)}=(\sigma_x^{(i)},\sigma_y^{(i)},\sigma_z^{(i)})$. The first term corresponds to the Zeeman splitting induced by the external magnetic field; the second term is the transverse microwave control, written in quadrature as
\begin{equation}
g(t)=\sum_{i=1}^N\big[I_i(t)\cos\omega_i t + Q_i(t)\sin\omega_i t\big];
\end{equation}
and the third term modulates the exchange couplings with $J_{ij}(t)$. Herein, we consider both linear and 2D arrays. We enforce this connectivity by fixing $J_{ij}(t)=0$ for unconnected pairs of qubits. We define our qubits in the rotating frame of the first term. Using this Hamiltonian, we can emulate the effect of hardware pulses on the spin qubits using a Suzuki-Trotter algorithm~\cite{Berry2005Aug, Suzuki1990Jun} adapted to the Heisenberg Hamiltonian as in Refs.~\cite{long2025minimal,long_2025_17116352,Henrik}. Our emulator employs the rotating wave approximation; thus, our simulations are agnostic to the value of $\frac{1}{N}\sum_{i=1}^NB_i$ and only depend on the differences $B_{i+1}-B_i=\Delta B=10~\unit{\mega\hertz}$ \cite{Huang2024Mar}. Further,  we bound $0\le J_i(t)\le J_{\max}=10~\unit{\mega\hertz}$ \cite{Tanttu2024Nov, philips2022universal}, $\left|I_i(t)\right|\le I_{\max}$, and $\left|Q_i(t)\right|\le Q_{\max}$, where $I_{\max}=Q_{\max}=4~\unit{\mega\hertz}$, in accordance with state-of-the-art experimental data \cite{stano2022review}\nocite{Stano2021}.

Our emulator tailors hardware pulses to implement a desired gate. Using a classical optimizer, we can compress these pulses until they reach the silicon hardware's minimal evolution time (MET), that is, the smallest time within which the hardware can implement a target gate within a target fidelity~\cite{long2025minimal}. 
To calculate the silicon METs, we split a pulse of duration $T$ into $M$ segments of equal duration $\Delta t=T/M$, which we then optimize. $J_i(t)$, $I_i(t)$, and $Q_i(t)$ are taken to be constant over each segment---\textit{i.e.}, they are piecewise constant functions. We collect the $m$th segment's pulse amplitudes [\textit{i.e.,} Eq. \eqref{eq:H_full}'s free parameters]  into a vector $\vec x^{(m)}$, which makes up the $m$th column of our pulse-parameter matrix $\bm{x}$. That is, $\bm{x}$ describes the Hamiltonian parameters throughout the duration of a pulse. For a given $\bm{x}$ and set of drive frequencies $\vec\omega$, the Hamiltonian generates a time evolution  $U(T)$ via the Schr\"odinger equation:
\begin{equation}
\dv{t}U(t)=-iH(t;\bm{x},\vec\omega)U(t),
\end{equation}
where $U(0)=I$. We minimize $T$ and optimize $\bm{x}$ using the \texttt{QuGrad} \cite{QuGrad,long2025minimal} implementation of GRAPE \cite{KHANEJA2005296} with respect to a target operator infidelity. In our work, we set the MET to the shortest pulse duration which still achieves a physical-level infidelity  of $\leq 0.1\%$.

To make pulse optimization computationally tractable at the magic-state-factory scale, we adopt a compartmentalized pulse-compression strategy. Each logical sub-circuit (\textit{e.g.}, syndrome measurements and magic-state injections~\cite{li2015magic, lao2022magic}) is decomposed into segments acting on at most four physical qubits and each segment is pulse compressed. For a circuit decomposed as
\begin{equation}\label{concate}
\mathcal{C}=\mathcal{S}_1\circ\mathcal{S}_2\circ\cdots\circ\mathcal{S}_K,
\end{equation}
each sub-circuit $\mathcal{S}_k$ is compiled into a locally optimized pulse with propagator $U_k = U\left(T^{MET}_k;\bm{x}^{(k)},\vec{\omega}^{(k)}\right)$, and concatenated in time to produce the logical circuit. This procedure allows us to quantify the cumulative error of concatenated compressed pulses. 

\begin{figure}
    \centering
    \includegraphics[width=\linewidth]{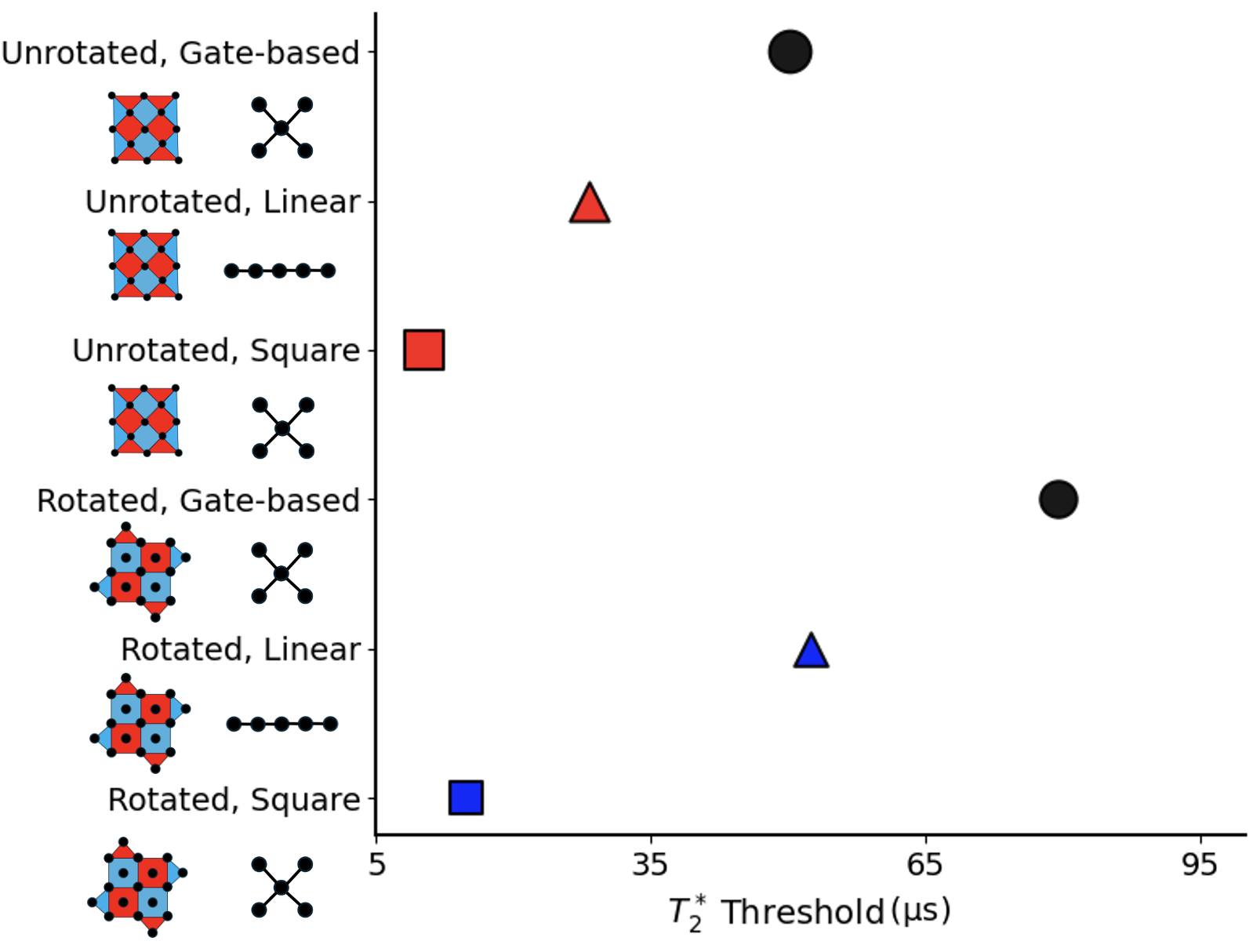}
    \caption{\textbf{Coherence-time thresholds for Z memory experiments.} The minimum $T^*_2$ coherence time (horizontal axis) for QEC protocols to improve the performance of an Z memory experiment with increased code distance. Various QEC schemes (vertical axis) are evaluated with or without pulse optimization. 
    The rotated surface code is plotted in blue and the unrotated surface code  in red. Linear connectivities are represented by triangles and square connectivities by squares.  The black dots indicate the baseline (gate-based operation without pulse-optimization) values with square connectivity.}
    
    \label{memory-errors}
\end{figure}

\subsection{Pulse optimization: results}
\label{sec:memory_expt}

Here, we summarize and interpret our memory experiments used to benchmark the effect of pulse compression and to characterize logical-memory error rates under different connectivity and coding choices. A memory experiment has three components. First, we prepare a logical basis state. Second, we let the logical qubits idle for a chosen time window whilst syndrome checks are applied to the physical qubits. Third,  we  measure and decode the logical state. The observed logical infidelity is related to the underlying physical error processes through the specific decoding-and-syndrome-extraction schedules used in each configuration. 

In this section, we limit our analysis to the standard surface code. Logical qubits are encoded in the joint $(+1)$-eigenvalued eigenspace of two sets of commuting stabilizers: Z- and X-type stabilizers, which can detect bit-flip and phase-flip errors, respectively. When implementing error correction throughout quantum computation, errors are detected by measuring both stabilizer types in syndrome extraction cycles. However, when benchmarking QEC codes or when quantifying error rates, one typically studies logical bit-flip and phase-flip errors individually. This can be done with a Z or X memory experiment. A Z memory experiment prepares a logical state $\ket{0}_L$ and measures how well it is preserved under repeated syndrome cycles, thereby quantifying logical X errors. An X memory experiment quantifies and detects Z errors.

To simplify the analysis, and to achieve tractable simulation times, we conduct our memory-experiment analysis with respect to local stochastic (Markovian) noise. Our noise model is specified in Sec.~\ref{markovian}. We study two connectivity layouts  (1D chain and square-grid nearest-neighbor) and two surface-code geometries (rotated and unrotated). The main difference in the connectivity layouts comes from the fact that CNOT operations between non-neighboring physical qubits require SWAP chains in the linear architecture. In the square-grid architecture, which is utilized in the dense and patched architectures, there are next-neighbor interactions between all qubits in a local syndrome check. We compare the pulse optimized emulations with a gate-based baseline where the syndrome circuits are compiled into standard gates. 

Figure~\ref{memory-errors} reports the threshold for Z-type memory experiments. To align our results with experimental parameters, we calculate our error thresholds in terms of the qubits' $T_2^*$ decoherence times: if a silicon processor has longer $T_2^*$ times than the threshold, then the logical error rate can be suppressed by increasing the code distance. We compare the rotated and unrotated surface codes. The unrotated surface code has a lower $T_2^*$ threshold. This is because the unrotated surface code has fewer minimum-weight logical error paths (the smallest number of physical errors needed to form a chain stretching from one boundary of the code to the opposite boundary, flipping the logical qubit's state)~\cite{o2025compare}. However, the rotated surface code improves more with pulse optimization, yielding pulse-optimized $T_2^*$ thresholds that are almost comparable. Moreover, the rotated surface code requires 25–30\% fewer physical qubits to achieve the same logical error rate as the unrotated surface code at physical error rates less than $10^{-3}$ \cite{o2025compare}. Despite the reduced qubit count of the rotated surface code, we proceed with the unrotated surface code for the rest of our analysis, because of the lower $T_2^*$ threshold.

\begin{figure*}
    \centering
    \includegraphics[width=\linewidth]{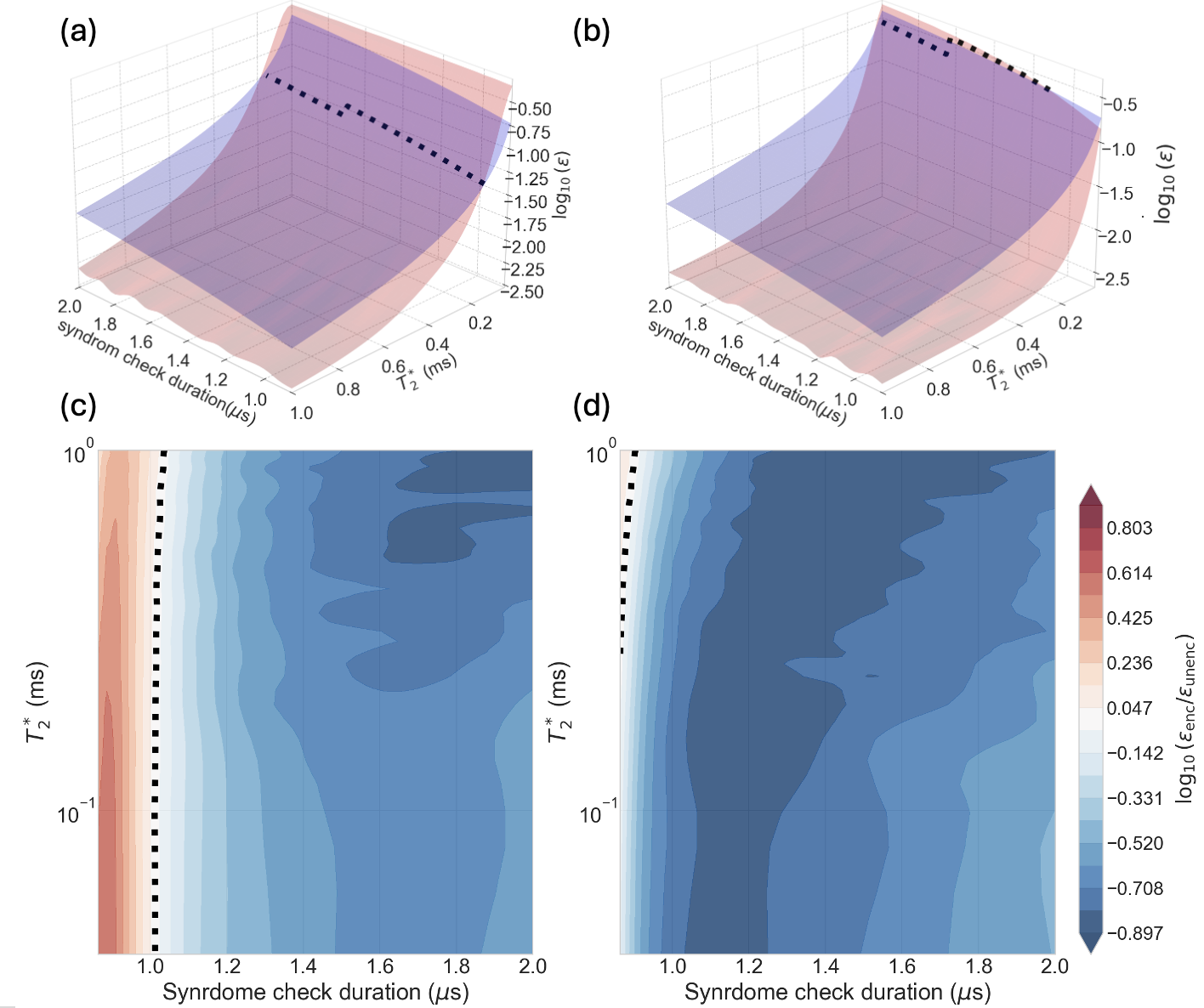}
    \caption{\textbf{Logical vs physical memory performance.} The data presented in the left panels [(a) and (c)] was calculated using a linear connectivity, while the data presented in the right panels [(b) and (d)] was calculated using a square connectivity. In both (a) and (b), the semi-transparent blue sheet corresponds to the error on a physical state. The red sheet represents the error on the logical state encoded with a $d=5$ code using 49 physical qubits. The black dotted line denotes the crossing point between parameter regions where the logical qubit outperforms the physical one. The bottom panels presents the same data in a contour plot.  The region where the logical qubits have greater error probabilities than the physical ones are highlighted in blue. Conversely, in the red regions the logical qubits have smaller error probabilities than the physical ones. The black dotted line indicates the crossover regime.}
    \label{fig:memory}
\end{figure*}

In addition to the memory experiments, we showcase the benefits of pulse optimization on logical states' coherence times throughout dephasing channels. In particular, we compare the infidelity $\epsilon$ of an idling single physical qubit ($\ket{1}$) with a logical qubit ($\ket{1}_L$) that undergoes pulse-optimized stabilizer cycles. The top panel of Fig.~\ref{fig:memory} shows the (log) infidelity as a function of the $T_2^*$ time and the syndrome-check duration, which we have set to be larger than the MET. (The lower panel shows a contour version of the upper one.) The red and blue sheets represent the infidelities of the logical and physical qubits, respectively. The dotted black lines highlight the cross-over parameters after which the logical qubit has a lower infidelity than the physical one (i.e., the red sheet lies below the blue one).  A clear trend emerges from the figure: when $T_2^*$ increases or syndrome-check duration decreases, the performance of the logical qubit relative to the physical one improves. This observation highlights the benefits of pulse compression. MET-limited pulse optimization does not merely shorten gate sequences but can actively extend the effective logical coherence window by minimizing exposure to noise. 
The left and right plots show the linear and square connectivity, respectively. Clearly, compared to the linear connectivity, the square one results in the logical qubits outperforming the physical qubits for a larger range of parameter values. This is because the syndrome checks in the linear connectivity rely on SWAP operations which introduce more noise.

\section{Noise modeling}\label{noise}

Most previous works on quantum-computational resource and overhead estimation, have utilized hardware agnostic and simplified noise models, for example, limiting the analysis to depolarizing noise. Here, we go beyond simple models. We tailor a silicon-specific noise model to align with real experiments. Our noise model builds on four components: (i) a Markovian error mechanism based on experimental relaxation and dephasing times; (ii) a non-Markovian mechanism obtained from a device-specific Hamiltonian and the filter-function formalism; (iii) a defect model appropriate for architectures with leakage and charge noise; and (iv) a model of the noise incurred when shuttling. Below, we specify each of these models. 

\subsection{Markovian error model} 
\label{markovian}

Most of the results presented in this work relies on a realistic silicon-tailored noise model. However, for computational tractability, we used a simplified Markovian noise model in Sec.~\ref{sec:memory_expt}. In our simplified model, noise is characterized by the  relaxation time $T_1$ and dephasing time $T_2^*$ as in Refs. \cite{tomita2014low, ghosh2012surface, sarvepalli2009asymmetric, silva2008scalable}. We used the following mapping to estimate the Pauli-error probabilities of single-qubit operations:
\begin{eqnarray}
p_X &=& p_Y = \frac{1-e^{-t/T_1}}{4}, \label{eq_probx}\\
p_Z &=& \frac{1-e^{-t/T_2^*}}{2} - \frac{1-e^{-t/T_1}}{4}. \label{eq_probz}
\end{eqnarray}
Intuitively, population relaxation contributes equally to bit-flip-type errors, while the additional dephasing component increases the effective Z-error probability.
Two-qubit errors are modeled similarly, but with more convoluted formulae \cite{tomita2014low, ghosh2012surface, sarvepalli2009asymmetric}.  In the simulations of Sec.~\ref{sec:memory_expt}, we applied these noise channels after each pulse operation [Eq.~\eqref{concate}] with the time  $t$ set by the pulse duration and $T_1$ and $T_2^*$ as input parameters. We then used approximately fitted Pauli errors to estimate error-correction thresholds using \texttt{STIM}~\cite{gidney2021stim}.

\subsection{Non-Markovian error model}
\label{nonmark}

Here we present the main noise model of our work. The model is targeted towards  the $1/f$ noise that dominates spin-qubit systems~\cite{dutta1981low, machlup1954noise, fleetwood20151}. This noise is caused by charge fluctuation in the semiconductor heterostructures  and stray magnetic field gradients, spatial variations of the $g$-factor, or interface-induced spin-orbit interaction~\cite{shehata2023modeling,ager2005high}. When two-level charge systems stochastically switch between their levels, they produce a local electric-field fluctuation that shifts the quantum dot potential.  The combination of such fluctuations, spanning timescales from nanoseconds to microseconds, gives rise to a noise spectrum dominated by slow components and approximately proportional to $1/f$, where $f$ denotes the transition frequency. 
To model temporally correlated noise, like the $1/f$ noise, we augment the Hamiltonian-level description  (Sec. \ref{sec:pulse_optimization}) with the filter-function formalism \cite{hangleiter2021filter}. Below, we summarize the methodology through a simple example. Detailed derivations are provided in App.~\ref{non-markovian}, where we rederive the relation between the infidelity, $\epsilon$, and the noise spectrum:
\begin{equation}
\label{NMfilter}
\epsilon=\frac{1}{D+1}
\sum_\alpha
\int_{-\infty}^{\infty}\frac{d\omega}{2\pi}
S_\alpha(\omega)
F_\alpha(\omega) .
\end{equation}
Here, $D$ is the dimensionality of the Hamiltonian of interest, $F_\alpha(\omega)$ is the filter function generated by the Hamiltonian, and $S_\alpha(\omega)$ is the noise spectral density function. 

As a simple illustration of the filter-function treatment of noise, consider a single qubit driven by a control Hamiltonian $H_c=\frac{\Omega}{2}\sigma_x$ and subject to fluctuating dephasing noise $H_n = b(t) \sigma_z$. The noiseless ideal unitary evolution of the system is $
U_c(t) = e^{-i(\Omega t/2)\sigma_x}$. However, $H_n$ will corrupt this evolution. To see how, we first define the noise operators in the interaction picture as $\tilde{B}_\alpha(t) = U_c^\dagger(t) B_\alpha U_c(t)$ where $B_\alpha$ is related to $H_n$ as shown in App.~\ref{non-markovian}. In our simple example, there is only one noise term generated by a Pauli-Z operator:
\begin{align}
    \tilde{B}_z(t) &= e^{i(\Omega t/2)\sigma_x} \sigma_z e^{-i(\Omega t/2)\sigma_x}\\
    & = \cos(\Omega t) \sigma_z - \sin(\Omega t) \sigma_y.
\end{align}
We can then expand the noise operator within the interaction picture in the Pauli basis:
\begin{equation}
    \tilde{H}_n(t) = b(t) \sum_k \tilde{B}_{z,k}(t) \sigma_k.
\end{equation}
The coefficients are given by $\tilde{B}_{z,k}(t) = \frac{1}{2}\text{tr}[\tilde{B}_z(t) \sigma_k]$:
\begin{align}
    \tilde{B}_{z,\mathds{1}}(t) &= \frac{1}{2}\text{tr}[\tilde{B}_z(t) \mathds{1}] = 0, \\
    \tilde{B}_{z,x}(t) &= \frac{1}{2}\text{tr}[\tilde{B}_z(t) \sigma_x] = 0, \\
    \tilde{B}_{z,y}(t) &= \frac{1}{2}\text{tr}[\tilde{B}_z(t) \sigma_y]  = -\sin(\Omega t), \\
    \tilde{B}_{z,z}(t) &= \frac{1}{2}\text{tr}[\tilde{B}_z(t) \sigma_z] =  \cos(\Omega t).
\end{align}
Next, we define the finite-time Fourier transforms $\tilde{B}_{z,k}(\omega) = \int_0^\tau dt \tilde{B}_{z,k}(t) e^{i\omega t}$ such that
\begin{align}
    \tilde{B}_{z,z}(\omega) &= \int_0^\tau \cos(\Omega t) e^{i\omega t} dt \\&= \frac{1}{2} \left[ I(\omega+\Omega, \tau) + I(\omega-\Omega, \tau) \right], \\
    \tilde{B}_{z,y}(\omega) &= -\int_0^\tau \sin(\Omega t) e^{i\omega t} dt\\&= \frac{i}{2} \left[ I(\omega+\Omega, \tau) - I(\omega-\Omega, \tau) \right],
\end{align}
where $I(\omega \pm \Omega, \tau) = \int_0^\tau e^{i(\omega \pm \Omega) t} dt$. The filter function is given by the sum of the squares of the magnitudes of these coefficients:
\begin{equation}\label{fz}
    F_z(\omega) = |\tilde{B}_{z,z}(\omega)|^2 + |\tilde{B}_{z,y}(\omega)|^2.
\end{equation}
Evaluating Eq.~\eqref{fz}, we find that
\begin{align}
    F_z(\omega) &= \left| \frac{1}{2} (I_+ + I_-) \right|^2 + \left| \frac{i}{2} (I_+ - I_-) \right|^2 \\
    &=\frac{1}{2} \left( |I_+|^2 + |I_-|^2 \right),
\end{align}
where $I_+ = I(\omega+\Omega, \tau)$ and $I_- = I(\omega-\Omega, \tau)$ and
\begin{equation}
    |I(a, \tau)|^2 = \left| \frac{e^{i a \tau} - 1}{i a} \right|^2 = \frac{4 \sin^2\left( \frac{a\tau}{2} \right)}{a^2}.
\end{equation}
Thus, the final expression for our simple example's filter function is
\begin{equation}
F(\omega)
=
\frac{\tau^2}{2}
\left[
\text{sinc}^2\!\left(\frac{(\omega+\Omega)\tau}{2}\right)
+
\text{sinc}^2\!\left(\frac{(\omega-\Omega)\tau}{2}\right)
\right].
\end{equation}
In this particular example, we see, via Eq. \eqref{NMfilter}, that the system is sensitive to noise around its Rabi frequency $\Omega$: the filter function $F(\omega)$ is peaked around $\omega=\pm\Omega$. The actual device Hamiltonian of our system [Eq. \eqref{eq:H_full}] is more complicated than that of this example. However, the calculation of the relevant filter function follows a similar derivation.

Next, we describe how we model the  spectral density function $S_{\alpha}(\omega)$. Experimental observations of spin qubits show that  the power spectrum of charge and magnetic noise follows a $1/f$ behavior in frequency, with some cutoffs~\cite{dutta1981low, machlup1954noise, fleetwood20151}.  We therefore construct a spectral density that satisfies the following requirements:
\begin{enumerate}
\item $1/f$ scaling over an intermediate frequency window.
\item Suppression of unphysical divergences at both low and high frequencies.
\end{enumerate}
These considerations lead to the phenomenological equation
\begin{equation}
S(\omega)
=
\frac{A}
{\max(|\omega|,\omega_{\mathrm{low}})
\left[1+\left(\frac{\omega}{\omega_{\mathrm{high}}}\right)^2\right]} .
\label{eq:1f_spectrum}
\end{equation}
Here, $A$ sets the overall noise strength, $\omega_{\mathrm{low}}$ is a low-frequency cutoff, and $\omega_{\mathrm{high}}$ is a high-frequency cutoff. For example, when $|\omega| \ll \omega_{\mathrm{low}}$, $S(\omega) \approx \frac{A}{\omega_{\mathrm{low}}}$, corresponding to a white-noise plateau at low frequencies. Physically, $\omega_{\mathrm{low}} \sim 1/T_{\mathrm{max}}$ is set by the maximum experimental observation time $T_{\mathrm{max}}$. 
In the intermediate regime $\omega_{\mathrm{low}} \ll |\omega| \ll \omega_{\mathrm{high}}$ we obtain the characteristic form $S(\omega) \approx \frac{A}{|\omega|}$. 

Using the numerical pulse-evolution methods described in previous sections, we can construct estimates of operational infidelities in the following way.
\begin{enumerate}
\item We propagate a tomographically complete set of states
$\{|g\rangle\}$ under the optimized control pulse to obtain
the noiseless states $\{ \ket{\psi_g(t_i)} \}$ at discrete time steps $t_i$. From these states, we can reconstruct the ideal unitary $U_c(t_i)$.
\item For each noise channel $B_\alpha$ ($H_n$ in the previous example) labeled by $\alpha$, we calculate the interaction-picture noise sensitivity $\tilde{B}_\alpha(t)$ at time $t$.
\item We perform finite-time Fourier transforms of $\tilde{B}_{\alpha k}(t)$ to obtain
$\tilde B_{\alpha k}(\omega)$ and construct the filter functions
$
F_\alpha(\omega)
$.
\item Using Eq. \eqref{NMfilter}, we evaluate the spectral overlap between the filter function and the noise power spectral density
$S_\alpha(\omega)$ to obtain the average gate infidelity $\epsilon$.
\item We numerically fit a Pauli channel to the channel generated by the non-Markovian noise under the constraint that both channels yield the same infidelity (details in App.~\ref{map}).
\end{enumerate}
In this way, we map the ideal pulse history onto a fully non-Markovian noise model and quantitatively predict the resulting operation (in)fidelities.

\subsection{Defect error model}
\label{defects}

Spin-qubit architectures that rely on physical electron shuttling between operational zones will naturally suffer from localized fabrication defects~\cite{newman2000efficient,siegel2025snakes}. Such defects can include charge traps or interface impurities~\cite{yoneda2018quantum, culcer2009dephasing, elsayed2024low}, which may render specific quantum dots temporarily unsuitable for qubit manipulation or transport. We model these defects as randomly blocked vertices on the device connectivity graph. Each node is independently defective with probability $\epsilon_{\mathrm{defect}}$.

We make the optimistic assumption that syndrome checks and efficient decoding can inform a quantum-computer user of the location of defects. Then, defect mitigation becomes a task of optimizing the shuttling routes such that defects are avoided.  In our simulations, we compute the shortest viable shuttling path using breadth-first search~\cite{cormen2022introduction} on the subgraph obtained by removing all defective vertices from the entire grid. The total shuttling time for a qubit transfer is given by $T_{\mathrm{shuttle}} = t_{\mathrm{step}} \cdot L_{\mathrm{BFS}}$, where $t_{\mathrm{step}}$ is the time required to shuttle a qubit across a single edge (typically $t_{\mathrm{step}} \sim 10~\unit{\nano\second}$ for conveyor-mode shuttling over $100~\unit{\nano\meter}$), and $L_{\mathrm{BFS}}$ is the hop count of the shortest defect-free path. The Manhattan distance serves as a reference metric for the ideal, defect-free case.

\begin{figure}
    \centering
    \includegraphics[width=\linewidth]{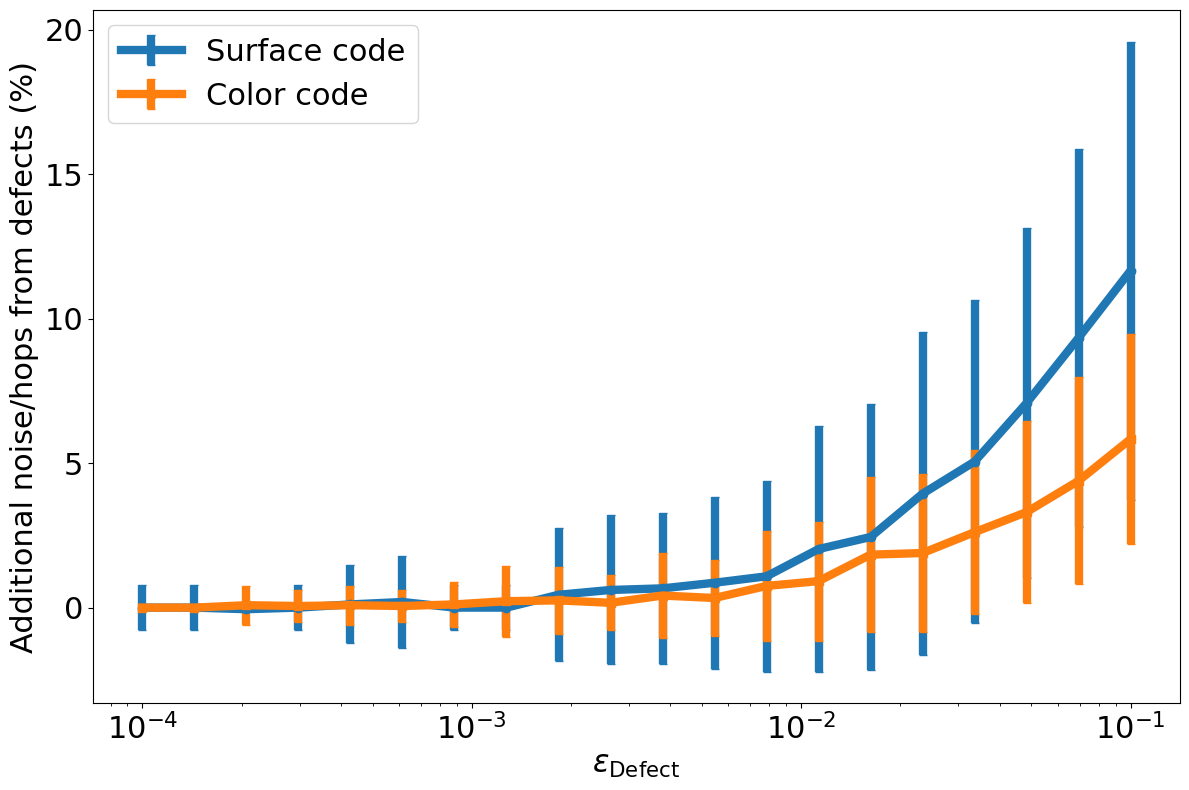}
    \caption{\textbf{Noise impact of defects in shuttling-based architectures.} The additional amount of noise or number of hops is plotted as a function of the probability of defect errors $\epsilon_{\mathrm{defect}}$.  The orange and blue lines denote data obtained from surface-code and color-code simulations, respectively.  While fabrication defects typically increase shuttling distances by forcing routes around unavailable sites, the effective path length $L_{\mathrm{BFS}}$ can in principle be \textit{shorter} than the ideal Manhattan distance: When the intended source or destination site is defective, the routing algorithm selects the nearest non-defective site. If both endpoints shift  toward each other, the effective separation can decrease. Here, the simulation scale is restricted to a $20 \times 20$ lattice for illustrative purposes only.}
    \label{fig:defects}
\end{figure}

Combined with gate and readout errors, the shuttling infidelity $\epsilon_{\mathrm{shuttle}}$ contributes additively to the overall physical error budget. In Fig~\ref{fig:defects}, we quantify how increased defect densities lead to longer shuttling detours (increasing $L_{\mathrm{BFS}}$) and degrades the qubit fidelities.

\subsection{Shuttling error model}
\label{sec: shuttle}

Our simulations also account for errors induced by the motion of the electron-spin qubits during shuttling. Errors occur primarily as a result of field fluctuations and the spin-orbit interaction at the point of acceleration and deceleration~\cite{flindt2006spin, levitov2002dynamical, rashba2003orbital}. Experimental investigations quote shuttling errors on the order of $0.01\%$ when shuttling over $10 \unit{\micro \meter}$. We inject the shuttling noise into our analysis simply as a source of error when qubits are shuttled.

\section{Space-time modeling for magic-state distillation}
\label{sec:st_modeling}

In this section, we outline our methodology for producing estimates of the space-time volume required to produce high-fidelity logical magic $T$ states. We focus our study on `grow-and-distill' MSD protocols~\cite{beverland2021cost}.  First, we briefly summarize the distillation protocols. Second, we outline the logical components that contribute to errors. Third, we specify the mathematical model we use to convert per-operation logical error rates to end-to-end distillation-acceptance probabilities and output-fidelity estimates. Finally, we show how the per-round acceptance probabilities feed into an iterative, multi-round MSD process and outline how we calculate the total resources needed for MSD in terms of physical qubit count, runtime and space-time overhead.

\begin{figure}[hbt!]
    \centering
    \includegraphics[width=\linewidth]{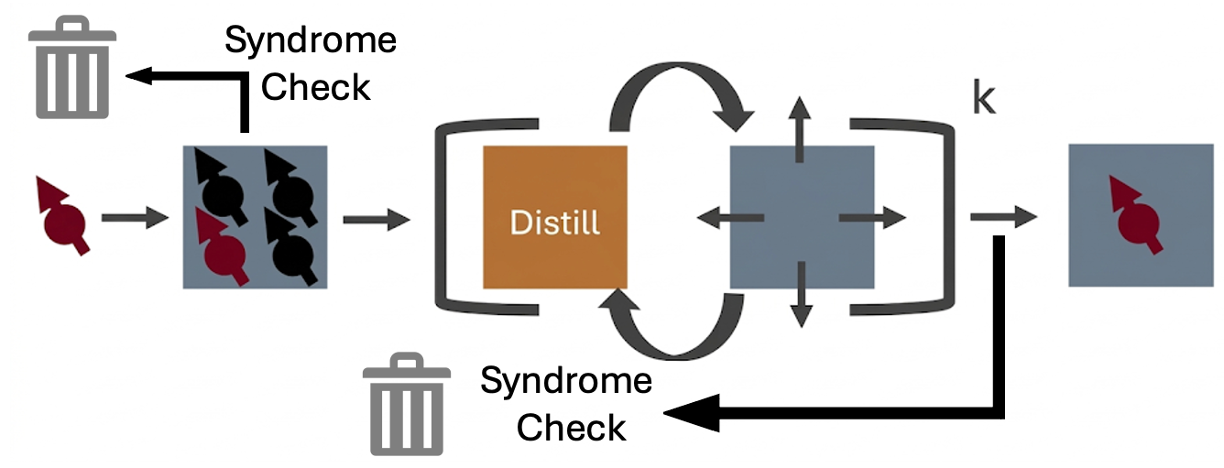}
    \caption{\textbf{Magic-state-distillation protocol.}  First, physical spin qubits (red) prepared in noisy magic states are injected into an underlying QEC code. This produces a low-fidelity logical magic state (first square patch, from the left). A distillation code (second square patch, from the left) is utilized to carry out syndrome checks of the logical magic states. If the syndrome checks fail, the MSD round is discarded. If the syndrome checks are successful, the QEC code distance is expanded (third square patch, from the left). The distillation process is repeated $k$ times, after which a high-fidelity logical magic state (fourth square patch, from the left) is output.}
    \label{fig:baseline}
\end{figure}

The MSD protocols, shown in Fig.~\ref{fig:baseline}, typically consist of two stages: (i) injection of a noisy physical magic state into a base quantum error-correcting code [[$N, k, d$]]~\cite{kim2024magic,lao2022magic} and (ii) logical distillation using a distillation code to suppress errors. This process is usually repeated iteratively, with each distillation round (ii) producing a single, high-fidelity logical magic state. Collectively, (i) and (ii) form what is known as an MSD factory. To further enhance efficiency, the logical code distance $d$ can be increased after each iteration, enabling the factory to suppress noise further and produce a higher-quality output with reduced overhead in subsequent rounds.

The distillation framework considered here is slightly different from protocols where one directly expands the code distance by a large interval~\cite{litinski2019magic}. Practically, we can improve resource efficiency by gradually increasing the code distance. Recent work has proposed several techniques that could substantially reduce the overhead of MSD, including magic-state cultivation protocols~\cite{gidney2024magic} and improved encoding schemes~\cite{claes2025lower,higgott2021optimal}. While these approaches show promising reductions in resource requirements, many are still under active development and lack comprehensive resource analyses across realistic hardware architectures. Therefore, in this work, we focus on the conventional MSD pipelines described above, which provide a well-established baseline for evaluating architectural tradeoffs.

\subsection{MSD protocols}
\label{protocols}
We first introduce the popular 15$\to$1 MSD protocol. It is based on Haah--Hastings construction~\cite{hastings2018distillation} and uses the transversality of $T$ gates on the quantum Reed--Muller code~\cite{steane2002quantum}. Concretely: if we prepare some stabilizer states with $n=16$ logical qubits, then we can inject 15 logical $T$ states via teleportation and output one higher-fidelity $T$ state. Under ideal conditions, and for an input infidelity of $q$, the Reed--Muller code yields an output infidelity  $q_{\rm out}\propto q^{3}$. Noisy Clifford operations (preparation, idle and CNOT) result in actual implementations suffering from additional rejections and failure contributions. These issues are accounted for by the effective logical-operation error model described below.

A compact alternative to the 15$\to$1  MSD protocol is the 5$\to$1 protocol~\cite{laflamme1996perfect}. The 5$\to$1 protocol uses a small, ``perfect'' $[[5,1,3]]$ code for both the encoding and decoding. A Clifford circuit maps five noisy input $T$ states to one higher-fidelity $T$ state. The leading-order power-law suppression of input-$T$ errors under ideal conditions is given by $q_{\rm out} \propto q^{2}$. In this work, we briefly compare the 5$\to$1 protocol to the 15$\to$1 protocol. However, the main parts of our analysis focuses on the 15$\to$1 circuit for which we have numerically simulated the explicit mappings from the physical error rates to the logical ones.

To quantify the errors incurred during an MSD protocol, one must study its individual logical components. These logical components are implemented with an underlying QEC code using repeated QEC cycles or rounds. Each of these logical components have failure and rejection probabilities that depend on the physical-level noise strength $p$ and on the code distance $d$ used to protect the logical patch. The dominant sources of logical errors in MSD protocols are:
\begin{itemize}
  \item \textit{Logical state preparation} ($\ket{\overline{0}},\ket{\overline{+}}$) fails with probability $\overline{p}_{\rm prep}(p,d)$.
  \item \textit{Logical idle} (one QEC cycle of idling) fails with probability $\overline{p}_{\rm idle}(p,d)$.
  \item \textit{Logical CNOT operations} fails with probability $\overline{p}_{\rm CNOT}(p,d)$.
  \item \textit{Initialization of low-fidelity $T$ states} fails with failure probability $p_{\rm fail}^{\rm T}(q, p, d)$ and are rejected with probability $p_{\rm rej}^{\rm T}(q, p,d)$.
\end{itemize}
 Given these per-operation failure probabilities, we model the performance of a logical MSD circuit as follows. Let $p_{\rm rej}^{\rm dist}(q,p,d)$ and $p_{\rm fail}^{\rm dist}(q,p,d)$ denote the rejection and failure probabilities of a single round of the logical distillation circuit (\textit{i.e.}, the probability that the protocol aborts because syndrome checks indicate an error, and the probability that it accepts a faulty output, respectively). These probabilities can be approximated, to leading order, in terms of the aforementioned logical errors~\cite{beverland2021cost, jochym2012robustness}. We extract the linear coefficients using the following approach: (i) set all sources of logical errors except the one under test to zero, (ii) run Monte Carlo trials of the complete logical operation or distillation circuit, (iii) measure the fraction of trials leading to rejection or failure, and (iv) fit the measured rejection or failure probability to a linear function of the logical error parameter under test. Repeating this for each source of logical error and combining the results yields
\begin{align}
p_{\rm rej}^{\rm dist}(q,p,d) =&  p_{\rm rej}^{T}(q,d) + 12.3\overline{p}_{\rm prep} \nonumber \\ &+ (466 + 4.13 d)\overline{p}_{\rm idle}+ 51.7\:\overline{p}_{\rm CNOT},\label{eq:prj_decomp}\\
p_{\rm fail}^{\rm dist}(q,p,d) =& p_{\rm fail}^{T}(q,d) + 16.9\overline{p}_{\rm idle} + 1.93\overline{p}_{\rm CNOT}.\label{eq:pfail_decomp}
\end{align}

To produce our resource estimates for MSD, we must gauge the effect of Eqs. \eqref{eq:prj_decomp} and \eqref{eq:pfail_decomp} on the MSD protocols. We use \texttt{STIM} to simulate the iterative MSD processes. (As \texttt{STIM} is a stabilizer simulator, we follow standard practice~\cite{gidney2024magic, dasu2025breaking}  and simulate the distillation of $S\ket{+}$ states rather than $T$ states.)
The output infidelity is obtained by iterating the map
\begin{equation}
q^{(i)} = p_{\rm fail}^{\rm dist}\big(q^{(i-1)},p,d^{(i)}\big),
\qquad q^{(0)} = p_{\rm fail}^{\rm init}(p),
\end{equation}
until the desired target infidelity $q^{(k)}\le p_{\rm target}$ is reached on the $k$th iteration~\cite{o2017quantum}.
The expected number of \textit{physical} qubits required by a $k$-round protocol is then
\begin{equation}\label{eq:NSD_surf_repeat}
  N_{\rm phy} = \max_{i=1,\dots,k} \Bigg[ \alpha^{(i)} N\big(d^{(i)}\big) \prod_{j=i}^{k}\frac{R^{(j)}}{r^{(j)}} \Bigg].
\end{equation}
Here, \begin{equation}
r^{(i)} = 1 - p_{\rm rej}^{\rm dist}\big(q^{(i-1)},p,d^{(i)}\big),
\end{equation} denotes the $i$th round's acceptance probability.  $R^{(i)}$ denotes the number of input logical $T$ states required in the $i$th round ($R^{(i)}=15$ for the 15$\to$1 protocol). $\alpha^{(i)}$ denotes the number of logical qubits consumed per input magic state ($\alpha^{(0)}=15$ and $\alpha^{(i>0)}=(16+15)/15$ for the  15$\to$1 protocol). Finally, $N(d)$ denotes the number of \textit{physical} qubits required to implement a single distance-$d$ logical qubit.

The protocol runtime is
\begin{equation}
  \tau= \tau_{\rm init} + \sum_{i=1}^k \tau_{\rm dist}\big(d^{(i)}\big).
\end{equation}
Here, $\tau_{\rm init}$ is the initialization time and $\tau_{\rm dist}(d)$ the time to execute a single logical distillation iteraion at distance $d$.  The total space-time overhead is the product
$N_{\rm phy}\times\tau$.

\subsection{MSD in silicon devices}
\label{MSD-in-silicon}
\begin{figure*}
    \centering
    \includegraphics[width=0.8\linewidth]{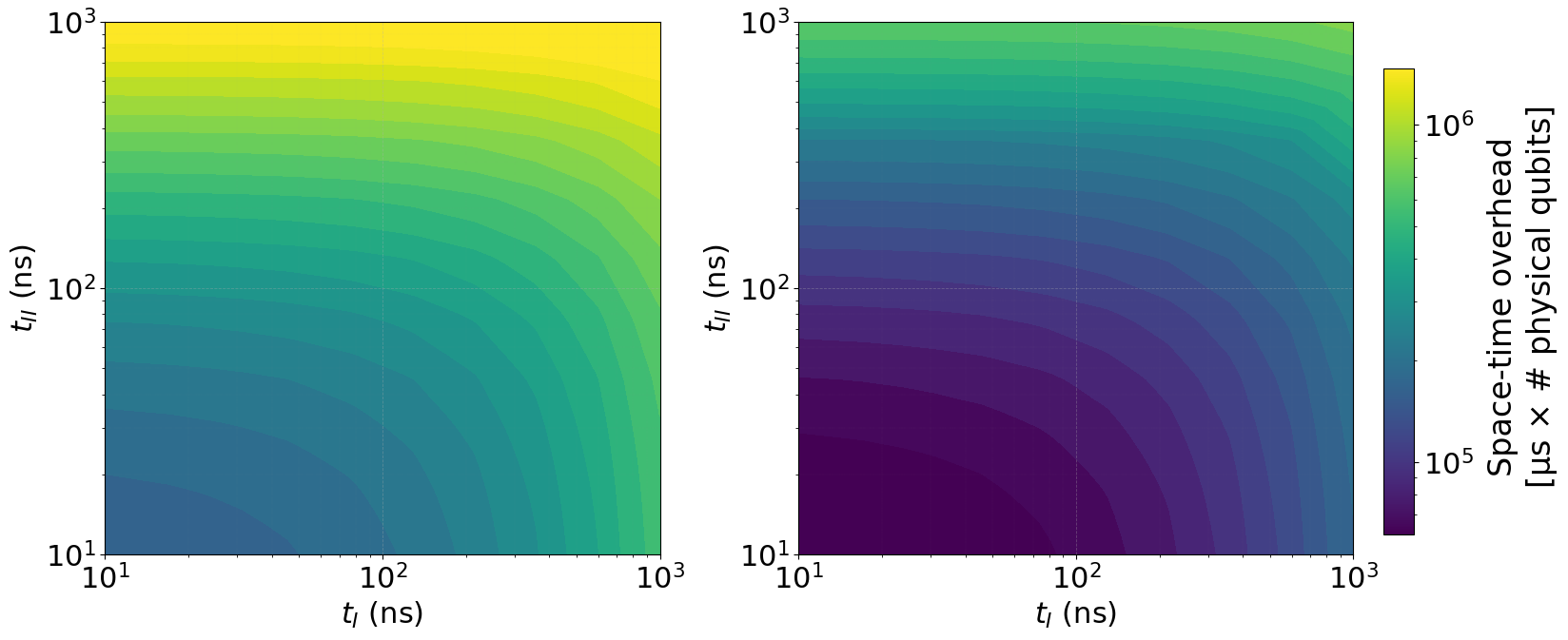}
    \includegraphics[width=0.8\linewidth]{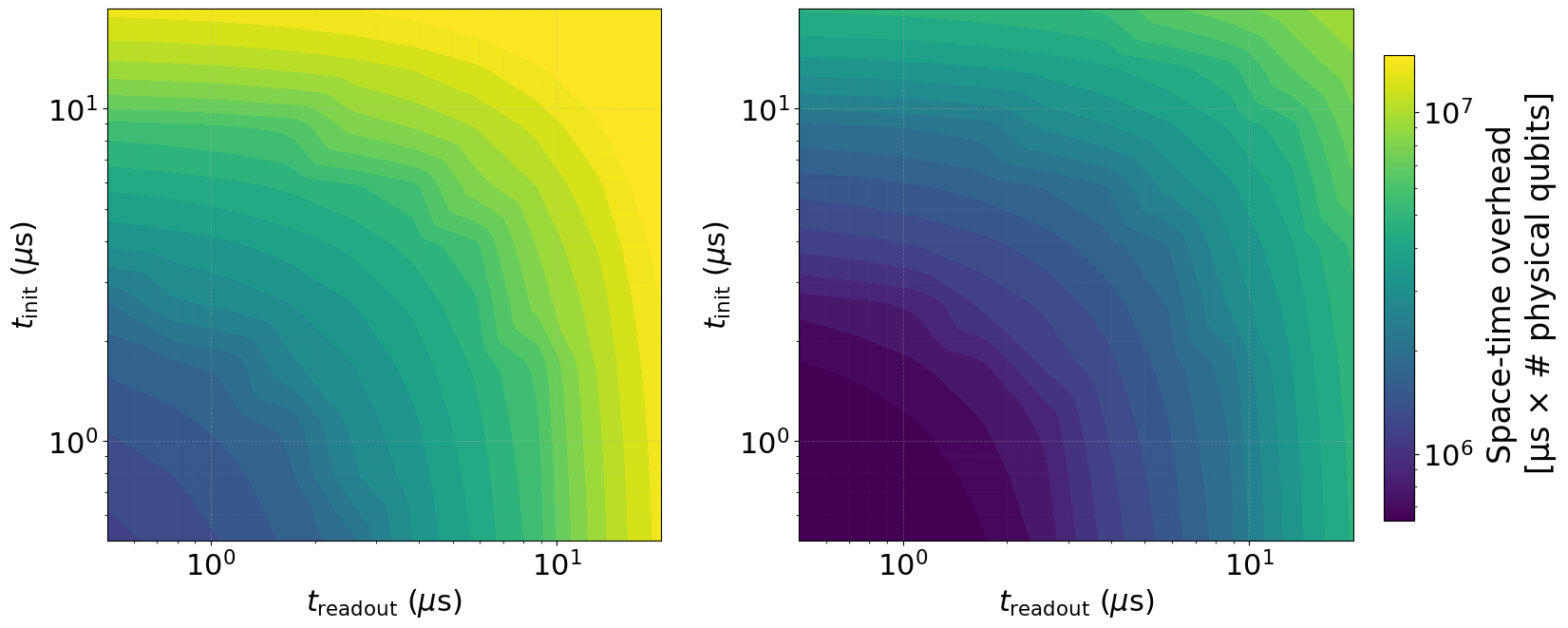}
    \includegraphics[width=0.8\linewidth]{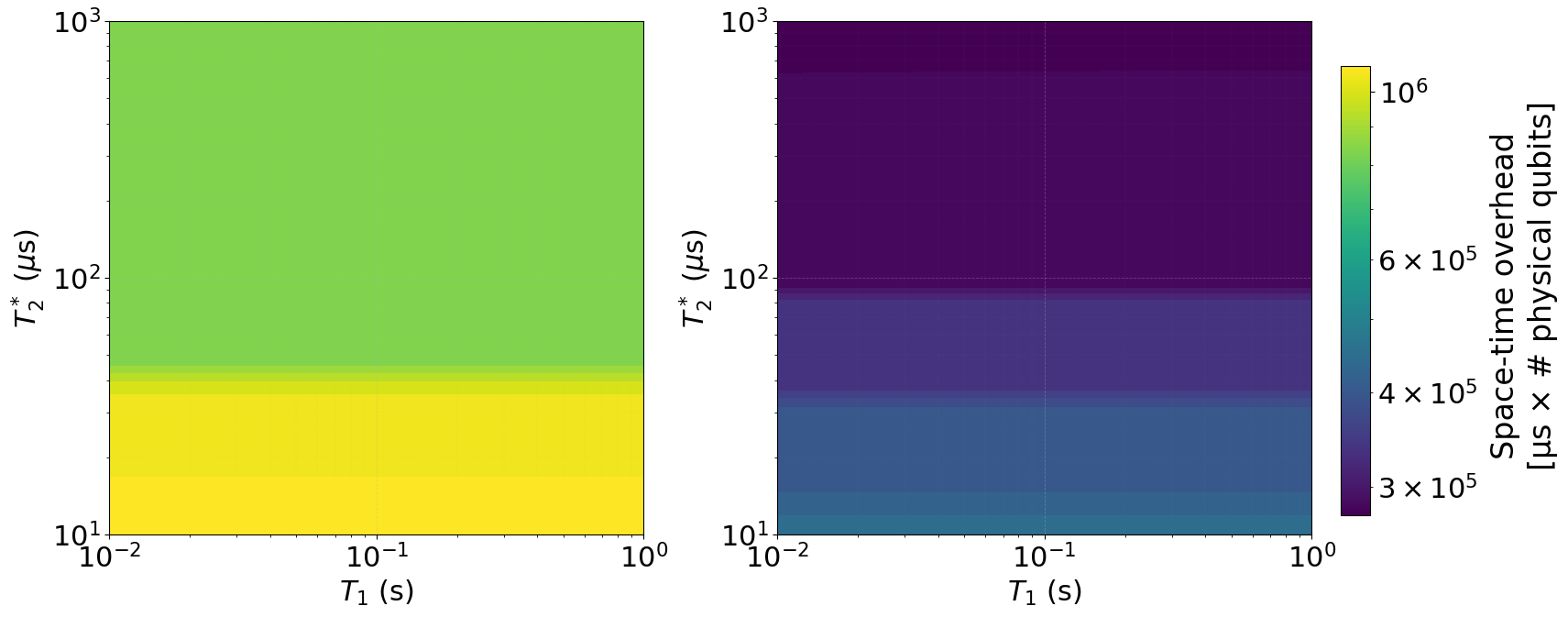}
    \caption{ \textbf{Space-time overhead for the production of one magic state.} The color bar indicates the space-time overhead (number of physical qubits times the wall-clock time in units of $\unit{\micro \second}$) as a function of gate times (top panel), readout and initialization times (middle panel), and  coherence times (bottom panel) for the dense architecture. The left  and right plots show data for the surface code with lattice-surgery operations and the color code with transversal operations, respectively. The logical magic state was produced using the $15\to1$ MSD protocol with a target infidelity of $10^{-12}$.}
    \label{fig:sweep}
\end{figure*}

We now summarize our analysis of MSD protocols operated on silicon hardware. We start by investigating the effect of coherence and gate times as well as initialization and measurement errors on the resource overheads in the dense architecture. The patched and sparse architectures follow similar qualitative trends (App.~\ref{complementary}). Our results for the $15\to1$ protocol with lattice surgery and with transversal gates are presented in Figs~\ref{fig:sweep}. The left column displays the results from simulations  of MSD with the surface code and lattice-surgery operations; the right column presents analogous results from simulations with the color code and transversal operations. In the first two rows, we quantitatively demonstrate that faster measurements, initializations, and single- and two-qubit Clifford operation times reduces the space-time resources for MSD. Similarly, the last row shows that the space-time resources decrease with improved $T_2^*$ times. For example, a $T_2^*$ increase of an order of magnitude generates a space-time--volume decrease by about a factor of 2 to 3. The noise bias in silicon ($T_2^* \ll T_1$) means that the resource improvements with increased $T_1$ times are less prominent.

\begin{figure*}
    \centering
    \includegraphics[width=\linewidth]{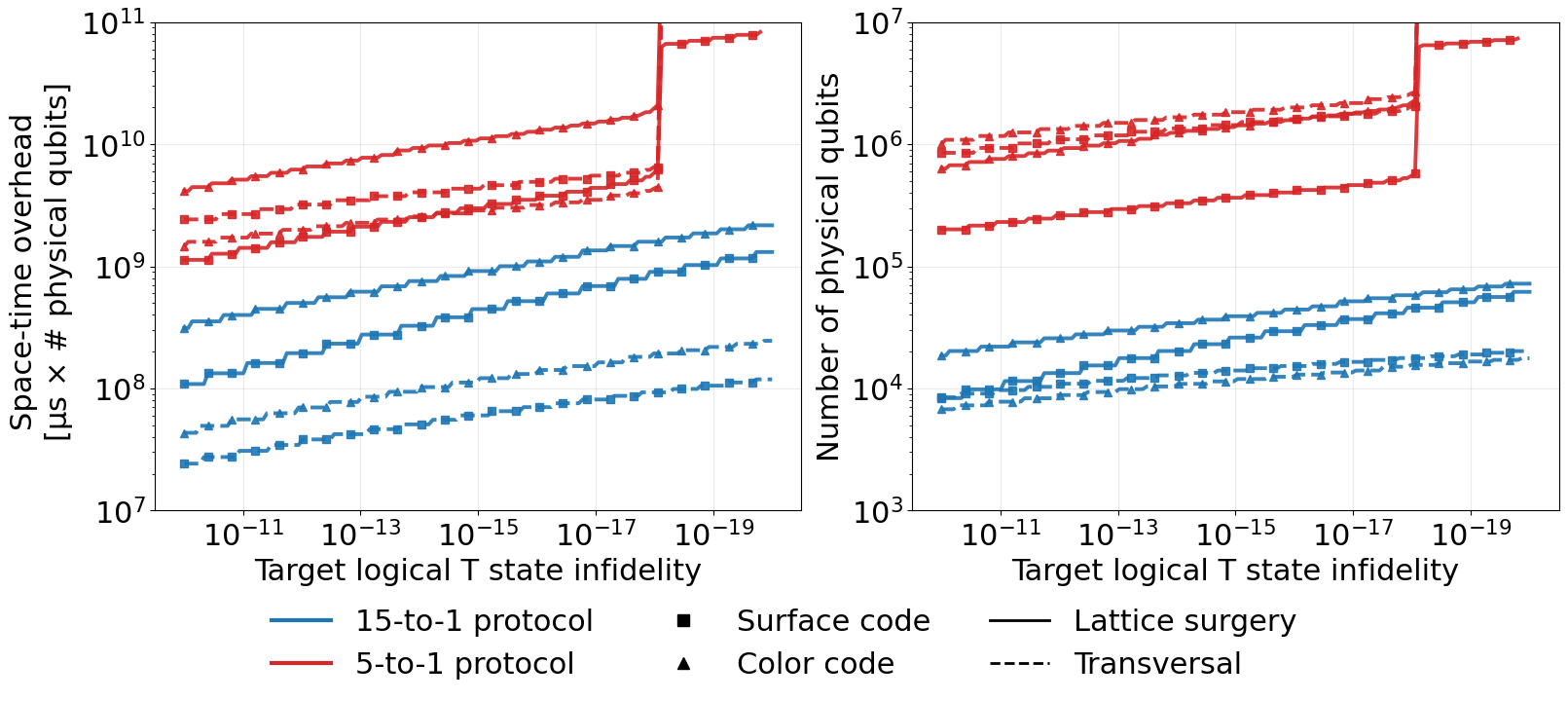}
    \caption{\textbf{Resources \textit{vs.} logical magic-state infidelity.} The space-time overhead (number of qubits times wall-clock time in $\unit{\micro \second}$) is plotted as a function of magic-state infidelity. The data is obtained for the sparse architecture. The noise model from Sec.~\ref{nonmark} is chosen in accordance with Table \ref{tab:parameter_summary}. Eight different setups are plotted:  the $5\to1$  and $15\to1$ distillation protocols are plotted in red and blue, respectively;  transversal operations and lattice-surgery operations are plotted with dotted and solid lines, respectively; and the surface code and color code are highlighted with triangular and circular markers, respectively. The space-time overhead is plotted on the left plot's vertical axis; the number of physical qubits is plotted on the right plot's vertical axis. The target magic-state infidelity is plotted on the horizontal axis.  }
    \label{fig:sparse}
\end{figure*}

We also compared the $5\to1$ and $15\to1$ protocols operating with either surface or color codes, and either lattice surgery or transversal gates. In Fig.~\ref{fig:sparse}, we present the sparse architecture's space-time overhead and qubit footprint as a function of $T$-state target infidelity. Naturally, the overheads increase with decreased target infidelities. We observe that the $5\to1$ protocol has a higher resource cost compared to the $15\to1$ protocol. Among all configuration we analyzed, we find that the  $15\to1$ protocol implemented with the surface code and transversal operation achieves the lowest space-time overheads. The lowest qubit footprint is achieved by the $15\to1$ protocol implemented with the color code and transversal operations. The $5\to1$ curve experience a large resource overhead divergence at a target infidelity of around $10^{-18}$. Similarly, we observe a large increase in the overheads of the $15\to1$ protocol at target infidelities below $10^{-22}$. These observation can be explained by the competition of resources for the lowering of $T$ state infidelities and the increase of Clifford-operation fidelities.  Initially, one can suppress both errors by increasing the QEC code distance, which generates the step-wise increase. However, for higher fidelities, it is cheaper, in terms of resources, to reduce the errors on the logical Clifford operations. Then, the input $T$-state infidelity becomes the bottleneck for the target distillation infidelity. Consequently, in addition to increasing the code distance, one must apply extra rounds of distillation. Due to the efficiency in suppressing the errors with $15\to1$ distillation protocol, the jump is not seen for the range of infidelity plotted. However, for the $5\to1$ protocol with only quadratic suppression in the infidelity, the jump is resources happens at infidelities around $10^{-18}$.

Finally, we demonstrate how pulse compression (Sec. \ref{sec:pulse_optimization}) can reduce the MSD overhead in the dense architecture. Our findings are summarized in Tab.~\ref{tab:compress}. The top block displays data from simulations of MSD implemented with standardized gates. The parameters are taken from Tab.~\ref{tab:parameter_summary}. The different settings for the fault-tolerant implementation naturally leads to different requirements on the number of physical qubits and the wall-clock time needed for an MSD procedure. The bottom block of the table presents the same analysis but with data from emulations with compressed and optimized hardware pulses. (The pulses where compressed in time to reach their MET at a cutoff infidelity of $10^{-4}$.) Clearly, pulse compression reduces the physical footprint and shortens the wall-clock time for $T$-state generation. The fastest to perform distillation is achieved with the surface code and transversal operations. The color code counterpart has less stringent requirements on the physical-qubit count.

\begin{table*}[t]
\bgroup
\def\arraystretch{1.2}
\begin{tabular}{|c|c|c|c|c|c|c|c|}
\hline
\multicolumn{8}{|c|}{\textbf{Gate-based operation}} \\ \hline
\multicolumn{4}{|c|}{surface} & \multicolumn{4}{c|}{color} \\ \hline
\multicolumn{2}{|c|}{surgery} & \multicolumn{2}{c|}{transversal} & \multicolumn{2}{c|}{surgery} & \multicolumn{2}{c|}{transversal} \\ \hline
15-1 & 5-1 & 15-1 & 5-1 & 15-1 & 5-1 & 15-1 & 5-1 \\ \hline
8 k (422 $\unit{\micro \second}$) & 200 k (182 $\unit{\micro \second}$) & 7 k (82 $\unit{\micro \second}$) & 850 k (92 $\unit{\micro \second}$)& 20 k (567 $\unit{\micro \second}$) & 630 k (212 $\unit{\micro \second}$) & 6 k (208 $\unit{\micro \second}$) &1086 k (473 $\unit{\micro \second}$) \\ \hline
\multicolumn{8}{|c|}{\textbf{Pulse-based operation (MET)}} \\ \hline
\multicolumn{4}{|c|}{surface} & \multicolumn{4}{c|}{color} \\ \hline
\multicolumn{2}{|c|}{surgery} & \multicolumn{2}{c|}{transversal} & \multicolumn{2}{c|}{surgery} & \multicolumn{2}{c|}{transversal} \\ \hline
15-1 & 5-1 & 15-1 & 5-1 & 15-1 & 5-1 & 15-1 & 5-1 \\ \hline
6 k (380 $\unit{\micro \second}$) & 131 k (140 $\unit{\micro \second}$) & 5 k (67 $\unit{\micro \second}$) & 687 k (90 $\unit{\micro \second}$)& 12 k (450 $\unit{\micro \second}$) & 426 k (190 $\unit{\micro \second}$) & 5 k (171 $\unit{\micro \second}$) &767 k (458 $\unit{\micro \second}$) \\ \hline
\end{tabular}
\egroup
\caption{\textbf{Summary of MSD factory resources.} Number of qubits (and time) needed to generate one magic state with infidelity below $10^{-12}$. The number of physical qubits and wall-clock time needed to distill one magic $T$ state is given for eight logical configurations (left to right) and for gate-based and pulse-optimized implementations (top and bottom). }
\label{tab:compress}
\end{table*}

\subsection{Resource Estimation modeling}\label{RE}

Finally, we outline the analysis used to produce the estimation of the resources used to implement archetypal quantum algorithms on silicon hardware. Our results are summarized in Fig.~\ref{fig:tradeoff}.  In large, we follow the strategies and conventions outlined in Ref.~\cite{beverland2022assessing}, but from a silicon-specific perspective. The data in Fig.~\ref{fig:tradeoff} is calculated with respect to the standard surface code for dense and sparse architectures and the XZZX surface code for the sparse architecture.

Resource estimation for large-scale fault-tolerant quantum algorithms is commonly performed by reducing all operations to the Cliffords + $T$ gate set. Under QEC, Clifford gates are cheap and $T$ gates are expensive, requiring costly MSD. Our workflow begins by expressing a target algorithm as modular subroutines and compiling these subroutines into a circuit model built from Toffoli gates, Clifford gates, and single-qubit rotations.  Depending on the algorithm, we consider either the native T-gate count or the Toffoli count. Then, we convert all Toffoli gates via a standard fault-tolerant construction that uses a small number of $T$ gates and Clifford gates~\cite{beverland2022assessing, harrigan2024qualtran}.
Next, we approximate the single-qubit rotations by short sequences of Clifford gates and $T$ gates chosen to meet a target precision~\cite{kliuchnikov2023shorter}. This conversion is efficient: the required circuit length of an approximation grows logarithmically with the desired accuracy. Thus, we calculate the total number of $T$ gates needed for the target algorithm by simply summing the contribution from the Toffoli gates and the single-qubit-gate approximations. Finally, we calculate the spatial and temporal resource cost of implementing the $T$ gates needed for the target algorithm. We do so by simulating the performance of the MSD factories described in Secs. \ref{protocols} and \ref{MSD-in-silicon}.

We adopt the standard $15\to1$ distillation protocol, as we found it to outperform the $5\to1$ protocol at comparable target errors for larger quantum algorithms. 
We study three quantum algorithms. Together, they cover the different scientific and commercial areas where quantum computers are expected to be useful: quantum-dynamics simulation~\cite{kim2023evidence,hatano2005finding}, quantum chemistry~\cite{babbush2018encoding, filippov2025architecting}, and integer factoring~\cite{gidney2021factor, shor_factoring}. The three applications showcase different algorithmic structures and resource bottlenecks, highlighting complementary challenges for large-scale fault-tolerant quantum computation. Next, we give brief summaries of the these algorithms. 

The first benchmark algorithm simulates the quantum dynamics of a two-dimensional transverse-field Ising model with 100 spins. This task is one of the smallest scientifically meaningful problems believed to be intractable for classical computers \cite{kim2023evidence}. The Ising Hamiltonian is
\begin{equation}
    H = -J\sum_{\langle j,k\rangle} Z_j Z_k + g\sum_j X_j,
    \label{eq:ising-hamiltonian}
\end{equation}
where $\langle j,k\rangle$ denotes nearest-neighbor pairs on a $\sqrt{N}\times\sqrt{N}$ lattice with $N=100$, $J$ is the coupling strength, and $g$ is the transverse field. Time evolution for 20 steps is implemented, with an overall error capped at 0.1\%, using a fourth-order Trotter-Suzuki product formula~\cite{hatano2005finding}. 

The second benchmark algorithm computes the activation energy level of the ZnS compound~\cite{filippov2025architecting}. The calculation uses quantum phase estimation combined with the double-factorized qubitization algorithm \cite{low2019hamiltonian,von2021quantum}. The algorithm constructs an exact block-encoding of the electronic structure Hamiltonian $H$ and then uses qubitization~\cite{low2019hamiltonian} to form a quantum walk operator $W(H) = e^{i\arccos(H/\lambda)}$, whose eigenphases encode the eigenvalues of the Hamiltonian. And $\lambda$ here is defined as the norm of the system Hamiltonian. Since this operator can be implemented exactly, applying quantum phase estimation to $W(H)$ recovers the spectrum with optimal query complexity. To design circuits whose T-gate cost for implementing $W(H)$ scales well, the Hamiltonian is decomposed into a linear combination of unitaries with efficiently implementable \textsc{SELECT} and \textsc{PREPARE} oracles \cite{childs2012hamiltonian}. When combined with phase estimation, this yields a simulation algorithm whose overall T-gate complexity scales linearly in system size and inverse precision.

The third benchmark algorithm applies Shor’s period finding \cite{shor_factoring, gidney2025factor, gidney2021factor, cain2026shor} to factor a 2048-bit RSA task. This benchmark serves as a canonical benchmark for fault-tolerant quantum architectures, directly related to the long-term security of widely deployed public-key cryptography systems. Our analysis relied on the logical construction from~\cite{gidney2021factor}. More advanced recent techniques~\cite{cain2026shor,babbush2026securing, webster2026pinnacle,luo2026space} may lead to an overhead reduction by two orders of magnitude. 

We work under the assumption that the algorithms are implemented on quantum hardware on which the MSD factories produce $T$ states at a sufficiently fast rate to not limit the algorithms' clock-speeds. With respect to a specific target algorithm, let $N^{\mathrm{phy}}_{\mathrm{q}}$ denote the number of physical qubits, and $N_{\mathrm{t}}$ the number of logical time cycles needed to generate all the $T$ states required by the algorithm. Naturally, one can reduce the runtime of an algorithm by parallelization. However, this will come at the cost of more physical qubits. On the other hand,  one can reduce the spatial footprint by sequentially injecting the minimal number of magic states needed at any one time. In such a scenario, the temporal cost is high. The space-time tradeoff is selected from a set of candidates as follows. 

We illustrate our methodology with a calculation. If the total number of physical qubits is $N^{\mathrm{phy}}_{\mathrm{q}}=N^{\text{phys}}_{\text{T}}m$, then we can produce $m$ magic states with $N_{\text{cycle}}$ cycles. Suppose the algorithm requires $M$ magic $T$ states; we can produce all $M$ in a $\lceil N_{\text{cycle}}M/m \rceil$ logical cycles with $N^{\text{phys}}_{\text{T}}m$ physical qubits. If our algorithm takes $N_{\mathrm{t}}$ logical cycles, then we need $N_{\mathrm{t}}\ge \lceil N_{\text{cycle}}M/m \rceil$ to ensure that the $T$ state production keeps up with the algorithm's $T$-state demands. We can minimize the number of physical qubits $N^{\text{phys}}_{\text{T}}m$ by imposing the equality: $N_{\mathrm{t}}=\lceil N_{\text{cycle}}M/m \rceil=\lceil N_{\text{cycle}}N^{\text{phys}}_{\text{T}}M/N_{\mathrm{q}}\rceil$. Finally, by slowing down the execution of our algorithm (increasing $N_{\mathrm{t}}$) we can reduce the number of physical qubits required for magic state distillation. 
In our resource-estimation routine, we calculate the total space-time volume needed to execute the target algorithms. We also study how the spatial footprint and algorithmic runtime changes as a function of the fraction of the quantum hardware that is allocated to the MSD factories. 

\section{Discussion}
\label{sec: discussion}

Our resource analysis provides a hardware-aware evaluation of magic-state distillation for silicon spin-qubit architectures, bridging theoretical QEC and practical device constraints. By modeling three distinct connectivity regimes (sparse SpinBus, patched, and dense nearest-neighbor architectures) we have quantified the tradeoffs between connectivity, shuttling overheads, and space-time volume for high-fidelity production of logical $T$ states. Several insights emerge.

First, connectivity is a dominant factor in distillation efficiency. Dense layouts consistently outperform sparse and patched architectures in space-time volume, primarily by eliminating shuttling latency and minimizing idle errors. However, the dense regime remains aspirational due to unresolved wiring challenges. The patched architecture offers a pragmatic compromise, achieving overhead reduction by localizing error correction within dense patches while using shuttling only for inter-patch communication. This suggests that near-term experimental efforts should focus on developing modular, patch-based designs that balance connectivity with fabrication feasibility. Second, pulse-level optimization substantively enhances logical coherence and reduces resource overhead. Our memory experiments demonstrate that pulse compression can extend logical coherence further beyond that of the underlying physical qubits.  Integrating hardware-native pulse optimization into syndrome extraction and distillation circuits should therefore be a priority for experimental implementations aiming to minimize space-time costs. Third, code and operation choices interact strongly with architecture. While surface codes with lattice surgery are widely adopted for their local connectivity, we find that both the surface code and the color code with transversal operations can achieve lower space-time overhead in dense regimes. This advantage becomes less obvious in sparse layouts where shuttling latency dominates. Thus, the optimal code choice is not universal, but depends on the underlying hardware connectivity and control capabilities. In addition, biased XZZX-surface code could lead to further resource reduction as it inherently fits the silicon spin-qubit platform where dephasing errors dominate. Finally, our hardware-tailored resource analysis---mapping target logical infidelities back to required physical parameters---provides actionable guidance for experimentalists. For example, in a dense surface-code layout with a target logical $T$-state infidelity of $10^{-12}$, an order-of-magnitude improvement in gate speed or in $T_2^*$ coherence time would reduce the space-time overhead by factors of approximately $6$ and $2$, respectively. These benchmarks help prioritize improvements in fabrication, control, and materials.
A major bottleneck for the Sparse and Patched architectures is the shuttling fidelity. Our pipeline clarified the inability to implement basic QEC codes unless the current best-case shuttling fidelities are improved by an order of magnitude. A more detailed analysis would benefit the experimental development of improved shuttling channels. Such an analysis, which we leave for future work, should include the treatment of dynamic decoherence during shuttling~\cite{nguyen2026suppressing}, spin--orbit effects and valley excitations.

This work opens several avenues for future research. For example, our work could be extended to quantify the resource improvement available via MSD without measurements~\cite{heussen2025magic}. Integrating recent advances in qLDPC codes~\cite{chadwick2026cablecar}, constant-overhead magic-state injection, and distributed quantum computing architectures may also provide new pathways to reduce both qubit counts and runtimes. Moreover, as indicated by our XZZX analysis, the design of bias- and hardware-tailored QEC codes could further lower the resource overheads. Future works should also aim to tailor the decoders to the silicon hardware's correlated noise profile. Such results may also lower overheads by enabling, \textit{e.g.,} single-shot QEC~\cite{bombin2016resilience, liu2024non}. More broadly, the resource-estimation pipeline developed in this work provides a general framework for systematically quantifying the impact of these future advances. As improved codes, decoders, state-injection protocols, and hardware architectures emerge, they can be incorporated into our workflow  to evaluate their effects on logical error rates, qubit requirements, and execution times under realistic noise assumptions. In this sense, our framework is not limited to the present study, but can serve as a reusable benchmark platform for guiding future fault-tolerant quantum computing design.

\bibliography{Bibliography}

@article{kunne2024spinbus,
  title={The SpinBus architecture for scaling spin qubits with electron shuttling},
  author={K{\"u}nne, Matthias and Willmes, Alexander and Oberl{\"a}nder, Max and Gorjaew, Christian and Teske, Julian D and Bhardwaj, Harsh and Beer, Max and Kammerloher, Eugen and Otten, Ren{\'e} and Seidler, Inga and others},
  journal={Nature Communications},
  volume={15},
  number={1},
  pages={4977},
  year={2024},
  publisher={Nature Publishing Group UK London}
}

@article{kliuchnikov2023shorter,
  title={Shorter quantum circuits via single-qubit gate approximation},
  author={Kliuchnikov, Vadym and Lauter, Kristin and Minko, Romy and Paetznick, Adam and Petit, Christophe},
  journal={Quantum},
  volume={7},
  pages={1208},
  year={2023},
  publisher={Verein zur F{\"o}rderung des Open Access Publizierens in den Quantenwissenschaften}
}

@article{lacroix2025scaling,
  title={Scaling and logic in the color code on a superconducting quantum processor},
  author={Lacroix, Nathan and Bourassa, Alexandre and Heras, Francisco JH and Zhang, Lei M and Bausch, Johannes and Senior, Andrew W and Edlich, Thomas and Shutty, Noah and Sivak, Volodymyr and Bengtsson, Andreas and others},
  journal={Nature},
  pages={1--3},
  year={2025},
  publisher={Nature Publishing Group UK London}
}

@article{butt2024fault,
  title={Fault-tolerant code-switching protocols for near-term quantum processors},
  author={Butt, Friederike and Heu{\ss}en, Sascha and Rispler, Manuel and M{\"u}ller, Markus},
  journal={PRX Quantum},
  volume={5},
  number={2},
  pages={020345},
  year={2024},
  publisher={APS}
}

@article{daguerre2025code,
  title={Code switching revisited: Low-overhead magic state preparation using color codes},
  author={Daguerre, Lucas and Kim, Isaac H},
  journal={Physical Review Research},
  volume={7},
  number={2},
  pages={023080},
  year={2025},
  publisher={APS}
}

@article{zhang2024facilitating,
  title={Facilitating practical fault-tolerant quantum computing based on color codes},
  author={Zhang, Jiaxuan and Wu, Yu-Chun and Guo, Guo-Ping},
  journal={Physical Review Research},
  volume={6},
  number={3},
  pages={033086},
  year={2024},
  publisher={APS}
}

@article{wan2024constant,
  title={Constant-time magic state distillation},
  author={Wan, Kwok Ho},
  journal={arXiv preprint arXiv:2410.17992},
  year={2024}
}

@article{litinski2019magic,
  title={Magic state distillation: Not as costly as you think},
  author={Litinski, Daniel},
  journal={Quantum},
  volume={3},
  pages={205},
  year={2019},
  publisher={Verein zur F{\"o}rderung des Open Access Publizierens in den Quantenwissenschaften}
}

@article{yoshida2025low,
  title={Low depth color code circuits with CXSWAP gate},
  author={Yoshida, Satoshi and Gidney, Craig and McEwen, Matt and Zalcman, Adam},
  journal={arXiv preprint arXiv:2510.00370},
  year={2025}
}

@article{erew2025pre,
  title={Pre-Distillation of Magic States via Composite Schemes},
  author={Erew, Muhammad and Goldstein, Moshe and Oz, Yaron and Suchowski, Haim},
  journal={arXiv preprint arXiv:2510.00804},
  year={2025}
}

@article{vandersypen2017interfacing,
  title={Interfacing spin qubits in quantum dots and donors—hot, dense, and coherent},
  author={Vandersypen, Lieven MK and Bluhm, Hendrik and Clarke, John S and Dzurak, Andrew S and Ishihara, Reza and Morello, Andrea and Reilly, David J and Schreiber, Laurens R and Veldhorst, Menno},
  journal={npj Quantum Information},
  volume={3},
  number={1},
  pages={34},
  year={2017},
  publisher={Nature Publishing Group UK London}
}

@article{gutierrez2019transversality,
  title={Transversality and lattice surgery: Exploring realistic routes toward coupled logical qubits with trapped-ion quantum processors},
  author={Guti{\'e}rrez, M and M{\"u}ller, M and Berm{\'u}dez, Alejandro},
  journal={Physical Review A},
  volume={99},
  number={2},
  pages={022330},
  year={2019},
  publisher={APS}
}

@article{daguerre2025experimental,
  title={Experimental demonstration of high-fidelity logical magic states from code switching},
  author={Daguerre, Lucas and Blume-Kohout, Robin and Brown, Natalie C and Hayes, David and Kim, Isaac H},
  journal={Physical Review X},
  volume={15},
  number={4},
  pages={041008},
  year={2025},
  publisher={APS}
}

@article{old2025fault,
  title={Fault-tolerant stabilizer measurements in surface codes with three-qubit gates},
  author={Old, Josias and Tasler, Stephan and Hartmann, Michael J and M{\"u}ller, Markus},
  journal={Physical Review Letters},
  volume={135},
  number={24},
  pages={240601},
  year={2025},
  publisher={APS}
}

@article{fowler2012surface,
  title={Surface codes: Towards practical large-scale quantum computation},
  author={Fowler, Austin G and Mariantoni, Matteo and Martinis, John M and Cleland, Andrew N},
  journal={Physical Review A},
  volume={86},
  number={3},
  pages={032324},
  year={2012},
  publisher={APS}
}

@article{google2025quantum,
  title={Quantum error correction below the surface code threshold},
  journal={Nature},
  volume={638},
  number={8052},
  pages={920--926},
  year={2025},
  publisher={Nature Publishing Group UK London}
}

@article{tasler2025optimizing,
  title={Optimizing Superconducting Three-Qubit Gates for Surface-Code Error Correction},
  author={Tasler, Stephan and Old, Josias and Heunisch, Lukas and Feulner, Verena and Eckstein, Timo and M{\"u}ller, Markus and Hartmann, Michael J},
  journal={arXiv preprint arXiv:2506.09028},
  year={2025}
}

@article{gidney2024magic,
  title={Magic state cultivation: growing T states as cheap as CNOT gates},
  author={Gidney, Craig and Shutty, Noah and Jones, Cody},
  journal={arXiv preprint arXiv:2409.17595},
  year={2024}
}

@article{rosenfeld2025magic,
  title={Magic state cultivation on a superconducting quantum processor},
  author={Rosenfeld, Emma and Gidney, Craig and Roberts, Gabrielle and Morvan, Alexis and Lacroix, Nathan and Kafri, Dvir and Marshall, Jeffrey and Li, Ming and Sivak, Volodymyr and Abanin, Dmitry and others},
  journal={arXiv preprint arXiv:2512.13908},
  year={2025}
}

@article{vaknin2025magic,
  title={Magic State Cultivation on the Surface Code},
  author={Vaknin, Yotam and Jacoby, Shoham and Grimsmo, Arne and Retzker, Alex},
  journal={arXiv preprint arXiv:2502.01743},
  year={2025}
}

@article{tomita2014low,
  title={Low-distance surface codes under realistic quantum noise},
  author={Tomita, Yu and Svore, Krysta M},
  journal={arXiv preprint arXiv:1404.3747},
  year={2014}
}

@article{chen2025efficient,
  title={Efficient Magic State Cultivation on $\mathbb{RP}^2$},
  author={Chen, Zi-Han and Chen, Ming-Cheng and Lu, Chao-Yang and Pan, Jian-Wei},
  journal={arXiv preprint arXiv:2503.18657},
  year={2025}
}

@article{escofet2025quantum,
  title={Quantum Reverse Mapping: Synthesizing an Optimal Spin Qubit Shuttling Bus Architecture for the Surface Code},
  author={Escofet, Pau and Alarc{\'o}n, Eduard and Abadal, Sergi and Semenov, Andrii and Murphy, Niall and Blokhina, Elena and Almud{\'e}ver, Carmen G},
  journal={arXiv preprint arXiv:2510.17689},
  year={2025}
}

@article{ryan2022implementing,
  title={Implementing fault-tolerant entangling gates on the five-qubit code and the color code},
  author={Ryan-Anderson, C and Brown, NC and Allman, MS and Arkin, B and Asa-Attuah, G and Baldwin, C and Berg, J and Bohnet, JG and Braxton, S and Burdick, N and others},
  journal={arXiv preprint arXiv:2208.01863},
  year={2022}
}

@article{stano2022review,
    author={Stano, Peter
    and Loss, Daniel},
    title={Review of performance metrics of spin qubits in gated semiconducting nanostructures},
    journal={Nature Reviews Physics},
    year={2022},
    month={Oct},
    day={01},
    volume={4},
    number={10},
    pages={672-688},
    issn={2522-5820},
    doi={10.1038/s42254-022-00484-w},
    note={{(The preprint \cite{Stano2021} is regularly updated with new data.)}}
}

@misc{Stano2021,
    title={Review of performance metrics of spin qubits in gated semiconducting nanostructures}, 
    author={Stano, Peter
    and Loss, Daniel},
    year={2021},
    month={06},
    day={14},
    eprint={2107.06485},
    archivePrefix={arXiv},
    primaryClass={quant-ph},
    doi={10.48550/arXiv.2107.06485},
    note={{(This preprint is a regularly updated version of Ref.~\cite{stano2022review}. The latest update at the time of writing is from the 24$^{\text{th}}$ of March 2025.)}}
}

@article{meier2012magic,
  title={Magic-state distillation with the four-qubit code},
  author={Meier, Adam M and Eastin, Bryan and Knill, Emanuel},
  journal={arXiv preprint arXiv:1204.4221},
  year={2012}
}

@article{long2025minimal,
  title={Minimal state-preparation times for silicon spin qubits},
  author={Long, Christopher K and Mayhall, Nicholas J and Economou, Sophia E and Barnes, Edwin and Barnes, Crispin HW and Martins, Frederico and Arvidsson-Shukur, David RM and Mertig, Normann},
  journal={npj Quantum Information},
  volume={11},
  number={1},
  pages={113},
  year={2025},
  publisher={Nature Publishing Group UK London}
}

@misc{BenchQ,
  title = {{BenchQ}},
  howpublished = {\url{https://github.com/zapatacomputing/benchq }},
  note = {Accessed: 2026-03-03}
}

@misc{Henrik,
    title = {{Pulse-optimised circuit elements for scalable and noise-resilient quantum chemistry}},
    author = {Gothen, Henrik and Long, Christopher K. and Hiller, Djamila and Qian, Yunming and Barnes, Crispin H. W. and Mertig, Normann and Arvidsson-Shukur, David R. M.},
    note = {In preparation}
}

@article{kim2023evidence,
  title={Evidence for the utility of quantum computing before fault tolerance},
  author={Kim, Youngseok and Eddins, Andrew and Anand, Sajant and Wei, Ken Xuan and Van Den Berg, Ewout and Rosenblatt, Sami and Nayfeh, Hasan and Wu, Yantao and Zaletel, Michael and Temme, Kristan and others},
  journal={Nature},
  volume={618},
  number={7965},
  pages={500--505},
  year={2023},
  publisher={Nature Publishing Group UK London}
}

@misc{long_2025_17116352,
  author       = {Long, Christopher K. and Barnes, Crispin H. W. and Mertig, Normann},
  title        = {Suzuki-Trotter-Evolver (v1.1.0)},
  month        = sep,
  year         = 2025,
  publisher    = {Zenodo},
  version      = {v1.1.0},
  doi          = {10.5281/zenodo.17116352},
  note         = {\newline\href{https://doi.org/10.5281/zenodo.17116352}{https://doi.org/10.5281/zenodo.17116352}},
  url          = {https://doi.org/10.5281/zenodo.17116352},
}

@article{tuckett2020fault,
  title={Fault-tolerant thresholds for the surface code in excess of 5\% under biased noise},
  author={Tuckett, David K and Bartlett, Stephen D and Flammia, Steven T and Brown, Benjamin J},
  journal={Physical review letters},
  volume={124},
  number={13},
  pages={130501},
  year={2020},
  publisher={APS}
}

@article{roffe2023bias,
  title={Bias-tailored quantum LDPC codes},
  author={Roffe, Joschka and Cohen, Lawrence Z and Quintavalle, Armanda O and Chandra, Daryus and Campbell, Earl T},
  journal={Quantum},
  volume={7},
  pages={1005},
  year={2023},
  publisher={Verein zur F{\"o}rderung des Open Access Publizierens in den Quantenwissenschaften}
}

@article{san2023cellular,
  title={A cellular automaton decoder for a noise-bias tailored color code},
  author={San Miguel, Jonathan F and Williamson, Dominic J and Brown, Benjamin J},
  journal={Quantum},
  volume={7},
  pages={940},
  year={2023},
  publisher={Verein zur F{\"o}rderung des Open Access Publizierens in den Quantenwissenschaften}
}

@article{loss1998quantum,
  title={Quantum computation with quantum dots},
  author={Loss, Daniel and DiVincenzo, David P},
  journal={Physical Review A},
  volume={57},
  number={1},
  pages={120},
  year={1998},
  publisher={APS}
}

@misc{harrigan2024qualtran,
    title={Expressing and Analyzing Quantum Algorithms with Qualtran},
    author={Matthew P. Harrigan and Tanuj Khattar
        and Charles Yuan and Anurudh Peduri and Noureldin Yosri
        and Fionn D. Malone and Ryan Babbush and Nicholas C. Rubin},
    year={2024},
    eprint={2409.04643},
    archivePrefix={arXiv},
    primaryClass={quant-ph},
    doi={10.48550/arXiv.2409.04643},
    url={https://arxiv.org/abs/2409.04643},
}

@article{jochym2012robustness,
  title={The robustness of magic state distillation against errors in Clifford gates},
  author={Jochym-O'Connor, Tomas and Yu, Yafei and Helou, Bassam and Laflamme, Raymond},
  journal={arXiv preprint arXiv:1205.6715},
  year={2012}
}

@article{berry2007efficient,
  title={Efficient quantum algorithms for simulating sparse Hamiltonians},
  author={Berry, Dominic W and Ahokas, Graeme and Cleve, Richard and Sanders, Barry C},
  journal={Communications in Mathematical Physics},
  volume={270},
  number={2},
  pages={359--371},
  year={2007},
  publisher={Springer}
}

@article{lee2023evaluating,
  title={Evaluating the evidence for exponential quantum advantage in ground-state quantum chemistry},
  author={Lee, Seunghoon and Lee, Joonho and Zhai, Huanchen and Tong, Yu and Dalzell, Alexander M and Kumar, Ashutosh and Helms, Phillip and Gray, Johnnie and Cui, Zhi-Hao and Liu, Wenyuan and others},
  journal={Nature communications},
  volume={14},
  number={1},
  pages={1952},
  year={2023},
  publisher={Nature Publishing Group UK London}
}

@article{dong2022ground,
  title={Ground-state preparation and energy estimation on early fault-tolerant quantum computers via quantum eigenvalue transformation of unitary matrices},
  author={Dong, Yulong and Lin, Lin and Tong, Yu},
  journal={PRX quantum},
  volume={3},
  number={4},
  pages={040305},
  year={2022},
  publisher={APS}
}

@article{sarvepalli2009asymmetric,
  title={Asymmetric quantum codes: constructions, bounds and performance},
  author={Sarvepalli, Pradeep Kiran and Klappenecker, Andreas and R{\"o}tteler, Martin},
  journal={Proceedings of the Royal Society A: Mathematical, Physical and Engineering Sciences},
  volume={465},
  number={2105},
  pages={1645--1672},
  year={2009},
  publisher={The Royal Society London}
}

@article{ghosh2012surface,
  title={Surface code with decoherence: An analysis of three superconducting architectures},
  author={Ghosh, Joydip and Fowler, Austin G and Geller, Michael R},
  journal={Physical Review A—Atomic, Molecular, and Optical Physics},
  volume={86},
  number={6},
  pages={062318},
  year={2012},
  publisher={APS}
}

@book{cormen2022introduction,
  title={Introduction to algorithms},
  author={Cormen, Thomas H and Leiserson, Charles E and Rivest, Ronald L and Stein, Clifford},
  year={2022},
  publisher={MIT press}
}

@article{newman2000efficient,
  title={Efficient Monte Carlo algorithm and high-precision results for percolation},
  author={Newman, MEJ and Ziff, Robert M},
  journal={Physical Review Letters},
  volume={85},
  number={19},
  pages={4104},
  year={2000},
  publisher={APS}
}

@article{bravyi2005universal,
  title={Universal quantum computation with ideal Clifford gates and noisy ancillas},
  author={Bravyi, Sergey and Kitaev, Alexei},
  journal={Physical Review A—Atomic, Molecular, and Optical Physics},
  volume={71},
  number={2},
  pages={022316},
  year={2005},
  publisher={APS}
}

@article{itogawa2024even,
  title={Even more efficient magic state distillation by zero-level distillation},
  author={Itogawa, Tomohiro and Takada, Yugo and Hirano, Yutaka and Fujii, Keisuke},
  journal={arXiv preprint arXiv:2403.03991},
  year={2024}
}

@article{jones2012multilevel,
  title={Multilevel distillation of magic states for quantum computing},
  author={Jones, Cody},
  journal={arXiv preprint arXiv:1210.3388},
  year={2012}
}

@article{reichardt2005quantum,
  title={Quantum universality from magic states distillation applied to CSS codes},
  author={Reichardt, Ben W},
  journal={Quantum Information Processing},
  volume={4},
  number={3},
  pages={251--264},
  year={2005},
  publisher={Springer}
}

@article{reichardt2004improved,
  title={Improved magic states distillation for quantum universality},
  author={Reichardt, Ben W},
  journal={arXiv preprint quant-ph/0411036},
  year={2004}
}

@article{reichardt2006quantum,
  title={Quantum universality by state distillation},
  author={Reichardt, Ben W},
  journal={arXiv preprint quant-ph/0608085},
  year={2006}
}

@article{laflamme1996perfect,
  title={Perfect quantum error correcting code},
  author={Laflamme, Raymond and Miquel, Cesar and Paz, Juan Pablo and Zurek, Wojciech Hubert},
  journal={Physical Review Letters},
  volume={77},
  number={1},
  pages={198},
  year={1996},
  publisher={APS}
}

@article{bermudez2017assessing,
  title={Assessing the progress of trapped-ion processors towards fault-tolerant quantum computation},
  author={Bermudez, Alejandro and Xu, Xiaosi and Nigmatullin, Ramil and O'Gorman, Joe and Negnevitsky, Vlad and Schindler, Philipp and Monz, Thomas and Poschinger, Ulrich G and Hempel, Cornelius and Home, Jonathan and others},
  journal={Physical Review X},
  volume={7},
  number={4},
  pages={041061},
  year={2017},
  publisher={APS}
}

@article{hirano2025efficient,
  title={Efficient magic state cultivation with lattice surgery},
  author={Hirano, Yutaka and Toshio, Riki and Itogawa, Tomohiro and Fujii, Keisuke},
  journal={arXiv preprint arXiv:2510.24615},
  year={2025}
}

@article{sahay2025error,
  title={Error Correction of Transversal CNOT Gates for Scalable Surface-Code Computation},
  author={Sahay, Kaavya and Lin, Yingjia and Huang, Shilin and Brown, Kenneth R and Puri, Shruti},
  journal={PRX quantum},
  volume={6},
  number={2},
  pages={020326},
  year={2025},
  publisher={APS}
}

@article{landahl2014quantum,
  title={Quantum computing by color-code lattice surgery},
  author={Landahl, Andrew J and Ryan-Anderson, Ciaran},
  journal={arXiv preprint arXiv:1407.5103},
  year={2014}
}

@article{thomsen2024low,
  title={Low-overhead quantum computing with the color code},
  author={Thomsen, Felix and Kesselring, Markus S and Bartlett, Stephen D and Brown, Benjamin J},
  journal={Physical Review Research},
  volume={6},
  number={4},
  pages={043125},
  year={2024},
  publisher={APS}
}

@article{fazio2025low,
  title={Low-overhead Magic State Circuits with Transversal CNOTs},
  author={Fazio, Nicholas and Webster, Mark and Cai, Zhenyu},
  journal={arXiv preprint arXiv:2501.10291},
  year={2025}
}

@article{cai2023looped,
  title={Looped pipelines enabling effective 3D qubit lattices in a strictly 2D device},
  author={Cai, Zhenyu and Siegel, Adam and Benjamin, Simon},
  journal={PRX Quantum},
  volume={4},
  number={2},
  pages={020345},
  year={2023},
  publisher={APS}
}

@article{lee2025low,
  title={Low-overhead magic state distillation with color codes},
  author={Lee, Seok-Hyung and Thomsen, Felix and Fazio, Nicholas and Brown, Benjamin J and Bartlett, Stephen D},
  journal={PRX Quantum},
  volume={6},
  number={3},
  pages={030317},
  year={2025},
  publisher={APS}
}

@article{chadwick2025manufacturable,
  title={A manufacturable surface code architecture for spin qubits with fast transversal logic},
  author={Chadwick, Jason D and Yang, Willers and Viszlai, Joshua and Chong, Frederic T},
  journal={arXiv preprint arXiv:2512.07131},
  year={2025}
}

@article{cywinski2008enhance,
  title={How to enhance dephasing time in superconducting qubits},
  author={Cywi{\'n}ski, {\L}ukasz and Lutchyn, Roman M and Nave, Cody P and Das Sarma, S},
  journal={Physical Review B},
  volume={77},
  number={17},
  pages={174509},
  year={2008},
  publisher={APS}
}

@article{beverland2022assessing,
  title={Assessing requirements to scale to practical quantum advantage},
  author={Beverland, Michael E and Murali, Prakash and Troyer, Matthias and Svore, Krysta M and Hoefler, Torsten and Kliuchnikov, Vadym and Low, Guang Hao and Soeken, Mathias and Sundaram, Aarthi and Vaschillo, Alexander},
  journal={arXiv preprint arXiv:2211.07629},
  year={2022}
}

@article{zhou2025opportunities,
  title={Opportunities in full-stack design of low-overhead fault-tolerant quantum computation},
  author={Zhou, Hengyun and Cain, Madelyn and Lukin, Mikhail D},
  journal={Nature Computational Science},
  volume={5},
  number={12},
  pages={1110--1119},
  year={2025},
  publisher={Nature Publishing Group}
}

@article{ciroth2025numerical,
  title={Numerical simulation of coherent spin-shuttling in a QuBus with charged defects},
  author={Ciroth, Nils and Sala, Arnau and Xue, Ran and Ermoneit, Lasse and Koprucki, Thomas and Kantner, Markus and Schreiber, Lars R},
  journal={arXiv preprint arXiv:2512.03588},
  year={2025}
}

@article{veldhorst2017silicon,
  title={Silicon CMOS architecture for a spin-based quantum computer},
  author={Veldhorst, Menno and Eenink, Harm GJ and Yang, Chih-Hwan and Dzurak, Andrew S},
  journal={Nature communications},
  volume={8},
  number={1},
  pages={1766},
  year={2017},
  publisher={Nature Publishing Group UK London}
}

@article{nielsen2002simple,
  title={A simple formula for the average gate fidelity of a quantum dynamical operation},
  author={Nielsen, Michael A},
  journal={Physics Letters A},
  volume={303},
  number={4},
  pages={249--252},
  year={2002},
  publisher={Elsevier}
}

@article{kimmel2014robust,
  title={Robust extraction of tomographic information via randomized benchmarking},
  author={Kimmel, Shelby and da Silva, Marcus P and Ryan, Colm A and Johnson, Blake R and Ohki, Thomas},
  journal={Physical Review X},
  volume={4},
  number={1},
  pages={011050},
  year={2014},
  publisher={APS}
}

@article{green2013arbitrary,
  title={Arbitrary quantum control of qubits in the presence of universal noise},
  author={Green, Todd J and Sastrawan, Jarrah and Uys, Hermann and Biercuk, Michael J},
  journal={New Journal of Physics},
  volume={15},
  number={9},
  pages={095004},
  year={2013},
  publisher={IOP Publishing}
}

@article{cerfontaine2021filter,
  title={Filter functions for quantum processes under correlated noise},
  author={Cerfontaine, Pascal and Hangleiter, Tobias and Bluhm, Hendrik},
  journal={Physical Review Letters},
  volume={127},
  number={17},
  pages={170403},
  year={2021},
  publisher={APS}
}

@article{bolsmann2025fast,
  title={Fast Native Three-Qubit Gates and Fault-Tolerant Quantum Error Correction with Trapped Rydberg Ions},
  author={Bolsmann, Katrin and Guedes, Thiago LM and Li, Weibin and Wilkinson, Joseph WP and Lesanovsky, Igor and M{\"u}ller, Markus},
  journal={arXiv preprint arXiv:2512.16641},
  year={2025}
}

@article{crawford2023compilation,
  title={Compilation and scaling strategies for a silicon quantum processor with sparse two-dimensional connectivity},
  author={Crawford, O and Cruise, JR and Mertig, N and Gonzalez-Zalba, MF},
  journal={npj Quantum Information},
  volume={9},
  number={1},
  pages={13},
  year={2023},
  publisher={Nature Publishing Group UK London}
}

@article{ginzel2024scalable,
  title={Scalable parity architecture with a shuttling-based spin qubit processor},
  author={Ginzel, Florian and Fellner, Michael and Ertler, Christian and Schreiber, Lars R and Bluhm, Hendrik and Lechner, Wolfgang},
  journal={Physical Review B},
  volume={110},
  number={7},
  pages={075302},
  year={2024},
  publisher={APS}
}

@article{beverland2021cost,
  title={Cost of universality: A comparative study of the overhead of state distillation and code switching with color codes},
  author={Beverland, Michael E and Kubica, Aleksander and Svore, Krysta M},
  journal={PRX Quantum},
  volume={2},
  number={2},
  pages={020341},
  year={2021},
  publisher={APS}
}

@article{seidler2022conveyor,
  title={Conveyor-mode single-electron shuttling in Si/SiGe for a scalable quantum computing architecture},
  author={Seidler, Inga and Struck, Tom and Xue, Ran and Focke, Niels and Trellenkamp, Stefan and Bluhm, Hendrik and Schreiber, Lars R},
  journal={npj Quantum information},
  volume={8},
  number={1},
  pages={100},
  year={2022},
  publisher={Nature Publishing Group UK London}
}

@article{langrock2023blueprint,
  title={Blueprint of a scalable spin qubit shuttle device for coherent mid-range qubit transfer in disordered Si/SiGe/SiO 2},
  author={Langrock, Veit and Krzywda, Jan A and Focke, Niels and Seidler, Inga and Schreiber, Lars R and Cywi{\'n}ski, {\L}ukasz},
  journal={PRX Quantum},
  volume={4},
  number={2},
  pages={020305},
  year={2023},
  publisher={APS}
}

@article{li2015magic,
  title={A magic state’s fidelity can be superior to the operations that created it},
  author={Li, Ying},
  journal={New Journal of Physics},
  volume={17},
  number={2},
  pages={023037},
  year={2015},
  publisher={IOP Publishing}
}

@incollection{hatano2005finding,
  title={Finding exponential product formulas of higher orders},
  author={Hatano, Naomichi and Suzuki, Masuo},
  booktitle={Quantum annealing and other optimization methods},
  pages={37--68},
  year={2005},
  publisher={Springer}
}

@article{von2021quantum,
  title={Quantum computing enhanced computational catalysis},
  author={von Burg, Vera and Low, Guang Hao and H{\"a}ner, Thomas and Steiger, Damian S and Reiher, Markus and Roetteler, Martin and Troyer, Matthias},
  journal={Physical Review Research},
  volume={3},
  number={3},
  pages={033055},
  year={2021},
  publisher={APS}
}

@article{low2019hamiltonian,
  title={Hamiltonian simulation by qubitization},
  author={Low, Guang Hao and Chuang, Isaac L},
  journal={Quantum},
  volume={3},
  pages={163},
  year={2019},
  publisher={Verein zur F{\"o}rderung des Open Access Publizierens in den Quantenwissenschaften}
}

@article{mohseni2024build,
  title={How to build a quantum supercomputer: Scaling from hundreds to millions of qubits},
  author={Mohseni, Masoud and Scherer, Artur and Johnson, K Grace and Wertheim, Oded and Otten, Matthew and Aadit, Navid Anjum and Alexeev, Yuri and Bresniker, Kirk M and Camsari, Kerem Y and Chapman, Barbara and others},
  journal={arXiv preprint arXiv:2411.10406},
  year={2024}
}

@article{pataki2025compiling,
  title={Compiling the surface code to crossbar spin qubit architectures},
  author={Pataki, D{\'a}vid and P{\'a}lyi, Andr{\'a}s},
  journal={Physical Review B},
  volume={111},
  number={11},
  pages={115307},
  year={2025},
  publisher={APS}
}

@article{boter2022spiderweb,
  title={Spiderweb array: A sparse spin-qubit array},
  author={Boter, Jelmer M and Dehollain, Juan P and Van Dijk, Jeroen PG and Xu, Yuanxing and Hensgens, Toivo and Versluis, Richard and Naus, Henricus WL and Clarke, James S and Veldhorst, Menno and Sebastiano, Fabio and others},
  journal={Physical review applied},
  volume={18},
  number={2},
  pages={024053},
  year={2022},
  publisher={APS}
}

@article{siegel2024towards,
  title={Towards early fault tolerance on a 2$\times$ n array of qubits equipped with shuttling},
  author={Siegel, Adam and Strikis, Armands and Fogarty, Michael},
  journal={PRX Quantum},
  volume={5},
  number={4},
  pages={040328},
  year={2024},
  publisher={APS}
}

@article{otxoa2025spinhex,
  title={SpinHex: A low-crosstalk, spin-qubit architecture based on multi-electron couplers},
  author={Otxoa, Rub{\'e}n M and Martinez, Josu Etxezarreta and Schnabl, Paul and Mertig, Normann and Smith, Charles and Martins, Frederico},
  journal={arXiv preprint arXiv:2504.03149},
  year={2025}
}

@article{li2018crossbar,
  title={A crossbar network for silicon quantum dot qubits},
  author={Li, Ruoyu and Petit, Luca and Franke, David P and Dehollain, Juan Pablo and Helsen, Jonas and Steudtner, Mark and Thomas, Nicole K and Yoscovits, Zachary R and Singh, Kanwal J and Wehner, Stephanie and others},
  journal={Science advances},
  volume={4},
  number={7},
  pages={eaar3960},
  year={2018},
  publisher={American Association for the Advancement of Science}
}

@article{siegel2025snakes,
  title={Snakes on a Plane: mobile, low dimensional logical qubits on a 2D surface},
  author={Siegel, Adam and Cai, Zhenyu and Jnane, Hamza and Koczor, Balint and Pexton, Shaun and Strikis, Armands and Benjamin, Simon},
  journal={arXiv preprint arXiv:2501.02120},
  year={2025}
}

@article{wang2024operating,
  title={Operating semiconductor quantum processors with hopping spins},
  author={Wang, Chien-An and John, Valentin and Tidjani, Hanifa and Yu, C{\'e}cile X and Ivlev, Alexander S and D{\'e}prez, Corentin and van Riggelen-Doelman, Floor and Woods, Benjamin D and Hendrickx, Nico W and Lawrie, William IL and others},
  journal={Science},
  volume={385},
  number={6707},
  pages={447--452},
  year={2024},
  publisher={American Association for the Advancement of Science}
}

@article{elsayed2024low,
  title={Low charge noise quantum dots with industrial CMOS manufacturing},
  author={Elsayed, Asser and Shehata, MMK and Godfrin, Clement and Kubicek, Stefan and Massar, Shana and Canvel, Yann and Jussot, Julien and Simion, George and Mongillo, Massimo and Wan, Danny and others},
  journal={npj Quantum Information},
  volume={10},
  number={1},
  pages={70},
  year={2024},
  publisher={Nature Publishing Group UK London}
}

@article{culcer2009dephasing,
  title={Dephasing of Si spin qubits due to charge noise},
  author={Culcer, Dimitrie and Hu, Xuedong and Das Sarma, S},
  journal={Applied Physics Letters},
  volume={95},
  number={7},
  year={2009},
  publisher={AIP Publishing}
}

@article{dasu2025breaking,
  title={Breaking even with magic: demonstration of a high-fidelity logical non-Clifford gate},
  author={Dasu, Shival and Burton, Simon and Mayer, Karl and Amaro, David and Gerber, Justin A and Gilmore, Kevin and Gresh, Dan and DelVento, Davide and Potter, Andrew C and Hayes, David},
  journal={arXiv preprint arXiv:2506.14688},
  year={2025}
}

@article{claes2025lower,
  title={Lower-depth local encoding circuits for the surface code},
  author={Claes, Jahan},
  journal={arXiv preprint arXiv:2509.09779},
  year={2025}
}

@article{higgott2021optimal,
  title={Optimal local unitary encoding circuits for the surface code},
  author={Higgott, Oscar and Wilson, Matthew and Hefford, James and Dborin, James and Hanif, Farhan and Burton, Simon and Browne, Dan E},
  journal={Quantum},
  volume={5},
  pages={517},
  year={2021},
  publisher={Verein zur F{\"o}rderung des Open Access Publizierens in den Quantenwissenschaften}
}

@article{liu2024non,
  title={Non-Markovian noise suppression simplified through channel representation},
  author={Liu, Zhenhuan and Xiao, Yunlong and Cai, Zhenyu},
  journal={arXiv preprint arXiv:2412.11220},
  year={2024}
}

@article{bombin2016resilience,
  title={Resilience to time-correlated noise in quantum computation},
  author={Bomb{\'\i}n, H{\'e}ctor},
  journal={Physical Review X},
  volume={6},
  number={4},
  pages={041034},
  year={2016},
  publisher={APS}
}

@article{google2023suppressing,
  title={Suppressing quantum errors by scaling a surface code logical qubit},
  journal={Nature},
  volume={614},
  number={7949},
  pages={676--681},
  year={2023},
  publisher={Nature Publishing Group UK London}
}

@article{o2017quantum,
  title={Quantum computation with realistic magic-state factories},
  author={O'Gorman, Joe and Campbell, Earl T},
  journal={Physical Review A},
  volume={95},
  number={3},
  pages={032338},
  year={2017},
  publisher={APS}
}

@article{devitt2013quantum,
  title={Quantum error correction for beginners},
  author={Devitt, Simon J and Munro, William J and Nemoto, Kae},
  journal={Reports on Progress in Physics},
  volume={76},
  number={7},
  pages={076001},
  year={2013},
  publisher={IOP Publishing}
}

@article{steane2003overhead,
  title={Overhead and noise threshold of fault-tolerant quantum error correction},
  author={Steane, Andrew M},
  journal={Physical Review A},
  volume={68},
  number={4},
  pages={042322},
  year={2003},
  publisher={APS}
}

@article{gidney2021stim,
  title={Stim: a fast stabilizer circuit simulator},
  author={Gidney, Craig},
  journal={Quantum},
  volume={5},
  pages={497},
  year={2021},
  publisher={Verein zur F{\"o}rderung des Open Access Publizierens in den Quantenwissenschaften}
}

@article{blanes2009magnus,
  title={The Magnus expansion and some of its applications},
  author={Blanes, Sergio and Casas, Fernando and Oteo, Jose-Angel and Ros, Jos{\'e}},
  journal={Physics reports},
  volume={470},
  number={5-6},
  pages={151--238},
  year={2009},
  publisher={Elsevier}
}

@article{hastings2018distillation,
  title={Distillation with sublogarithmic overhead},
  author={Hastings, Matthew B and Haah, Jeongwan},
  journal={Physical review letters},
  volume={120},
  number={5},
  pages={050504},
  year={2018},
  publisher={APS}
}

@article{steane2002quantum,
  title={Quantum reed-muller codes},
  author={Steane, Andrew M},
  journal={IEEE Transactions on Information Theory},
  volume={45},
  number={5},
  pages={1701--1703},
  year={2002},
  publisher={IEEE}
}

@article{sales2025experimental,
  title={Experimental demonstration of logical magic state distillation},
  author={Sales Rodriguez, Pedro and Robinson, John M and Jepsen, Paul Niklas and He, Zhiyang and Duckering, Casey and Zhao, Chen and Wu, Kai-Hsin and Campo, Joseph and Bagnall, Kevin and Kwon, Minho and others},
  journal={Nature},
  volume={645},
  number={8081},
  pages={620--625},
  year={2025},
  publisher={Nature Publishing Group UK London}
}

@article{chamberland2020triangular,
  title={Triangular color codes on trivalent graphs with flag qubits},
  author={Chamberland, Christopher and Kubica, Aleksander and Yoder, Theodore J and Zhu, Guanyu},
  journal={New Journal of Physics},
  volume={22},
  number={2},
  pages={023019},
  year={2020},
  publisher={IOP Publishing}
}

@article{gong2022experimental,
  title={Experimental exploration of five-qubit quantum error-correcting code with superconducting qubits},
  author={Gong, Ming and Yuan, Xiao and Wang, Shiyu and Wu, Yulin and Zhao, Youwei and Zha, Chen and Li, Shaowei and Zhang, Zhen and Zhao, Qi and Liu, Yunchao and others},
  journal={National Science Review},
  volume={9},
  number={1},
  pages={nwab011},
  year={2022},
  publisher={Oxford University Press}
}

@article{zhang2025demonstration,
  title={Demonstration of quantum error detection in a silicon quantum processor},
  author={Zhang, Chunhui and Li, Chunhui and Tian, Zhen and Jiang, Yan and Xu, Feng and Zhang, Shihang and Wang, Hao and Zhang, Yu-Ning and Bai, Xuesong and Zhao, Baolong and others},
  journal={arXiv preprint arXiv:2509.24766},
  year={2025}
}

@article{gavriel2022transversal,
  title={Transversal injection: A method for direct encoding of ancilla states for non-Clifford gates using stabiliser codes},
  author={Gavriel, Jason and Herr, Daniel and Shaw, Alexis and Bremner, Michael J and Paler, Alexandru and Devitt, Simon J},
  journal={arXiv preprint arXiv:2211.10046},
  year={2022}
}

@article{choi2023fault,
  title={Fault tolerant non-clifford state preparation for arbitrary rotations},
  author={Choi, Hyeongrak and Chong, Frederic T and Englund, Dirk and Ding, Yongshan},
  journal={arXiv preprint arXiv:2303.17380},
  year={2023}
}

@article{volmer2025reduction,
  title={Reduction of the impact of the local valley splitting on the coherence of conveyor-belt spin shuttling in $\hat{28}$ Si/SiGe},
  author={Volmer, Mats and Struck, Tom and Tu, Jhih-Sian and Trellenkamp, Stefan and Esposti, Davide Degli and Scappucci, Giordano and Cywi{\'n}ski, {\L}ukasz and Bluhm, Hendrik and Schreiber, Lars R},
  journal={arXiv preprint arXiv:2510.03773},
  year={2025}
}

@inproceedings{shor_factoring,
  author = {Shor, Peter W.},
  title = {Algorithms for quantum computation: discrete logarithms and factoring},
  booktitle = {Proceedings of the 35th Annual Symposium on Foundations of Computer Science},
  pages = {124--134},
  year = {1994},
  doi = {10.1109/SFCS.1994.365700}
}

@article{gidney2021factor,
  title={How to factor 2048 bit RSA integers in 8 hours using 20 million noisy qubits},
  author={Gidney, Craig and Eker{\aa}, Martin},
  journal={Quantum},
  volume={5},
  pages={433},
  year={2021},
  publisher={Verein zur F{\"o}rderung des Open Access Publizierens in den Quantenwissenschaften}
}

@article{childs2018toward,
  title={Toward the first quantum simulation with quantum speedup},
  author={Childs, Andrew M and Maslov, Dmitri and Nam, Yunseong and Ross, Neil J and Su, Yuan},
  journal={Proceedings of the National Academy of Sciences},
  volume={115},
  number={38},
  pages={9456--9461},
  year={2018},
  publisher={National Academy of Sciences}
}

@article{lloyd1996universal,
  title={Universal quantum simulators},
  author={Lloyd, Seth},
  journal={Science},
  volume={273},
  number={5278},
  pages={1073--1078},
  year={1996},
  publisher={American Association for the Advancement of Science}
}

@article{babbush2018encoding,
  title={Encoding electronic spectra in quantum circuits with linear T complexity},
  author={Babbush, Ryan and Gidney, Craig and Berry, Dominic W and Wiebe, Nathan and McClean, Jarrod and Paler, Alexandru and Fowler, Austin and Neven, Hartmut},
  journal={Physical Review X},
  volume={8},
  number={4},
  pages={041015},
  year={2018},
  publisher={APS}
}

@inproceedings{gilyen2019optimizing,
  title={Optimizing quantum optimization algorithms via faster quantum gradient computation},
  author={Gily{\'e}n, Andr{\'a}s and Arunachalam, Srinivasan and Wiebe, Nathan},
  booktitle={Proceedings of the Thirtieth Annual ACM-SIAM Symposium on Discrete Algorithms},
  pages={1425--1444},
  year={2019},
  organization={SIAM}
}

@article{farhi2014quantum,
  title={A quantum approximate optimization algorithm},
  author={Farhi, Edward and Goldstone, Jeffrey and Gutmann, Sam},
  journal={arXiv preprint arXiv:1411.4028},
  year={2014}
}

@article{kim2024magic,
  title={Magic State Injection on IBM Quantum Processors Above the Distillation Threshold},
  author={Kim, Younghun and Sevior, Martin and Usman, Muhammad},
  journal={arXiv preprint arXiv:2412.01446},
  year={2024}
}

@inproceedings{lao2022magic,
  title={Magic state injection on the rotated surface code},
  author={Lao, Lingling and Criger, Ben},
  booktitle={Proceedings of the 19th ACM International Conference on Computing Frontiers},
  pages={113--120},
  year={2022}
}

@article{zhang2025constant,
  title={Constant-Overhead Magic State Injection into qLDPC Codes with Error Independence Guarantees},
  author={Zhang, Guo and Zhu, Yuanye and Yuan, Xiao and Li, Ying},
  journal={arXiv preprint arXiv:2505.06981},
  year={2025}
}

@article{heussen2025magic,
  title={Magic state distillation without measurements and post-selection},
  author={Heu{\ss}en, Sascha},
  journal={arXiv preprint arXiv:2504.17509},
  year={2025}
}

@article{o2025compare,
  title={Compare the pair: Rotated versus unrotated surface codes at equal logical error rates},
  author={O'Rourke, Anthony Ryan and Devitt, Simon},
  journal={Physical Review Research},
  volume={7},
  number={3},
  pages={033074},
  year={2025},
  publisher={APS}
}

@article{yoshida2025concatenate,
  title={Concatenate codes, save qubits},
  author={Yoshida, Satoshi and Tamiya, Shiro and Yamasaki, Hayata},
  journal={npj Quantum Information},
  volume={11},
  number={1},
  pages={88},
  year={2025},
  publisher={Nature Publishing Group UK London}
}

@article{wu2025bias,
  title={Bias-tailored single-shot quantum LDPC codes},
  author={Wu, Shixin and Brun, Todd A and Lidar, Daniel A},
  journal={arXiv preprint arXiv:2507.02239},
  year={2025}
}

@article{ruiz2025unfolded,
  title={Unfolded distillation: very low-cost magic state preparation for biased-noise qubits},
  author={Ruiz, Diego and Guillaud, J{\'e}r{\'e}mie and Vuillot, Christophe and Mirrahimi, Mazyar},
  journal={arXiv preprint arXiv:2507.12511},
  year={2025}
}

@article{goto2014step,
  title={Step-by-step magic state encoding for efficient fault-tolerant quantum computation},
  author={Goto, Hayato},
  journal={Scientific Reports},
  volume={4},
  number={1},
  pages={7501},
  year={2014},
  publisher={Nature Publishing Group UK London}
}

@article{gonzalez2021scaling,
  title={Scaling silicon-based quantum computing using CMOS technology},
  author={Gonzalez-Zalba, MF and De Franceschi, S and Charbon, E and Meunier, Tristan and Vinet, M and Dzurak, AS},
  journal={Nature Electronics},
  volume={4},
  number={12},
  pages={872--884},
  year={2021},
  publisher={Nature Publishing Group UK London}
}

@article{maurand2016cmos,
  title={A CMOS silicon spin qubit},
  author={Maurand, R and Jehl, X and Kotekar-Patil, D and Corna, Andrea and Bohuslavskyi, Heorhii and Lavi{\'e}ville, R and Hutin, L and Barraud, S and Vinet, M and Sanquer, M and others},
  journal={Nature communications},
  volume={7},
  number={1},
  pages={13575},
  year={2016},
  publisher={Nature Publishing Group UK London}
}

@article{zwerver2022qubits,
  title={Qubits made by advanced semiconductor manufacturing},
  author={Zwerver, AMJ and Kr{\"a}henmann, T and Watson, TF and Lampert, Lester and George, Hubert C and Pillarisetty, Ravi and Bojarski, SA and Amin, Payam and Amitonov, SV and Boter, JM and others},
  journal={Nature Electronics},
  volume={5},
  number={3},
  pages={184--190},
  year={2022},
  publisher={Nature Publishing Group UK London}
}

@article{swift2025large,
  title={Large-scale characterization of Single-Hole Transistors in 22-nm FDSOI CMOS Technology},
  author={Swift, Thomas H and Gomez-Saiz, Alberto and Ciriano-Tejel, Virginia N and Wise, David F and Noah, Grayson M and Morton, John JL and Gonzalez-Zalba, M Fernando and Johnson, Mark AI},
  journal={arXiv preprint arXiv:2507.21306},
  year={2025}
}

@article{swift2025superinductor,
  title={A superinductor in a deep sub-micron integrated circuit},
  author={Swift, Thomas H and Olivieri, Fabio and Aizpurua-Iraola, Gorka and Kirkman, James and Noah, Grayson M and de Kruijf, Mathieu and von Horstig, Felix-Ekkehard and Gomez-Saiz, Alberto and Morton, John JL and Gonzalez-Zalba, M Fernando},
  journal={arXiv preprint arXiv:2507.13202},
  year={2025}
}

@article{clarke2025spin,
  title={Spin Readout in a 22 nm Node Integrated Circuit},
  author={Clarke, Isobel C and Ciriano-Tejel, Virginia and Ibberson, David J and Noah, Grayson M and Swift, Thomas H and Johnson, Mark AI and Leon, Ross CC and Gomez-Saiz, Alberto and Morton, John JL and Gonzalez-Zalba, M Fernando},
  journal={arXiv preprint arXiv:2510.13674},
  year={2025}
}

@article{philips2022universal,
  title={Universal control of a six-qubit quantum processor in silicon},
  author={Philips, Stephan GJ and Madzik, Mateusz T and Amitonov, Sergey V and de Snoo, Sander L and Russ, Maximilian and Kalhor, Nima and Volk, Christian and Lawrie, William IL and Brousse, Delphine and Tryputen, Larysa and others},
  journal={Nature},
  volume={609},
  number={7929},
  pages={919--924},
  year={2022},
  publisher={Nature Publishing Group UK London}
}

@article{yoneda2018quantum,
  title={A quantum-dot spin qubit with coherence limited by charge noise and fidelity higher than 99.9\%},
  author={Yoneda, Jun and Takeda, Kenta and Otsuka, Tomohiro and Nakajima, Takashi and Delbecq, Matthieu R and Allison, Giles and Honda, Takumu and Kodera, Tetsuo and Oda, Shunri and Hoshi, Yusuke and others},
  journal={Nature nanotechnology},
  volume={13},
  number={2},
  pages={102--106},
  year={2018},
  publisher={Nature Publishing Group UK London}
}

@article{noiri2022fast,
  title={Fast universal quantum gate above the fault-tolerance threshold in silicon},
  author={Noiri, Akito and Takeda, Kenta and Nakajima, Takashi and Kobayashi, Takashi and Sammak, Amir and Scappucci, Giordano and Tarucha, Seigo},
  journal={Nature},
  volume={601},
  number={7893},
  pages={338--342},
  year={2022},
  publisher={Nature Publishing Group UK London}
}

@article{mills2022two,
  title={Two-qubit silicon quantum processor with operation fidelity exceeding 99\%},
  author={Mills, Adam R and Guinn, Charles R and Gullans, Michael J and Sigillito, Anthony J and Feldman, Mayer M and Nielsen, Erik and Petta, Jason R},
  journal={Science Advances},
  volume={8},
  number={14},
  pages={eabn5130},
  year={2022},
  publisher={American Association for the Advancement of Science}
}

@article{de2025high,
  title={High-fidelity single-spin shuttling in silicon},
  author={De Smet, Maxim and Matsumoto, Yuta and Zwerver, Anne-Marije J and Tryputen, Larysa and de Snoo, Sander L and Amitonov, Sergey V and Katiraee-Far, Sam R and Sammak, Amir and Samkharadze, Nodar and G{\"u}l, {\"O}nder and others},
  journal={Nature Nanotechnology},
  pages={1--7},
  year={2025},
  publisher={Nature Publishing Group UK London}
}

@article{mkadzik2022precision,
  title={Precision tomography of a three-qubit donor quantum processor in silicon},
  author={M{\k{a}}dzik, Mateusz T and Asaad, Serwan and Youssry, Akram and Joecker, Benjamin and Rudinger, Kenneth M and Nielsen, Erik and Young, Kevin C and Proctor, Timothy J and Baczewski, Andrew D and Laucht, Arne and others},
  journal={Nature},
  volume={601},
  number={7893},
  pages={348--353},
  year={2022},
  publisher={Nature Publishing Group UK London}
}

@article{micciche2025optimizing,
  title={Optimizing compilation of error correction codes for 2xN quantum dot arrays and its NP-hardness},
  author={Micciche, Anthony and Mian, Feroz Ahmed and Chatterjee, Anasua and McGregor, Andrew and Krastanov, Stefan},
  journal={arXiv preprint arXiv:2501.09061},
  year={2025}
}

@article{oakes2023fast,
  title={Fast high-fidelity single-shot readout of spins in silicon using a single-electron box},
  author={Oakes, GA and Ciriano-Tejel, VN and Wise, DF and Fogarty, MA and Lundberg, T and Lain{\'e}, C and Schaal, S and Martins, F and Ibberson, DJ and Hutin, L and others},
  journal={Physical Review X},
  volume={13},
  number={1},
  pages={011023},
  year={2023},
  publisher={APS}
}

@article{sunami2025transversal,
  title={Transversal surface-code game powered by neutral atoms},
  author={Sunami, Shinichi and Goban, Akihisa and Yamasaki, Hayata},
  journal={arXiv preprint arXiv:2506.18979},
  year={2025}
}

@article{campbell2026resource,
  title={Resource Estimation via Efficient Compilation of Key Quantum Primitives},
  author={Campbell, Colin and Rines, Rich and Omole, Victory and Oberoi, Tina and Goiporia, Palash and Roy, Rayat and Cline, R Peyton and Jones, Eric B and Tomesh, Teague},
  journal={arXiv preprint arXiv:2604.01376},
  year={2026}
}

@article{babbush2026securing,
  title={Securing elliptic curve cryptocurrencies against quantum vulnerabilities: Resource estimates and mitigations},
  author={Babbush, Ryan and Zalcman, Adam and Gidney, Craig and Broughton, Michael and Khattar, Tanuj and Neven, Hartmut and Bergamaschi, Thiago and Drake, Justin and Boneh, Dan},
  journal={arXiv preprint arXiv:2603.28846},
  year={2026}
}

@article{webster2026pinnacle,
  title={The Pinnacle Architecture: Reducing the cost of breaking RSA-2048 to 100 000 physical qubits using quantum LDPC codes},
  author={Webster, Paul and Berent, Lucas and Chandra, Omprakash and Hockings, Evan T and Baspin, Nou{\'e}dyn and Thomsen, Felix and Smith, Samuel C and Cohen, Lawrence Z},
  journal={arXiv preprint arXiv:2602.11457},
  year={2026}
}

@article{luo2026space,
  title={Space-Efficient Quantum Algorithm for Elliptic Curve Discrete Logarithms with Resource Estimation},
  author={Luo, Han and Yang, Ziyi and Wang, Ziruo and Su, Yuexin and Li, Tongyang},
  journal={arXiv preprint arXiv:2604.02311},
  year={2026}
}

@article{tripier2026fault,
  title={Fault-Tolerant Quantum Computing with Trapped Ions: The Walking Cat Architecture},
  author={Tripier, Felix and Chung, Woo Chang and Young, Jacob and Alam, Safwan and Bjork, Bryce and Brodutch, Aharon and Buessen, Finn Lasse and Coble, Nolan J and Dellaert, Thomas and Maslov, Dmitri and others},
  journal={arXiv preprint arXiv:2604.19481},
  year={2026}
}

@article{nguyen2026suppressing,
  title={Suppressing spin qubit decoherence during shuttling via confinement modulation},
  author={Nguyen, Daniel QL and Rimbach-Russ, Maximilian and Bosco, Stefano},
  journal={arXiv preprint arXiv:2605.00611},
  year={2026}
}

@article{litinski2018lattice,
  title={Lattice surgery with a twist: Simplifying Clifford gates of surface codes},
  author={Litinski, Daniel and von Oppen, Felix},
  journal={Quantum},
  volume={2},
  pages={62},
  year={2018},
  publisher={Verein zur F{\"o}rderung des Open Access Publizierens in den Quantenwissenschaften}
}

@article{chamberland2022circuit,
  title={Circuit-level protocol and analysis for twist-based lattice surgery},
  author={Chamberland, Christopher and Campbell, Earl T},
  journal={Physical Review Research},
  volume={4},
  number={2},
  pages={023090},
  year={2022},
  publisher={APS}
}

@article{ladd2026digitally,
  title={A digitally controlled silicon quantum processing unit},
  author={Ladd, Thaddeus and Reed, Matthew and Blumoff, Jacob},
  journal={arXiv preprint arXiv:2604.16216},
  year={2026}
}

@article{chadwick2026cablecar,
  title={CAbLECAR: efficiently scheduling QLDPC codes on a tileable spin qubit chip with shuttling},
  author={Chadwick, Jason D and Chong, Frederic T},
  journal={arXiv preprint arXiv:2604.24739},
  year={2026}
}

@article{moncy2026surface,
  title={Surface-Code Thresholds and Qubit Footprints in Shuttling-Based Spin-Qubit Railways},
  author={Moncy, Arun John and Dastbasteh, Reza and Martinez, Josu Etxezarreta and Nagai, Ryo and Crespo, Pedro M and Mertig, Normann and Smith, Charles and Otxoa, Ruben M},
  journal={arXiv preprint arXiv:2605.05881},
  year={2026}
}

@article{connors2020rapid,
  title={Rapid high-fidelity spin-state readout in Si/Si-Ge quantum dots via rf reflectometry},
  author={Connors, Elliot J and Nelson, JJ and Nichol, John M},
  journal={Physical Review Applied},
  volume={13},
  number={2},
  pages={024019},
  year={2020},
  publisher={APS}
}

@article{blumoff2022fast,
  title={Fast and high-fidelity state preparation and measurement in triple-quantum-dot spin qubits},
  author={Blumoff, Jacob Z and Pan, Andrew S and Keating, Tyler E and Andrews, Reed W and Barnes, David W and Brecht, Teresa L and Croke, Edward T and Euliss, Larken E and Fast, Jacob A and Jackson, Clayton AC and others},
  journal={PRX Quantum},
  volume={3},
  number={1},
  pages={010352},
  year={2022},
  publisher={APS}
}

@article{keith2019single,
  title={Single-shot spin readout in semiconductors near the shot-noise sensitivity limit},
  author={Keith, D and House, MG and Donnelly, MB and Watson, TF and Weber, Bent and Simmons, MY},
  journal={Physical Review X},
  volume={9},
  number={4},
  pages={041003},
  year={2019},
  publisher={APS}
}

@article{steinacker2025industry,
  title={Industry-compatible silicon spin-qubit unit cells exceeding 99\% fidelity},
  author={Steinacker, Paul and Dumoulin Stuyck, Nard and Lim, Wee Han and Tanttu, Tuomo and Feng, MengKe and Serrano, Santiago and Nickl, Andreas and Candido, Marco and Cifuentes, Jesus D and Vahapoglu, Ensar and others},
  journal={Nature},
  pages={1--7},
  year={2025},
  publisher={Nature Publishing Group UK London}
}

@article{wu2025simultaneous,
  title={Simultaneous high-fidelity single-qubit gates in a spin qubit array},
  author={Wu, Yi-Hsien and Camenzind, Leon C and B{\"u}tler, Patrick and Jin, Ik Kyeong and Noiri, Akito and Takeda, Kenta and Nakajima, Takashi and Kobayashi, Takashi and Scappucci, Giordano and Goan, Hsi-Sheng and others},
  journal={arXiv preprint arXiv:2507.11918},
  year={2025}
}

@article{flindt2006spin,
  title={Spin-orbit mediated control of spin qubits},
  author={Flindt, Christian and S{\o}rensen, Anders S and Flensberg, Karsten},
  journal={Physical review letters},
  volume={97},
  number={24},
  pages={240501},
  year={2006},
  publisher={APS}
}

@article{levitov2002dynamical,
  title={Dynamical spin-electric coupling in a quantum dot},
  author={Levitov, LS and Rashba, EI},
  journal={arXiv preprint cond-mat/0209507},
  year={2002}
}

@article{rashba2003orbital,
  title={Orbital mechanisms of electron-spin manipulation by an electric field},
  author={Rashba, EI and Efros, Al L},
  journal={Physical review letters},
  volume={91},
  number={12},
  pages={126405},
  year={2003},
  publisher={APS}
}

@article{tuckett2019tailoring,
  title={Tailoring surface codes for highly biased noise},
  author={Tuckett, David K and Darmawan, Andrew S and Chubb, Christopher T and Bravyi, Sergey and Bartlett, Stephen D and Flammia, Steven T},
  journal={Physical Review X},
  volume={9},
  number={4},
  pages={041031},
  year={2019},
  publisher={APS}
}

@article{tuckett2018ultrahigh,
  title={Ultrahigh error threshold for surface codes with biased noise},
  author={Tuckett, David K and Bartlett, Stephen D and Flammia, Steven T},
  journal={Physical review letters},
  volume={120},
  number={5},
  pages={050505},
  year={2018},
  publisher={APS}
}

@article{bonilla2021xzzx,
  title={The XZZX surface code},
  author={Bonilla Ataides, J Pablo and Tuckett, David K and Bartlett, Stephen D and Flammia, Steven T and Brown, Benjamin J},
  journal={Nature communications},
  volume={12},
  number={1},
  pages={2172},
  year={2021},
  publisher={Nature Publishing Group UK London}
}

@article{sun2026folded,
  title={A Folded Surface Code Architecture for 2D Quantum Hardware},
  author={Sun, Zhu and Cai, Zhenyu},
  journal={arXiv preprint arXiv:2601.19823},
  year={2026}
}

@article{franke2019rent,
  title={Rent’s rule and extensibility in quantum computing},
  author={Franke, David P and Clarke, James S and Vandersypen, Lieven MK and Veldhorst, Menno},
  journal={Microprocessors and Microsystems},
  volume={67},
  pages={1--7},
  year={2019},
  publisher={Elsevier}
}

@article{filippov2025architecting,
  title={Architecting Distributed Quantum Computers: Design Insights from Resource Estimation},
  author={Filippov, Dmitry and Yang, Peter and Murali, Prakash},
  journal={arXiv preprint arXiv:2508.19160},
  year={2025}
}

@article{aaronson2004improved,
  title={Improved simulation of stabilizer circuits},
  author={Aaronson, Scott and Gottesman, Daniel},
  journal={Physical Review A—Atomic, Molecular, and Optical Physics},
  volume={70},
  number={5},
  pages={052328},
  year={2004},
  publisher={APS}
}

@article{bravyi2012magic,
  title={Magic-state distillation with low overhead},
  author={Bravyi, Sergey and Haah, Jeongwan},
  journal={Physical Review A—Atomic, Molecular, and Optical Physics},
  volume={86},
  number={5},
  pages={052329},
  year={2012},
  publisher={APS}
}

@Article{Dalton2024,
author={Dalton, Kieran
and Long, Christopher K.
and Yordanov, Yordan S.
and Smith, Charles G.
and Barnes, Crispin H. W.
and Mertig, Normann
and Arvidsson-Shukur, David R. M.},
title={Quantifying the effect of gate errors on variational quantum eigensolvers for quantum chemistry},
journal={npj Quantum Information},
year={2024},
month={Jan},
day={27},
volume={10},
pages={18},
issn={2056-6387},
doi={10.1038/s41534-024-00808-x},
}

@Article{StilckFranca2021,
author={Stilck Fran{\c{c}}a, Daniel
and Garc{\'i}a-Patr{\'o}n, Raul},
title={Limitations of optimization algorithms on noisy quantum devices},
journal={Nature Physics},
year={2021},
month={Nov},
day={01},
volume={17},
number={11},
pages={1221-1227},
issn={1745-2481},
doi={10.1038/s41567-021-01356-3}
}

@article{ager2005high,
  title={High-purity, isotopically enriched bulk silicon},
  author={Ager III, JW and Beeman, JW and Hansen, WL and Haller, EE and Sharp, ID and Liao, C and Yang, A and Thewalt, MLW and Riemann, H},
  journal={Journal of the Electrochemical Society},
  volume={152},
  number={6},
  pages={G448--G451},
  year={2005},
  publisher={The Electrochemical Society, Inc.}
}

@article{DePalma2023,
  title = {Limitations of Variational Quantum Algorithms: A Quantum Optimal Transport Approach},
  author = {De Palma, Giacomo and Marvian, Milad and Rouz\'e, Cambyse and Fran\ifmmode \mbox{\c{c}}\else \c{c}\fi{}a, Daniel Stilck},
  journal = {PRX Quantum},
  volume = {4},
  issue = {1},
  pages = {010309},
  numpages = {30},
  year = {2023},
  month = {Jan},
  publisher = {American Physical Society},
  doi = {10.1103/PRXQuantum.4.010309},
}

@article{Fontana2021,
  title = {Evaluating the noise resilience of variational quantum algorithms},
  author = {Fontana, Enrico and Fitzpatrick, Nathan and Ramo, David Mu\~noz and Duncan, Ross and Rungger, Ivan},
  journal = {Phys. Rev. A},
  volume = {104},
  issue = {2},
  pages = {022403},
  numpages = {19},
  year = {2021},
  month = {Aug},
  publisher = {American Physical Society},
  doi = {10.1103/PhysRevA.104.022403}
}

@article{messinger2025fault,
  title={Fault-tolerant quantum computing with the parity code and biased-noise qubits},
  author={Messinger, Anette and Torggler, Valentin and Klaver, Berend and Fellner, Michael and Lechner, Wolfgang},
  journal={Physical Review Applied},
  volume={23},
  number={4},
  pages={044032},
  year={2025},
  publisher={APS}
}

@article{etxezarreta2026leveraging,
  title={Leveraging biased noise for more efficient quantum error correction at the circuit level with two-level qubits},
  author={Etxezarreta Martinez, Josu and Schnabl, Paul and Oliva del Moral, Javier and Dastbasteh, Reza and Crespo, Pedro M and Otxoa, Ruben M},
  journal={Physical Review Applied},
  volume={25},
  number={1},
  pages={014021},
  year={2026},
  publisher={APS}
}

@article{steane1996error,
  title={Error correcting codes in quantum theory},
  author={Steane, Andrew M},
  journal={Physical Review Letters},
  volume={77},
  number={5},
  pages={793},
  year={1996},
  publisher={APS}
}

@article{campbell2017roads,
  title={Roads towards fault-tolerant universal quantum computation},
  author={Campbell, Earl T and Terhal, Barbara M and Vuillot, Christophe},
  journal={Nature},
  volume={549},
  number={7671},
  pages={172--179},
  year={2017},
  publisher={Nature Publishing Group UK London}
}

@article{terhal2015quantum,
  title={Quantum error correction for quantum memories},
  author={Terhal, Barbara M},
  journal={Reviews of Modern Physics},
  volume={87},
  number={2},
  pages={307--346},
  year={2015},
  publisher={APS}
}

@article{knill1998resilient,
  title={Resilient quantum computation},
  author={Knill, Emanuel and Laflamme, Raymond and Zurek, Wojciech H},
  journal={Science},
  volume={279},
  number={5349},
  pages={342--345},
  year={1998},
  publisher={American Association for the Advancement of Science}
}

@inproceedings{gottesman2010introduction,
  title={An introduction to quantum error correction and fault-tolerant quantum computation},
  author={Gottesman, Daniel},
  booktitle={Quantum information science and its contributions to mathematics, Proceedings of Symposia in Applied Mathematics},
  volume={68},
  pages={13--58},
  year={2010}
}

@article{hendrickx2021four,
  title={A four-qubit germanium quantum processor},
  author={Hendrickx, Nico W and Lawrie, William IL and Russ, Maximilian and Van Riggelen, Floor and De Snoo, Sander L and Schouten, Raymond N and Sammak, Amir and Scappucci, Giordano and Veldhorst, Menno},
  journal={Nature},
  volume={591},
  number={7851},
  pages={580--585},
  year={2021},
  publisher={Nature Publishing Group UK London}
}

@article{preskill2018quantum,
  title={Quantum computing in the NISQ era and beyond},
  author={Preskill, John},
  journal={Quantum},
  volume={2},
  pages={79},
  year={2018},
  publisher={Verein zur F{\"o}rderung des Open Access Publizierens in den Quantenwissenschaften}
}

@article{mcclean2018barren,
  title={Barren plateaus in quantum neural network training landscapes},
  author={McClean, Jarrod R and Boixo, Sergio and Smelyanskiy, Vadim N and Babbush, Ryan and Neven, Hartmut},
  journal={Nature communications},
  volume={9},
  number={1},
  pages={4812},
  year={2018},
  publisher={Nature Publishing Group UK London}
}

@article{bouland2019complexity,
  title={On the complexity and verification of quantum random circuit sampling},
  author={Bouland, Adam and Fefferman, Bill and Nirkhe, Chinmay and Vazirani, Umesh},
  journal={Nature Physics},
  volume={15},
  number={2},
  pages={159--163},
  year={2019},
  publisher={Nature Publishing Group UK London}
}

@inproceedings{van2023using,
  title={Using azure quantum resource estimator for assessing performance of fault tolerant quantum computation},
  author={van Dam, Wim and Mykhailova, Mariia and Soeken, Mathias},
  booktitle={Proceedings of the SC'23 Workshops of the International Conference on High Performance Computing, Network, Storage, and Analysis},
  pages={1414--1419},
  year={2023}
}

@article{litinski2019game,
  title={A game of surface codes: Large-scale quantum computing with lattice surgery},
  author={Litinski, Daniel},
  journal={Quantum},
  volume={3},
  pages={128},
  year={2019},
  publisher={Verein zur F{\"o}rderung des Open Access Publizierens in den Quantenwissenschaften}
}

@article{huggins2025fluid,
  title={The FLuid Allocation of Surface code Qubits (FLASQ) cost model for early fault-tolerant quantum algorithms},
  author={Huggins, William J and Khattar, Tanuj and Xu, Amanda and Harrigan, Matthew and Kang, Christopher and Low, Guang Hao and Fowler, Austin and Rubin, Nicholas C and Babbush, Ryan},
  journal={arXiv preprint arXiv:2511.08508},
  year={2025}
}

@article{ismail2025transversal,
  title={Transversal STAR architecture for megaquop-scale quantum simulation with neutral atoms},
  author={Ismail, Refaat and Chen, I and Zhao, Chen and Weiss, Ronen and Liu, Fangli and Zhou, Hengyun and Wang, Sheng-Tao and Sornborger, Andrew and Kornja{\v{c}}a, Milan},
  journal={arXiv preprint arXiv:2509.18294},
  year={2025}
}

@inproceedings{zhou2025resource,
  title={Resource analysis of low-overhead transversal architectures for reconfigurable atom arrays},
  author={Zhou, Hengyun and Duckering, Casey and Zhao, Chen and Bluvstein, Dolev and Cain, Madelyn and Kubica, Aleksander and Wang, Sheng-Tao and Lukin, Mikhail D},
  booktitle={Proceedings of the 52nd Annual International Symposium on Computer Architecture},
  pages={1432--1448},
  year={2025}
}

@article{litinski2023compute,
  title={How to compute a 256-bit elliptic curve private key with only 50 million Toffoli gates},
  author={Litinski, Daniel},
  journal={arXiv preprint arXiv:2306.08585},
  year={2023}
}

@article{litinski2025blocklet,
  title={Blocklet concatenation: Low-overhead fault-tolerant protocols for fusion-based quantum computation},
  author={Litinski, Daniel},
  journal={arXiv preprint arXiv:2506.13619},
  year={2025}
}

@article{gidney2025factor,
  title={How to factor 2048 bit RSA integers with less than a million noisy qubits},
  author={Gidney, Craig},
  journal={arXiv preprint arXiv:2505.15917},
  year={2025}
}

@article{silva2008scalable,
  title={Scalable protocol for identification of correctable codes},
  author={Silva, Marcus and Magesan, Easwar and Kribs, David W and Emerson, Joseph},
  journal={Physical Review A—Atomic, Molecular, and Optical Physics},
  volume={78},
  number={1},
  pages={012347},
  year={2008},
  publisher={APS}
}

@article{zwanenburg2013silicon,
  title={Silicon quantum electronics},
  author={Zwanenburg, Floris A and Dzurak, Andrew S and Morello, Andrea and Simmons, Michelle Y and Hollenberg, Lloyd CL and Klimeck, Gerhard and Rogge, Sven and Coppersmith, Susan N and Eriksson, Mark A},
  journal={Reviews of modern physics},
  volume={85},
  number={3},
  pages={961--1019},
  year={2013},
  publisher={APS}
}

@article{burkard2023semiconductor,
  title={Semiconductor spin qubits},
  author={Burkard, Guido and Ladd, Thaddeus D and Pan, Andrew and Nichol, John M and Petta, Jason R},
  journal={Reviews of Modern Physics},
  volume={95},
  number={2},
  pages={025003},
  year={2023},
  publisher={APS}
}

@article{neyens2024probing,
  title={Probing single electrons across 300-mm spin qubit wafers},
  author={Neyens, Samuel and Zietz, Otto K and Watson, Thomas F and Luthi, Florian and Nethwewala, Aditi and George, Hubert C and Henry, Eric and Islam, Mohammad and Wagner, Andrew J and Borjans, Felix and others},
  journal={Nature},
  volume={629},
  number={8010},
  pages={80--85},
  year={2024},
  publisher={Nature Publishing Group UK London}
}

@article{hangleiter2021filter,
  title={Filter-function formalism and software package to compute quantum processes of gate sequences for classical non-Markovian noise},
  author={Hangleiter, Tobias and Cerfontaine, Pascal and Bluhm, Hendrik},
  journal={Physical Review Research},
  volume={3},
  number={4},
  pages={043047},
  year={2021},
  publisher={APS}
}

@article{shehata2023modeling,
  title={Modeling semiconductor spin qubits and their charge noise environment for quantum gate fidelity estimation},
  author={Shehata, M Mohamed El Kordy and Simion, George and Li, Ruoyu and Mohiyaddin, Fahd A and Wan, Danny and Mongillo, Massimo and Govoreanu, Bogdan and Radu, Iuliana and De Greve, Kristiaan and Van Dorpe, Pol},
  journal={Physical Review B},
  volume={108},
  number={4},
  pages={045305},
  year={2023},
  publisher={APS}
}

@article{machlup1954noise,
  title={Noise in semiconductors: spectrum of a two-parameter random signal},
  author={Machlup, Stefan},
  journal={Journal of Applied Physics},
  volume={25},
  number={3},
  pages={341--343},
  year={1954},
  publisher={American Institute of Physics}
}

@article{dutta1981low,
  title={Low-frequency fluctuations in solids: 1 f noise},
  author={Dutta, Pulak and Horn, PM},
  journal={Reviews of Modern physics},
  volume={53},
  number={3},
  pages={497},
  year={1981},
  publisher={APS}
}

@article{fleetwood20151,
  title={$1/f $ noise and defects in microelectronic materials and devices},
  author={Fleetwood, DM},
  journal={IEEE Transactions on Nuclear Science},
  volume={62},
  number={4},
  pages={1462--1486},
  year={2015},
  publisher={IEEE}
}

@article{struck2024spin,
  title={Spin-EPR-pair separation by conveyor-mode single electron shuttling in Si/SiGe},
  author={Struck, Tom and Volmer, Mats and Visser, Lino and Offermann, Tobias and Xue, Ran and Tu, Jhih-Sian and Trellenkamp, Stefan and Cywi{\'n}ski, {\L}ukasz and Bluhm, Hendrik and Schreiber, Lars R},
  journal={Nature Communications},
  volume={15},
  number={1},
  pages={1325},
  year={2024},
  publisher={Nature Publishing Group UK London}
}

@article{KHANEJA2005296,
title = {Optimal control of coupled spin dynamics: design of NMR pulse sequences by gradient ascent algorithms},
journal = {Journal of Magnetic Resonance},
volume = {172},
number = {2},
pages = {296-305},
year = {2005},
issn = {1090-7807},
doi = {https://doi.org/10.1016/j.jmr.2004.11.004},
url = {https://www.sciencedirect.com/science/article/pii/S1090780704003696},
author = {Navin Khaneja and Timo Reiss and Cindie Kehlet and Thomas Schulte-Herbrüggen and Steffen J. Glaser}
}

@software{QuGrad,
author = {Long, Christopher K. and Barnes, Crispin H. W. and Mertig, Normann},
doi = {10.5281/zenodo.17116721},
month = {10},
title = {QuGrad},
url = {https://github.com/Christopher-K-Long/QuGrad},
year = {2025}
}

@article{Lepage20,
  title = {Entanglement generation via power-of-swap operations between dynamic electron-spin qubits},
  author = {Lepage, Hugo V. and Lasek, Aleksander A. and Arvidsson-Shukur, David R. M. and Barnes, Crispin H. W.},
  journal = {Phys. Rev. A},
  volume = {101},
  issue = {2},
  pages = {022329},
  numpages = {11},
  year = {2020},
  month = {Feb},
  publisher = {American Physical Society},
  doi = {10.1103/PhysRevA.101.022329},
  url = {https://link.aps.org/doi/10.1103/PhysRevA.101.022329}
}

@article{ArvidssonShukur17,
  title = {Protocol for fermionic positive-operator-valued measures},
  author = {Arvidsson-Shukur, D. R. M. and Lepage, H. V. and Owen, E. T. and Ferrus, T. and Barnes, C. H. W.},
  journal = {Phys. Rev. A},
  volume = {96},
  issue = {5},
  pages = {052305},
  numpages = {7},
  year = {2017},
  month = {Nov},
  publisher = {American Physical Society},
  doi = {10.1103/PhysRevA.96.052305},
  url = {https://link.aps.org/doi/10.1103/PhysRevA.96.052305}
}

@article{Berry2005Aug,
author={Berry, Dominic W.
and Ahokas, Graeme
and Cleve, Richard
and Sanders, Barry C.},
title={Efficient Quantum Algorithms for Simulating Sparse Hamiltonians},
journal={Communications in Mathematical Physics},
year={2007},
month={Mar},
day={01},
volume={270},
number={2},
pages={359-371},
issn={1432-0916},
doi={10.1007/s00220-006-0150-x},
url={https://doi.org/10.1007/s00220-006-0150-x}
}

@article{Suzuki1990Jun,
	author = {Suzuki, Masuo},
	title = {{Fractal decomposition of exponential operators with applications to many-body theories and Monte Carlo simulations}},
	journal = {Phys. Lett. A},
	volume = {146},
	number = {6},
	pages = {319--323},
	year = {1990},
	month = jun,
	issn = {0375-9601},
	publisher = {North-Holland},
	doi = {10.1016/0375-9601(90)90962-N}
}

@article{Huang2024Mar,
	author = {Huang, Jonathan Y. and Su, Rocky Y. and Lim, Wee Han and Feng, MengKe and van Straaten, Barnaby and Severin, Brandon and Gilbert, Will and Dumoulin Stuyck, Nard and Tanttu, Tuomo and Serrano, Santiago and Cifuentes, Jesus D. and Hansen, Ingvild and Seedhouse, Amanda E. and Vahapoglu, Ensar and Leon, Ross C. C. and Abrosimov, Nikolay V. and Pohl, Hans-Joachim and Thewalt, Michael L. W. and Hudson, Fay E. and Escott, Christopher C. and Ares, Natalia and Bartlett, Stephen D. and Morello, Andrea and Saraiva, Andre and Laucht, Arne and Dzurak, Andrew S. and Yang, Chih Hwan},
	title = {{High-fidelity spin qubit operation and algorithmic initialization above 1 K}},
	journal = {Nature},
	volume = {627},
	pages = {772--777},
	year = {2024},
	month = mar,
	issn = {1476-4687},
	publisher = {Nature Publishing Group},
	doi = {10.1038/s41586-024-07160-2}
}

@article{Tanttu2024Nov,
	author = {Tanttu, Tuomo and Lim, Wee Han and Huang, Jonathan Y. and Dumoulin Stuyck, Nard and Gilbert, Will and Su, Rocky Y. and Feng, MengKe and Cifuentes, Jesus D. and Seedhouse, Amanda E. and Seritan, Stefan K. and Ostrove, Corey I. and Rudinger, Kenneth M. and Leon, Ross C. C. and Huang, Wister and Escott, Christopher C. and Itoh, Kohei M. and Abrosimov, Nikolay V. and Pohl, Hans-Joachim and Thewalt, Michael L. W. and Hudson, Fay E. and Blume-Kohout, Robin and Bartlett, Stephen D. and Morello, Andrea and Laucht, Arne and Yang, Chih Hwan and Saraiva, Andre and Dzurak, Andrew S.},
	title = {{Assessment of the errors of high-fidelity two-qubit gates in silicon quantum dots}},
	journal = {Nat. Phys.},
	volume = {20},
	pages = {1804--1809},
	year = {2024},
	month = nov,
	issn = {1745-2481},
	publisher = {Nature Publishing Group},
	doi = {10.1038/s41567-024-02614-w}
}

@article{losert2024strategies,
  title={Strategies for enhancing spin-shuttling fidelities in Si/Si Ge quantum wells with random-alloy disorder},
  author={Losert, Merritt P and Oberl{\"a}nder, Max and Teske, Julian D and Volmer, Mats and Schreiber, Lars R and Bluhm, Hendrik and Coppersmith, SN and Friesen, Mark},
  journal={PRX Quantum},
  volume={5},
  number={4},
  pages={040322},
  year={2024},
  publisher={APS}
}

@article{fujiu2026dense,
  title={Dense packing of the surface code: Code deformation procedures and hook-error-avoiding gate scheduling},
  author={Fujiu, Kohei and Nagayama, Shota and Nishio, Shin and Kawaguchi, Hideaki and Satoh, Takahiko},
  journal={Physical Review A},
  volume={113},
  number={4},
  pages={042412},
  year={2026},
  publisher={APS}
}

@article{xue2024si,
  title={Si/SiGe QuBus for single electron information-processing devices with memory and micron-scale connectivity function},
  author={Xue, Ran and Beer, Max and Seidler, Inga and Humpohl, Simon and Tu, Jhih-Sian and Trellenkamp, Stefan and Struck, Tom and Bluhm, Hendrik and Schreiber, Lars R},
  journal={Nature Communications},
  volume={15},
  number={1},
  pages={2296},
  year={2024},
  publisher={Nature Publishing Group UK London}
}

@article{childs2012hamiltonian,
  title={Hamiltonian simulation using linear combinations of unitary operations},
  author={Childs, Andrew M and Wiebe, Nathan},
  journal={arXiv preprint arXiv:1202.5822},
  year={2012}
}

@article{volmer2024mapping,
  title={Mapping of valley splitting by conveyor-mode spin-coherent electron shuttling},
  author={Volmer, Mats and Struck, Tom and Sala, Arnau and Chen, Bingjie and Oberl{\"a}nder, Max and Offermann, Tobias and Xue, Ran and Visser, Lino and Tu, Jhih-Sian and Trellenkamp, Stefan and others},
  journal={npj Quantum Information},
  volume={10},
  number={1},
  pages={61},
  year={2024},
  publisher={Nature Publishing Group UK London}
}

@article{cain2026shor,
  title={Shor's algorithm is possible with as few as 10,000 reconfigurable atomic qubits},
  author={Cain, Madelyn and Xu, Qian and King, Robbie and Picard, Lewis RB and Levine, Harry and Endres, Manuel and Preskill, John and Huang, Hsin-Yuan and Bluvstein, Dolev},
  journal={arXiv preprint arXiv:2603.28627},
  year={2026}
}

\pagebreak
\onecolumngrid
\newpage
\begin{center}
\textbf{\large Supplemental Material}
\end{center}

\setcounter{equation}{0}
\setcounter{figure}{0}
\setcounter{table}{0}
\setcounter{page}{1}
\makeatletter
\renewcommand{\theequation}{S\arabic{equation}}
\setcounter{section}{0}
\renewcommand{\thesection}{\Roman{section}}
\setcounter{Theorem}{0}
\renewcommand{\theTheorem}{S\arabic{Theorem}}

\section{Non-Markovian noise derivation}
\label{non-markovian}
In this appendix, we will derive the expression for Eq.~\ref{NMfilter}. We start by considering a system evolving under a time-dependent Hamiltonian
\begin{equation}
H(t) = H_c(t) + H_n(t), 
\qquad
H_n(t) = \sum_\alpha b_\alpha(t) B_\alpha ,
\end{equation}
where $H_c(t)$ is the control Hamiltonian as in Eq.~\eqref{eq:H_full} and $H_n(t)$ describes classical stochastic noise~\cite{hangleiter2021filter}. The stochastic processes $b_\alpha(t)$ are assumed to be zero mean, and the operators $B_\alpha$ are fixed Pauli strings acting on the noisy subsystem, which are time independent. We direct the interested readers to~\cite{cerfontaine2021filter} for the general time-dependent framework. We denote the control propagator by $U_c(t)$ satisfying Schr\"{o}dinger's equation $i\frac{d}{dt}U_c(t)=H_c(t)U_c(t)$, $U_c(0)=\mathbb I$. In the interaction picture, the noise Hamiltonian is
$$
\tilde H_n(t)=U_c^\dagger(t)H_n(t)U_c(t),
$$
and the interaction-picture propagator $\tilde U(t)=U_c^\dagger(t)U(t)$ obeys
$$
i\frac{d}{dt}\tilde U(t)=\tilde H_n(t)\tilde U(t),\qquad \tilde U(0)=\mathbb I.
$$
Applying the Magnus expansion~\cite{blanes2009magnus} over a gate interval $[0,\tau]$ we can write down the evolution operator,
$$
\tilde U(\tau)=\exp\big(-i\tau H_{\mathrm{eff}}\big),\qquad
H_{\mathrm{eff}}=\sum_{\mu\ge1}H_{\mathrm{eff};\mu},
$$
in terms of an effective Hamiltonian, $H_{\mathrm{eff}}$. Providing the noise strength $\xi \equiv \sum ||B_\alpha|| \sqrt{\langle b_\alpha(0)^2 \rangle} \tau << 1$ we can neglect higher order terms and just retain the first two:
\begin{align}
H_{\mathrm{eff};1}&=1/\tau \int_0^\tau dt\tilde H_n(t)\\
H_{\mathrm{eff};2}&=-\tfrac{i}{2\tau}\int_0^\tau dt_1\int_0^{t_1}dt_2[\tilde H_n(t_1),\tilde H_n(t_2)].
\end{align}
Taylor expanding the evolution to second order yields the map
$$
\tilde U\rho\tilde U^\dagger = \rho + \mathcal L^{(1)}(\rho) + \mathcal L^{(2)}(\rho) + O(\xi^3),
$$
where $\mathcal L^{(1)}(\rho) = -i\tau [H_{\mathrm{eff};1},\rho]$, and $\mathcal L^{(2)}(\rho) = -i\tau [H_{\mathrm{eff};2},\rho]
+ \tau^2 H_{\mathrm{eff};1}\rho H_{\mathrm{eff};1} - \tfrac12 \tau^2\{H_{\mathrm{eff};1}^2,\rho\}$. Because the stochastic processes $b_\alpha(t)$ are zero mean, any terms with an odd number of $\tilde H_n(t)$ will average to zero, such as $\mathbb E[H_{\mathrm{eff};1}]=0$ and consequently $\mathbb E[\mathcal L^{(1)}]=0$. Keeping terms to $O(\xi^2)$ and averaging over the noise yields the leading-order dissipative contribution
\begin{equation}
\label{eq_master_avg}
\frac{\mathbb E[\tilde U\rho\tilde U^\dagger]-\rho}{\tau}
= -i\mathbb E\big[[H_{\mathrm{eff};2},\rho]\big] \nonumber+ \tau\mathbb E\big[H_{\mathrm{eff};1}\rho H_{\mathrm{eff};1}\big]
- \tfrac{1}{2}\tau\mathbb E\big\{H_{\mathrm{eff};1}^2,\rho\big\} + O(\xi^3). 
\end{equation}
To obtain explicit expressions we expand $\tilde H_n$ in an orthonormal basis of the Hermitian operators, $\{\sigma_k\}$ (with $\Tr(\sigma_k\sigma_l)=D\delta_{kl}$):
\begin{align*}
\tilde H_n(t)&=\sum_{\alpha,k} b_\alpha(t)\tilde B_{\alpha k}(t)\sigma_k,\\
\tilde B_{\alpha k}(t)&=\Tr\big[U_c^\dagger(t)B_\alpha U_c(t)\sigma_k\big].
\end{align*}
Next, we define the second-order kernels, where the scalar $\tilde B_{\alpha k}(t)$ is time-dependent and the expectation value is taken over the stochastic process $b_\alpha(t)$:
\begin{equation}
\label{gamma}
\Gamma_{\alpha;kl}
\equiv
\int_0^\tau dt_1\int_0^\tau dt_2
\mathbb E[b_\alpha(t_1)b_\alpha(t_2)]
\tilde B_{\alpha k}(t_1)\tilde B_{\alpha l}(t_2).
\end{equation}
Then, we define the two-point autocorrelation function $C_\alpha(\tau)$, which depends only on the time difference,
\begin{equation}
C_\alpha(\tau)=C_\alpha(t_1-t_2)=\mathbb E[b_\alpha(t_1)b_\alpha(t_2)]
\end{equation}
so that
\begin{equation}
\Gamma_{\alpha;kl}
=
\int_0^\tau dt_1\int_0^\tau dt_2
C_\alpha(t_1-t_2)
\tilde B_{\alpha k}(t_1)\tilde B_{\alpha l}(t_2).
\label{gamma_time}
\end{equation}
Assuming a noise spectral density $S_\alpha(\omega)$ defined via
\begin{equation}
C_\alpha(\tau)
=
\int_{-\infty}^{\infty}\frac{d\omega}{2\pi}
S_\alpha(\omega)e^{-i\omega\tau},
\end{equation}
we substitute this representation into Eq.~\eqref{gamma_time} to obtain
\begin{align}
\Gamma_{\alpha;kl}
&=\int_0^\tau dt_1\int_0^\tau dt_2
\int_{-\infty}^{\infty}\frac{d\omega}{2\pi}
S_\alpha(\omega)
e^{-i\omega(t_1-t_2)}
\tilde B_{\alpha k}(t_1)\tilde B_{\alpha l}(t_2)\\
\nonumber
&=
\int_{-\infty}^{\infty}\frac{d\omega}{2\pi}
S_\alpha(\omega)
\left[
\int_0^\tau dt_1 e^{-i\omega t_1}\tilde B_{\alpha k}(t_1)
\right]
\left[
\int_0^\tau dt_2 e^{+i\omega t_2}\tilde B_{\alpha l}(t_2)
\right].
\label{gamma_freq_step}
\end{align}
Defining the finite-time Fourier transforms
\begin{equation}
\mathbb{\tilde B}_{\alpha k}(\omega)
\equiv
\int_0^\tau dt e^{i\omega t}\tilde B_{\alpha k}(t),
\end{equation}
we identify
\begin{equation}
\int_0^\tau dt_1 e^{-i\omega t_1}\tilde B_{\alpha k}(t_1)
=
\mathbb{\tilde B}_{\alpha k}^*(\omega),
\end{equation}
which finally yields
\begin{equation}
\Gamma_{\alpha;kl}
=
\int_{-\infty}^{\infty}\frac{d\omega}{2\pi}
S_\alpha(\omega)
\mathbb{\tilde B}_{\alpha k}^*(\omega)\mathbb{\tilde B}_{\alpha l}(\omega).
\end{equation}
Assuming independent noise channels $\mathbb E\left[ b_\alpha(t_1)b_\beta(t_2) \right]
=
\delta_{\alpha\beta}
\mathbb E\left[ b_\alpha(t_1)b_\alpha(t_2) \right]$, we obtain
\begin{align}
\tau^2\mathbb E\left[ H_{\mathrm{eff};1}\rho H_{\mathrm{eff};1} \right]
=\sum_{\alpha}
\sum_{k,l}
\int_0^\tau dt_1 \int_0^\tau dt_2
\mathbb E\left[ b_\alpha(t_1)b_\alpha(t_2) \right]
\tilde B_{\alpha k}(t_1)\tilde B_{\alpha l}(t_2)
\sigma_k \rho \sigma_l.\nonumber
\end{align}
We can also express it with the kernel
\begin{equation}
\tau^2\mathbb E\left[ H_{\mathrm{eff};1}\rho H_{\mathrm{eff};1} \right]
=
\sum_{\alpha,k,l}
\Gamma_{\alpha;kl}
\sigma_k \rho \sigma_l .
\label{eq:term1}
\end{equation}
Similarly,
\begin{align}
\mathbb E\left[ H_{\mathrm{eff};1}^2 \right]
&=
\frac{1}{\tau^2}
\sum_{\alpha,k,l}
\Gamma_{\alpha;kl}
\sigma_k \sigma_l .
\end{align}
Therefore,
\begin{equation}
\frac{1}{2}\tau^2
\mathbb E\left[ \{ H_{\mathrm{eff};1}^2,\rho \} \right]
=
\frac{1}{2}
\sum_{\alpha,k,l}
\Gamma_{\alpha;kl}
\{ \sigma_k \sigma_l,\rho \} .
\label{eq:term2}
\end{equation}
One then finds the averaged dissipator
\begin{align*}
\mathbb E[\tilde U\rho\tilde U^\dagger]-\rho
= \sum_\alpha\sum_{k,l}\Gamma_{\alpha;kl}
\Big(\sigma_k\rho\sigma_l-\tfrac12\{\sigma_k\sigma_l,\rho\}\Big)- i\tau\mathbb E\big[[H_{\mathrm{eff};2},\rho]\big]
+ O(\xi^3).
\end{align*}
Since the term containing $H_{\mathrm{eff};2}$, it does not contribute to the average gate fidelity of the quantum operation.Thus, we will drop this term from now on. To calculate the average gate fidelity, we express $\mathbb E[\tilde U\rho\tilde U^\dagger]$ as $\tilde {\mathcal E}[\rho]$, the noise-averaged quantum process. Then we can use the average gate fidelity formula $F = (\Tr[\tilde {\mathcal E}]+D)/(D^2+D)$~\cite{green2013arbitrary, kimmel2014robust, nielsen2002simple} where $D$ is the Hilbert-space dimension. Writing $\tilde {\mathcal E} = \mathds{1} + \mathcal D + O(\xi^3)$, with
\begin{equation}
\mathcal D(\rho)=\sum_{\alpha,k,l}
\Gamma_{\alpha;kl}
\Big(
\sigma_k\rho\sigma_l
-\tfrac12\{\sigma_k\sigma_l,\rho\}
\Big),
\end{equation}
and since $\Tr[\mathds{1}]=D$, the leading-order infidelity is
\begin{equation}
\epsilon= \frac{D^2-\Tr\left[
\mathcal D
\right]}{D(D+1)}.
\end{equation} 
Substituting the explicit form of $\mathcal D$ and decomposing it in the complete basis $\sigma_i$ yields
\begin{align}
\sum_j
\Tr\left[
\sigma_j \mathcal D(\sigma_j)
\right]= \sum_{\alpha,k,l}
\Gamma_{\alpha;kl}
\sum_j
\Tr\left[
\sigma_j
\Big(
\sigma_k\sigma_j\sigma_l
-\tfrac12\{\sigma_k\sigma_l,\sigma_j\}
\Big)
\right].\nonumber
\end{align}
Using completeness and orthonormality of the operator basis, the fidelity yields
\begin{equation}
\epsilon=\frac{1}{D+1}
\sum_{\alpha,k}
\Gamma_{\alpha;kk}.
\end{equation}
Using the frequency-domain representation of the kernel,
\begin{equation}
\Gamma_{\alpha;kk}=\int_{-\infty}^{\infty}\frac{d\omega}{2\pi}
S_\alpha(\omega)
\big|\mathbb{\tilde B}_{\alpha k}(\omega)\big|^2,
\end{equation}
we finally obtain Eq.~\ref{NMfilter}. For the noise spectral-density function considered as Eq.~\ref{eq:1f_spectrum}, the autocorrelation function becomes
\begin{align}
C(\tau) \approx \int_{\omega_{\mathrm{low}}}^{\omega_{\mathrm{high}}} \frac{d\omega}{\pi} \frac{\cos(\omega \tau)}{\omega} A = \frac{A}{\pi} \int_{\omega_{\mathrm{low}}}^{\omega_{\mathrm{high}}} \frac{\cos(\omega \tau)}{\omega} d\omega = \frac{A}{\pi} \left[ \mathrm{Ci}(\omega_{\mathrm{high}} \tau) - \mathrm{Ci}(\omega_{\mathrm{low}} \tau) \right],
\end{align}
where $\mathrm{Ci}(x)$ is the cosine integral function $\mathrm{Ci}(x) = -\int_x^{\infty} \frac{\cos t}{t} dt$. For intermediate times $\tau \gg 1/\omega_{\mathrm{high}}$, $\mathrm{Ci}(\omega_{\mathrm{high}} \tau) \to 0$ and $\mathrm{Ci}(\omega_{\mathrm{low}} \tau) \approx \gamma_E + \ln(1/(\omega_{\mathrm{low}}\tau))$, so
\begin{equation}
C(\tau) \approx \frac{A}{\pi} \left( \ln\left(\frac{1}{\omega_{\mathrm{high}} \tau}\right) - \gamma_E\right),
\end{equation}
where $\gamma_E$ is Euler's constant. The correlation function decays logarithmically in this regime, while we can show that at high frequencies $S(\omega) \approx \frac{A\omega_{\mathrm{high}}^2}{|\omega|^3}$, and the correlations decay quadratically, $C(\tau) \sim \frac{A}{\omega_{\mathrm{low}}}\frac{1}{\tau^2}$, reflecting the finite span of the correlation noise. This scaling is derived by approximating the integral in the $\tau \ll 1/\omega$, then
\begin{align}
C(\tau) &=\int_0^\infty \frac{d\omega}{\pi}\cos(\omega\tau) \frac{A\omega_{\mathrm{high}}^2}{\omega^3}\nonumber\\ &\approx \frac{A\omega_{\mathrm{high}}^2}{\pi}\int_{\omega_{\mathrm{high}}}^{\frac{1}{\tau}} d\omega (1-\frac{(\omega \tau)^2}{2}) \frac{1}{\omega^3} \sim \tau^2.
\end{align}

\clearpage
\newpage

\section{Mapping Non-Markovian noise onto Pauli Channel}
\label{map}

In this appendix, we explain how we mapped the correlated noise onto the Pauli channel. For a single qubit, the maximally entangled state $\ket{\Phi^+_2} = \frac{1}{\sqrt{2}}(\ket{00} + \ket{11})$ can be expressed in the Pauli basis as,
\begin{equation}
\ket{\Phi^+_2}\bra{\Phi^+_2} = \frac{1}{4} \sum_{P \in \{I,X,Y,Z\}} P \otimes P^T.
\end{equation}
For $n$ qubits, the maximally entangled state $\ket{\Phi^+_{2^n}}$ between subsystems $A$ and $B$ (each of dimension $2^n$) factorizes,
\begin{align}
\ket{\Phi^+_{2^n}}_{AB} &:= \sum_{i \in \{0,1\}^n} \ket{i}_A \otimes \ket{i}_B \nonumber \\
&= \sum_{i_1,\ldots,i_n \in \{0,1\}} \ket{i_1\ldots i_n}_A \otimes \ket{i_1\ldots i_n}_B \nonumber \\
&= \sum_{i_1,\ldots,i_n \in \{0,1\}} \ket{i_1}_A\ket{i_1}_B \otimes \cdots \otimes \ket{i_n}_A\ket{i_n}_B \nonumber \\
&= \left(\sum_{i_1 \in \{0,1\}} \ket{i_1}_A\ket{i_1}_B\right) \otimes \cdots \otimes \left(\sum_{i_n \in \{0,1\}} \ket{i_n}_A\ket{i_n}_B\right) \nonumber \\
&= \ket{\Phi^+_2}_{AB} \otimes \cdots \otimes \ket{\Phi^+_2}_{AB}.
\end{align}
Therefore, the density matrix is,
\begin{align}
\ket{\Phi^+_{2^n}}\bra{\Phi^+_{2^n}}_{AB} &= \ket{\Phi^+_2}\bra{\Phi^+_2}_{AB} \otimes \cdots \otimes \ket{\Phi^+_2}\bra{\Phi^+_2}_{AB} \nonumber \\
&= \left(\frac{1}{4} \sum_{P_1 \in \{I,X,Y,Z\}} P_{1,A} \otimes (P_{1,B})^T\right) \otimes \cdots \otimes \left(\frac{1}{4} \sum_{P_n \in \{I,X,Y,Z\}} P_{n,A} \otimes (P_{n,B})^T\right) \nonumber \\
&= \frac{1}{4^n} \sum_{P_1,\ldots,P_n \in \{I,X,Y,Z\}} P_{1,A} \otimes (P_{1,B})^T \otimes \cdots \otimes P_{n,A} \otimes (P_{n,B})^T \nonumber \\
&= \frac{1}{4^n} \sum_{P_1,\ldots,P_n \in \{I,X,Y,Z\}} [P_{1,A} \otimes \cdots \otimes P_{n,A}] \otimes [P_{1,B} \otimes \cdots \otimes P_{n,B}]^T \nonumber \\
&= \frac{1}{D^2} \sum_{P \in \{I,X,Y,Z\}^{\otimes n}} P_A \otimes (P_B)^T
\end{align}
where $D = 2^n$ and $D^2 = 4^n$. An $n$-qubit Pauli channel acts as,
\begin{equation}
\mathcal{E}_{\text{Pauli}}(\rho) = \sum_{i=0}^{D^2-1} \theta_i P_i \rho P_i^\dagger
\end{equation}
where $\{P_i\}$ are $n$-qubit Pauli operators. The Choi matrix is:
\begin{align}
J_{\text{Pauli}} &= (\mathcal{E}_{\text{Pauli}} \otimes \mathds{1})(\ket{\Phi^+_{2^n}}\bra{\Phi^+_{2^n}}) \nonumber \\
&= \sum_{i=0}^{D^2-1} \theta_i (P_i \otimes \mathds{1}) \left(\frac{1}{D^2} \sum_{P \in \{I,X,Y,Z\}^{\otimes n}} P \otimes P^T\right) (P_i^\dagger \otimes \mathds{1}).
\end{align}
Using the identity $\bra{\Phi^+}(A \otimes \mathds{1}) = \bra{\Phi^+}(\mathds{1} \otimes A^T)$, and since $P_i^T = \bar{P}_i$,
\begin{align}
J_{\text{Pauli}} &= \sum_{i=0}^{D^2-1} \theta_i (P_i \otimes \mathds{1}) \left(\frac{1}{D^2} \sum_{P \in \{I,X,Y,Z\}^{\otimes n}} P \otimes P^T\right) (\mathds{1} \otimes \bar{P}_i) \nonumber \\
&= \frac{1}{D^2} \sum_{i=0}^{D^2-1} \sum_{P \in \{I,X,Y,Z\}^{\otimes n}} \theta_i (P_i P) \otimes (\bar{P}_i \bar{P}).
\end{align}
The filter function channel including both dissipative and coherent contributions is
\begin{equation}
\mathcal{E}_{\text{FF}}(\rho) = \rho - i[H_{\text{coh}}, \rho] + \sum_{\alpha=1}^M \sum_{k,l=0}^{D^2-1} \Gamma_{\alpha,kl} \left(\sigma_k \rho \sigma_l - \frac{1}{2}\{\sigma_k\sigma_l, \rho\}\right),
\end{equation}
where $\Gamma_{\alpha,kl} = \int_0^\tau dt_1 \int_0^\tau dt_2\, C_\alpha(t_1 - t_2) \tilde B_{\alpha k}(t_1) \tilde B_{\alpha l}(t_2)$ is the symmetric part of the noise kernel, and the coherent Hamiltonian is induced by $\mathbb{E}[[H_{\mathrm{eff};2},\rho]]$.
The Choi matrix is
\begin{equation}
J_{\text{FF}} = (\mathcal{E}_{\text{FF}} \otimes \mathds{1})(\ket{\Phi^+_{2^n}}\bra{\Phi^+_{2^n}}).
\end{equation}
Express it in Pauli basis,
\begin{align}
J_{\text{FF}} = (\mathcal{E}_{\text{FF}} \otimes \mathds{1})\left(\frac{1}{D^2} \sum_{r=0}^{D^2-1} P_r \otimes \bar{P}_r\right)= \frac{1}{D^2} \sum_{r=0}^{D^2-1} \left[\mathcal{E}_{\text{FF}}(P_r) \otimes \bar{P}_r\right]
\end{align}
Using the definition of $\mathcal{E}_{\text{FF}}$
\begin{align}
\mathcal{E}_{\text{FF}}(P_r) &= P_r - i[H_{\text{coh}}, P_r] + \sum_{\alpha=1}^M \sum_{k,l=0}^{D^2-1} \Gamma_{\alpha,kl} \left(\sigma_k P_r \sigma_l - \frac{1}{2}\{\sigma_k\sigma_l, P_r\}\right)
\end{align}
so that
\begin{align}
J_{\text{FF}} = \frac{1}{D^2} \sum_{r=0}^{D^2-1} \Bigg[ P_r \otimes \bar{P}_r - i[H_{\text{coh}}, P_r] \otimes \bar{P}_r + \sum_{\alpha=1}^M \sum_{k,l=0}^{D^2-1} \Gamma_{\alpha,kl} \left(\sigma_k P_r \sigma_l \otimes \bar{P}_r - \frac{1}{2}\{\sigma_k\sigma_l, P_r\} \otimes \bar{P}_r\right) \Bigg].
\end{align}
We want to write $J_{\text{FF}}$ in the Pauli basis $\{P_p \otimes \bar{P}_q\}$
\begin{equation}
J_{\text{FF}} = \frac{1}{D^2} \sum_{p,q=0}^{D^2-1} J_{pq} (P_p \otimes \bar{P}_q).
\end{equation} 
Our goal is to seek Pauli channel parameters $\{\theta_i\}_{i=0}^{D^2-1}$ that minimize,
\begin{equation}
\mathcal{L}(\bm{\theta}, \lambda_1, \lambda_2) = d_F(J_{\text{FF}}, J_{\text{Pauli}}(\bm{\theta})) + \lambda_1 \left(\mathcal{F}_{\text{FF}} - \mathcal{F}_{\text{Pauli}}(\bm{\theta})\right) + \lambda_2 \left(\sum_{i=0}^{D^2-1} \theta_i - 1\right)
\end{equation}
where
\begin{itemize}
    \item $d_F$ is the Frobenius norm distance
    \item $\lambda_1, \lambda_2$ are Lagrange multipliers for constraints
    \item $\mathcal{F}_{\text{FF}}$ and $\mathcal{F}_{\text{Pauli}}$ are the process fidelities
\end{itemize}
The constraints are
\begin{align}
\mathcal{F}_{\text{FF}} - \mathcal{F}_{\text{Pauli}}(\bm{\theta}) &= 0 \quad \text{(fidelity matching)} \\
\sum_{i=0}^{D^2-1} \theta_i &= 1 \quad \text{(normalization)} 
\end{align}
We consider the first term,
\begin{equation}
d_F(J_1, J_2) = \|J_1 - J_2\|_F^2 = \tr[(J_1 - J_2)^\dagger (J_1 - J_2)].
\end{equation}
Expanding
\begin{equation}
d_F = \underbrace{\tr(J_{\text{FF}}^\dagger J_{\text{FF}})}_{T_1} + \underbrace{\tr(J_{\text{Pauli}}(\bm{\theta})^\dagger J_{\text{Pauli}}(\bm{\theta}))}_{T_2} - 2\underbrace{\Re[\tr(J_{\text{FF}}^\dagger J_{\text{Pauli}}(\bm{\theta}))]}_{T_3}.
\end{equation}
We do not explicitly calculate $T_1$ since it is a constant with respect to $\bm{\theta}$ though. Note that the Pauli strings are an orthonormal basis with repsect to the Hilbert–Schmidt inner product. And since it is generally hard to evaluate the commutation rules with Pauli strings to simplify the notation, we choose to write $\bm{\theta}$ in the Pauli basis directly,
\begin{align}
T_2 &= \tr\left[\left(\frac{1}{D^2} \sum_{i=0}^{D^2-1}  \theta_i  \right)^\dagger \left(\frac{1}{D^2} \sum_{j=0}^{D^2-1} \theta_j \right)\right] \nonumber \\
&= \frac{1}{D^4}(\sum_{i,j} \theta_i \theta_j \,\delta_{ij}) = \frac{1}{D^2}\sum_{i,j} \theta_i^2
\end{align}
For $T_3$
\begin{equation}
T_3 = \tr[J_{\text{FF}}^\dagger J_{\text{Pauli}}(\bm{\theta})] = \frac{1}{D^2}\sum_{i,j} \theta_i J_{ij}
\end{equation}
Now, from 
\begin{equation}
\mathcal{F}_{\text{FF}} = 1 - \frac{1}{d+1}\sum_{\alpha=1}^M \sum_{k=0}^{D^2-1} \Gamma_{\alpha,kk},
\end{equation}
we define
\begin{equation}
\gamma = \frac{1}{D}\sum_{\alpha=1}^M \sum_{k=0}^{D^2-1} \Gamma_{\alpha,kk}.
\end{equation}
Then
\begin{equation}
\mathcal{F}_{\text{FF}} = 1 - \frac{d}{d+1}\gamma \equiv 1 - \kappa\gamma.
\end{equation}
where $\kappa = \frac{d}{d+1}$. For the Pauli channel
\begin{align}
\mathcal{F}_{\text{Pauli}}(\bm{\theta}) &= \frac{1}{d(d+1)}\left(d + \sum_{i=0}^{D^2-1} \theta_i |\tr(P_i)|^2\right) \nonumber \\
&= \frac{1}{d(d+1)}\left(d + \theta_0 D^2\right) = \frac{1 + d\theta_0}{d+1}.
\end{align}
Using $\theta_0 = 1 - \sum_{i=1}^{D^2-1} \theta_i$
\begin{equation}
\mathcal{F}_{\text{Pauli}}(\bm{\theta}) = \frac{1 + d(1 - \sum_{i=1}^{D^2-1} \theta_i)}{d+1} = 1 - \kappa\sum_{i=1}^{D^2-1} \theta_i.
\end{equation}
Let us write down the full lagrangian
\begin{equation}
\mathcal{L}(\bm{\theta}, \lambda_1, \lambda_2) = \sum_{p,q} |J_{pq}|^2+\frac{1}{D^2}\sum_{i=0}^{D^2-1} \theta_i^2 + \frac{1}{D^2}\sum_{i,j=0}^{D^2-1} \theta_i J_{ij} + \lambda_1 \kappa\left(\sum_{i=1}^{D^2-1} \theta_i - \gamma\right) + \lambda_2\left(\sum_{i=0}^{D^2-1} \theta_i - 1\right).
\end{equation}
Next we want to calculate the stationary conditions. For $i = 0$,
\begin{equation}
\frac{\partial \mathcal{L}}{\partial \theta_0} = \frac{2}{D^2}\theta_0 + \frac{1}{D^2}J_{00}^* + \lambda_2 = 0 \implies \theta_0 = \frac{-1}{2} J_{00}^* - \frac{\lambda_2 D^2}{2}.
\end{equation}
For $i \geq 1$,
\begin{equation}
\frac{\partial \mathcal{L}}{\partial \theta_i} = \frac{2}{D^2}\theta_i + \frac{1}{D^2}J_{ii}^* + \lambda_1 \kappa + \lambda_2 = 0
\end{equation}
\begin{equation}
\implies \theta_i = \frac{-1}{2}J_{ii}^*  - D^2\frac{\lambda_1 \kappa + \lambda_2}{2}.
\end{equation}
Let $S = \sum_{i=1}^{D^2-1} \theta_i$. The normalization constraint $\sum_{i=0}^{D^2-1} \theta_i = 1$ gives
\begin{equation}
\theta_0 + S = 1 \implies \frac{-1}{2}\sum_{i}^{D^2-1} J_{ii}^*  - \frac{\lambda_2 D^2}{2} - (D^2-1) D^2\frac{\lambda_1 \kappa + \lambda_2}{2} = 1.
\end{equation}
The fidelity constraint $\mathcal{F}_{\text{FF}} = \mathcal{F}_{\text{Pauli}}(\bm{\theta})$ gives
\begin{equation}
1 - \kappa\gamma = 1 - \kappa S \implies S = \gamma,
\end{equation}
which leads to
\begin{equation}
\gamma = \frac{-1}{2}\sum_{i=1}^{D^2-1} J_{ii}^*  -  D^2(D^2-1)\frac{\lambda_1 \kappa + \lambda_2}{2}.
\end{equation}
Let $J_{\text{sum}}^* = \sum_{i=0}^{D^2-1} J_{ii}^*$. Then
\begin{equation}
-\frac{1}{2}J_{\text{sum}}^* - \frac{ D^2}{2}\lambda_2 - \frac{ D^2}{2}(D^2-1)(\lambda_1\kappa + \lambda_2) = 1.
\end{equation}
Now solve for $\lambda_1$ and $\lambda_2$
\begin{equation}
\lambda_1\kappa + \lambda_2 = -\frac{1}{ D^2(D^2-1)}\left(\sum_{i=1}^{D^2-1} J_{ii}^* + 2\gamma\right).
\end{equation}
Make a substitution
\begin{equation}
-\frac{1}{2}J_{\text{sum}}^* - \frac{D^2}{2}\lambda_2 - \frac{1}{2} D^2(D^2-1)\left[-\frac{1}{ D^2(D^2-1)}\left(\sum_{i=1}^{D^2-1} J_{ii}^* + 2\gamma\right)\right] = 1.
\end{equation}
Simplify the third term
\begin{equation}
-\frac{1}{2}J_{\text{sum}}^* - \frac{D^2}{2}\lambda_2 + \frac{1}{2}\sum_{i=1}^{D^2-1} J_{ii}^* + \gamma = 1.
\end{equation}
Note that $J_{\text{sum}}^* = J_{00}^* + \sum_{i=1}^{D^2-1} J_{ii}^*$, so,
\begin{equation}
-\frac{1}{2}J_{00}^* - \frac{D^2}{2}\lambda_2 + \gamma = 1.
\end{equation}
Solving for $\lambda_2$
\begin{equation}
\lambda_2  = (2\gamma-2 - J_{00}^*)/D^2 .
\end{equation}
Now for $\lambda_1$,
\begin{equation}
\lambda_1\kappa = -\frac{1}{ D^2(D^2-1)}\left(\sum_{i=1}^{D^2-1} J_{ii}^* + 2\gamma\right) - \lambda_2.
\end{equation}
Substitute $\lambda_2$,
\begin{equation}
\lambda_1 = -\frac{1}{\kappa D^2(D^2-1)}\left(\sum_{i=1}^{D^2-1} J_{ii}^* + 2\gamma\right) + \frac{2\gamma - J_{00}^* - 2}{D^2\kappa}.
\end{equation}
Now we can solve for theta using the setup here with the analytical form of $\lambda_1$ and $\lambda_2$.
\clearpage
\newpage

\section{Supplementary data}
\label{complementary}

\begin{figure*}[htb]
    \centering
    \includegraphics[width=\linewidth]{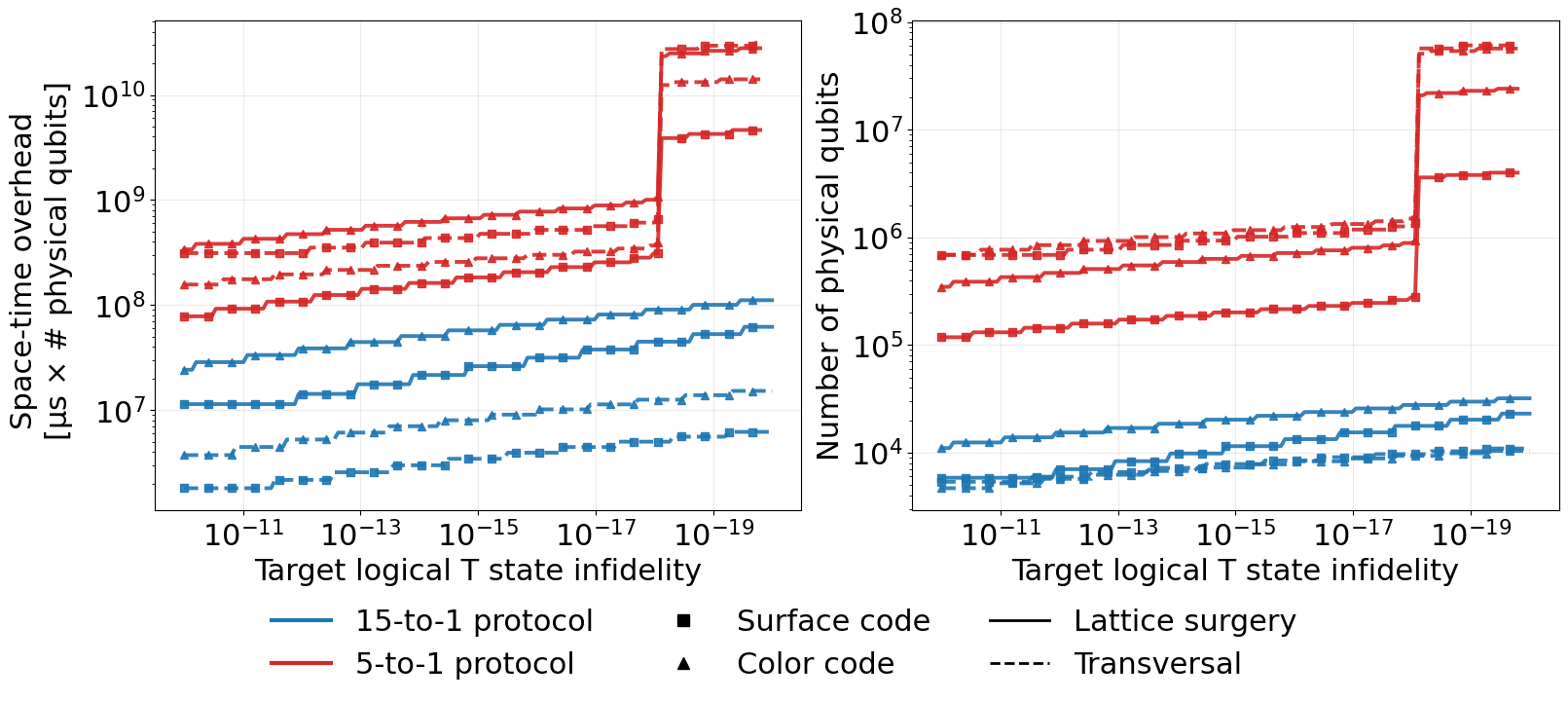}
    \caption{Experimental space-time overhead data plot for the dense layout where the label shows three different setups: red for the $5\to1$ distillation protocol and blue for the $15\to1$ distillation protocol; dotted line for transversal operation and solid line for lattice surgery-based operation; triangular marker for the surface code and circular marker for the color code. The noise parameters here are chosen according to Tab.~\ref{tab:parameter_summary}. On the vertical axis, we show the number of physical qubits (right) as well as the space-time overhead (left) with the time in units of $\unit{\micro \second}$. And on the horizontal axis, we have the targeted logical infidelity for the logical $T$ state. }
    \label{fig:dense_overhead}
\end{figure*}

In this appendix, we give all the data simulated in the Sec.~\ref{MSD-in-silicon}. First, we show the space-time overhead for the dense architecture in Fig.~\ref{fig:dense_overhead}. Compared to the sparse layout's space-time overhead, the trend is very similar, but we obtain about an order-of-magnitude reduction.

\clearpage
\newpage

Next, we present parameter sweeps of hardware values in the following figures for all three architectures. These data are for the interested reader to compare different architectural choices and QEC setups.

\subsection{Dense architecture data}
The figures on this page and the next show the space-time overhead for the dense layout with the $5\to1$ and $15\to1$ MSD protocol, respectively (complementary to Fig.~\ref{fig:sweep} in the main text).

\begin{figure*}[h]
        \centering
    \includegraphics[width=\linewidth]{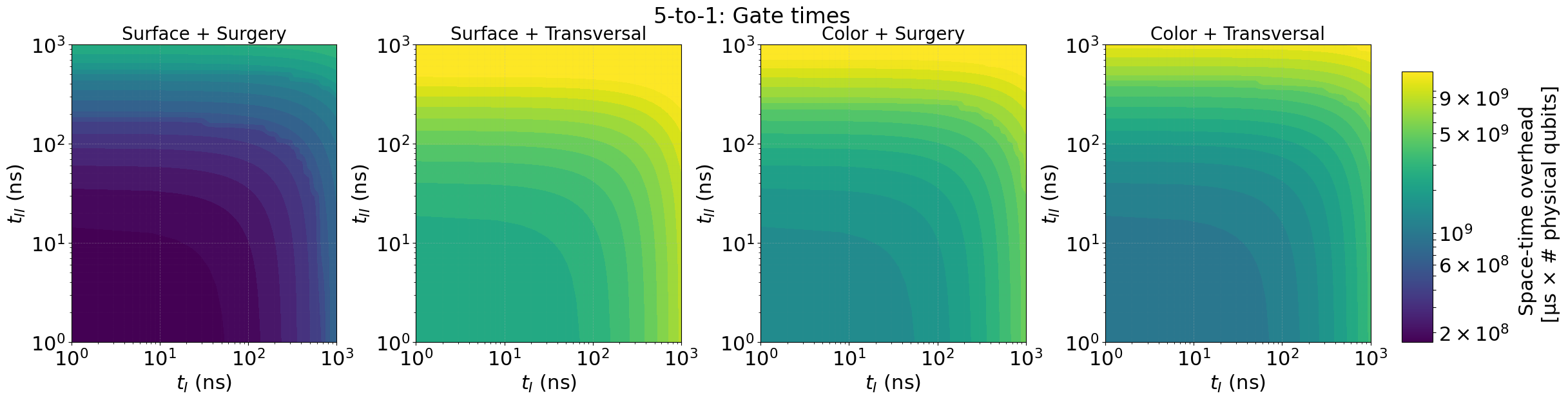}
    \includegraphics[width=\linewidth]{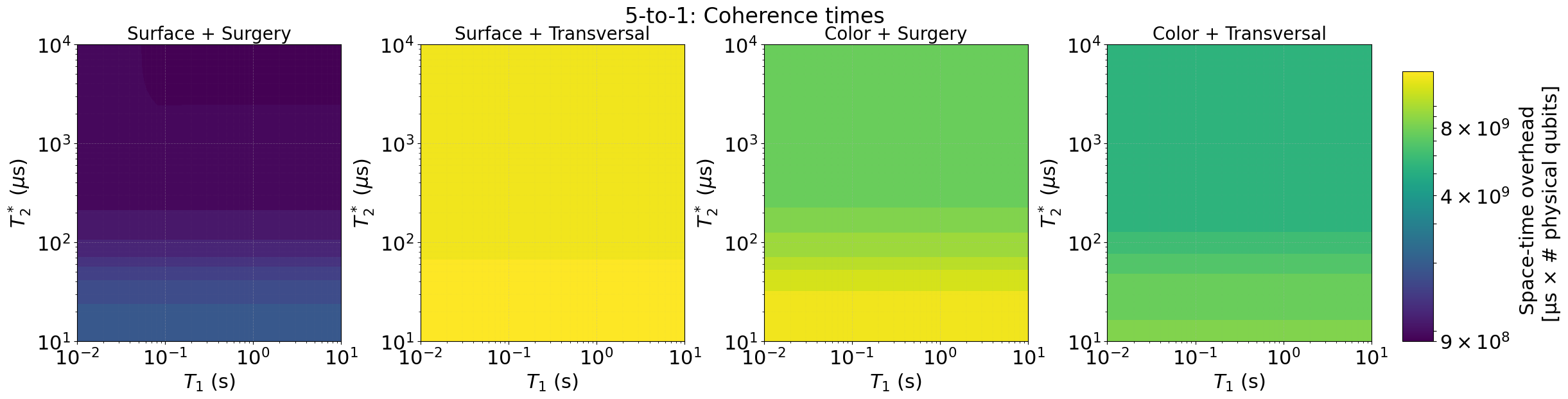}
    \includegraphics[width=\linewidth]{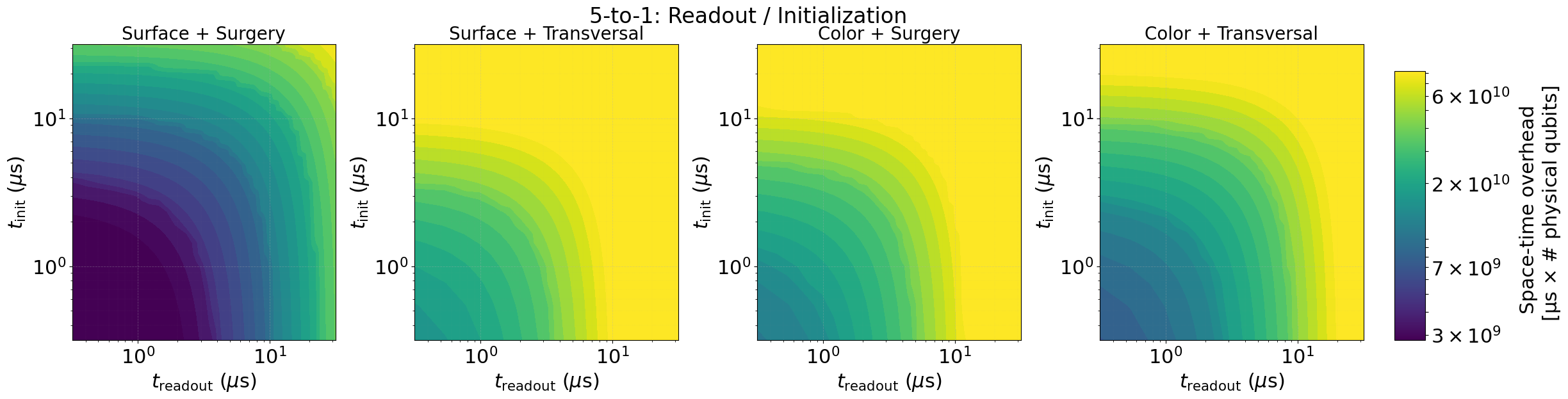}
    \caption{
    Space-time overhead for the dense architecture using $5\to1$ MSD as a function of gate times (top row), coherence times (middle row), and readout and initialization times (bottom row). The four columns from left to right correspond to the surface code with lattice surgery and transversal operations, and the color code with lattice surgery and transversal operations, respectively.}
\end{figure*}

\clearpage
\newpage

\begin{figure*}[h]
        \centering
    \includegraphics[width=\linewidth]{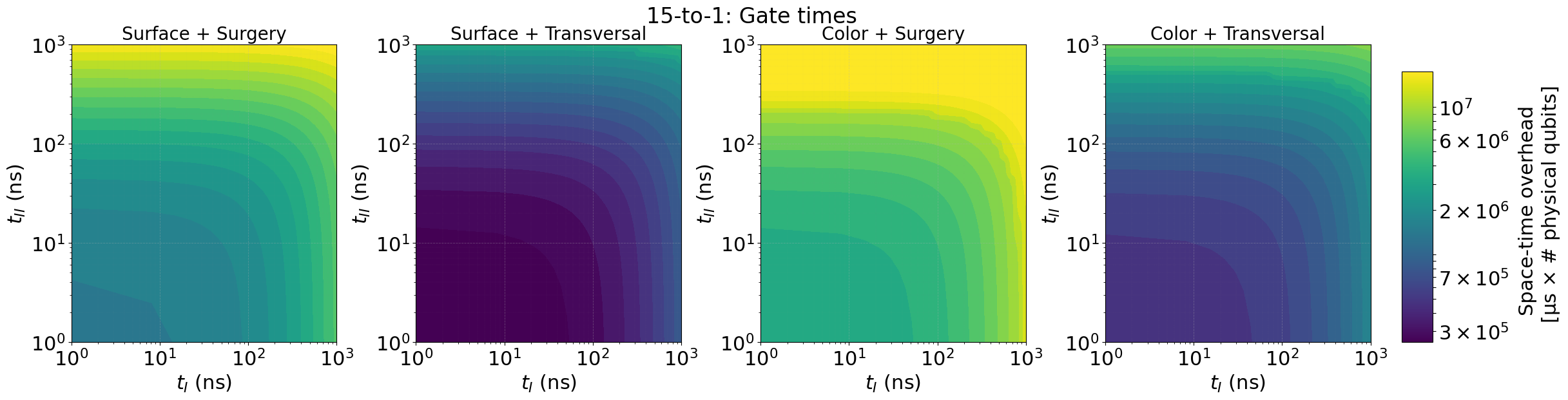}
     \includegraphics[width=\linewidth]{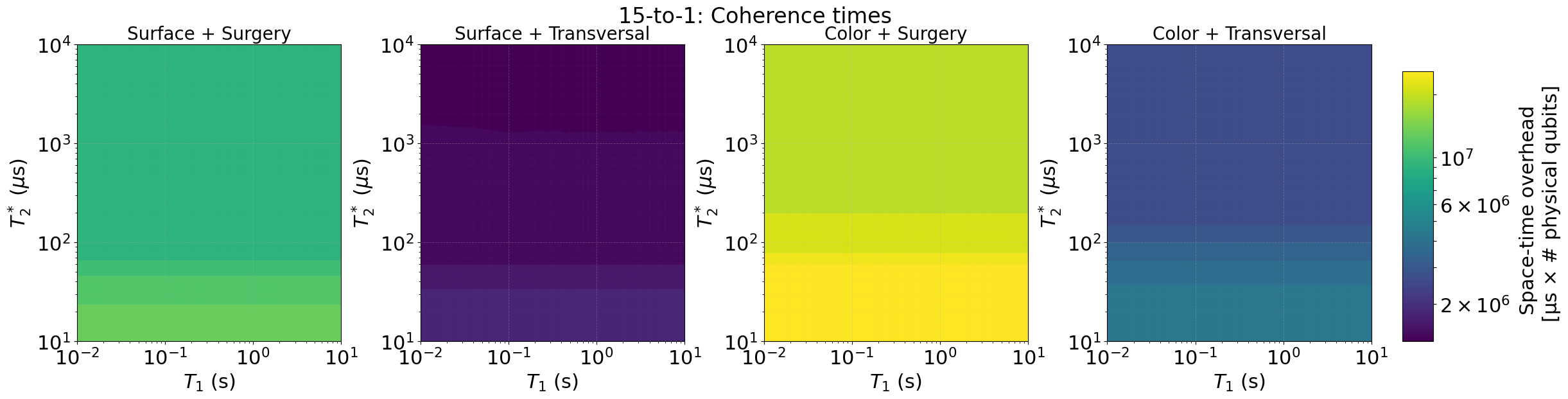}
 \includegraphics[width=\linewidth]{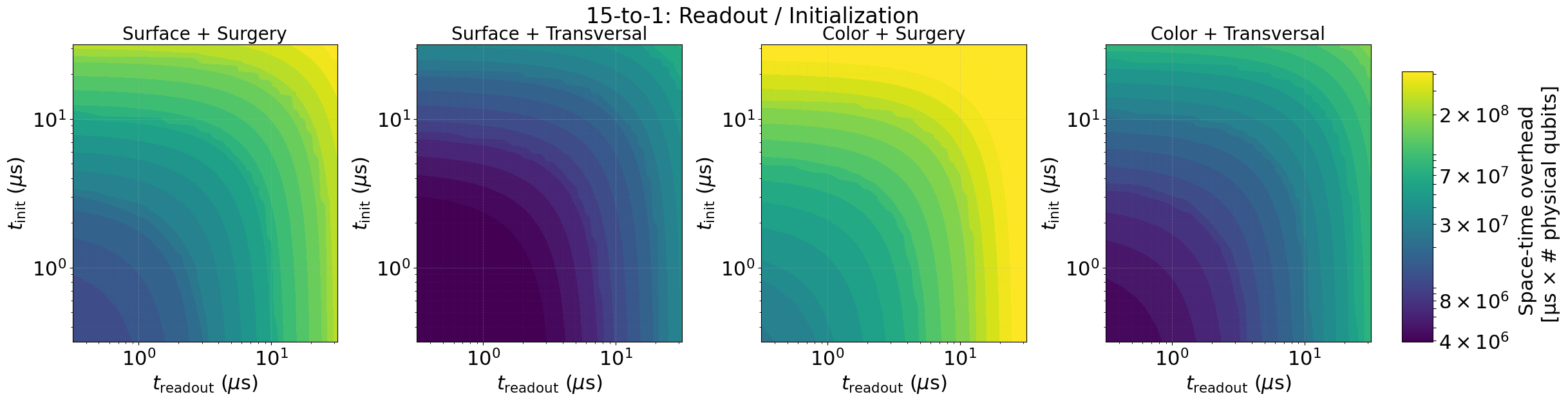}
    \caption{ Space-time overhead for the dense architecture using $15\to1$ MSD as a function of gate times (top row), coherence times (middle row), and readout and initialization times (bottom row). The four columns from left to right correspond to the surface code with lattice surgery and transversal operations, and the color code with lattice surgery and transversal operations, respectively.}
\end{figure*}

\clearpage
\newpage

\subsection{Sparse architecture data}
The figures on this page and the next show the space-time overhead for the sparse layout with the $5\to1$ and $15\to1$ MSD protocol, respectively (complementary to Fig.~\ref{fig:sweep} in the main text).
\begin{figure*}[h]
        \centering
    \includegraphics[width=\linewidth]{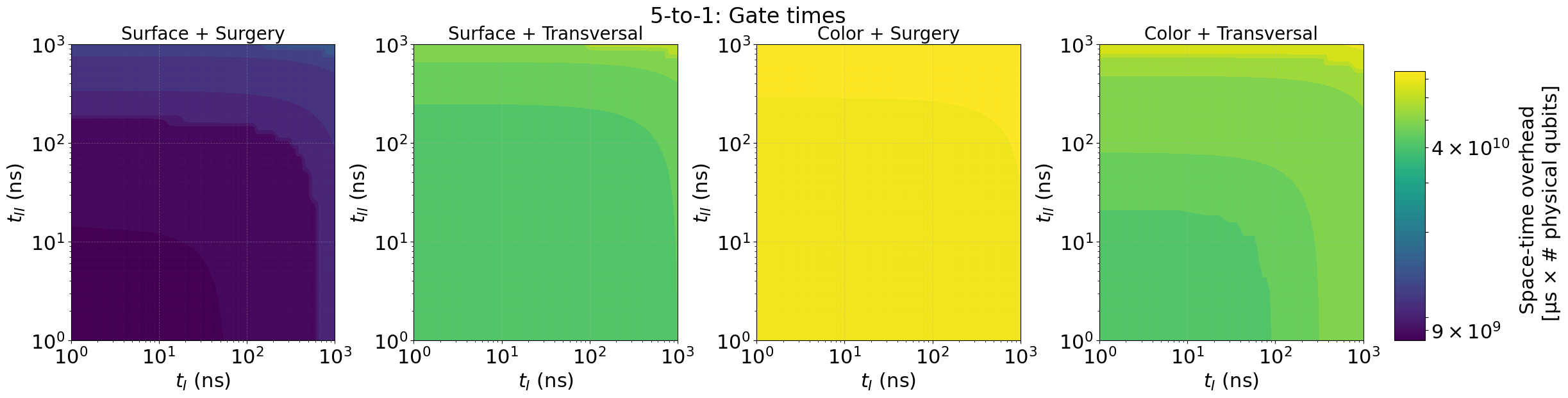}
    \includegraphics[width=\linewidth]{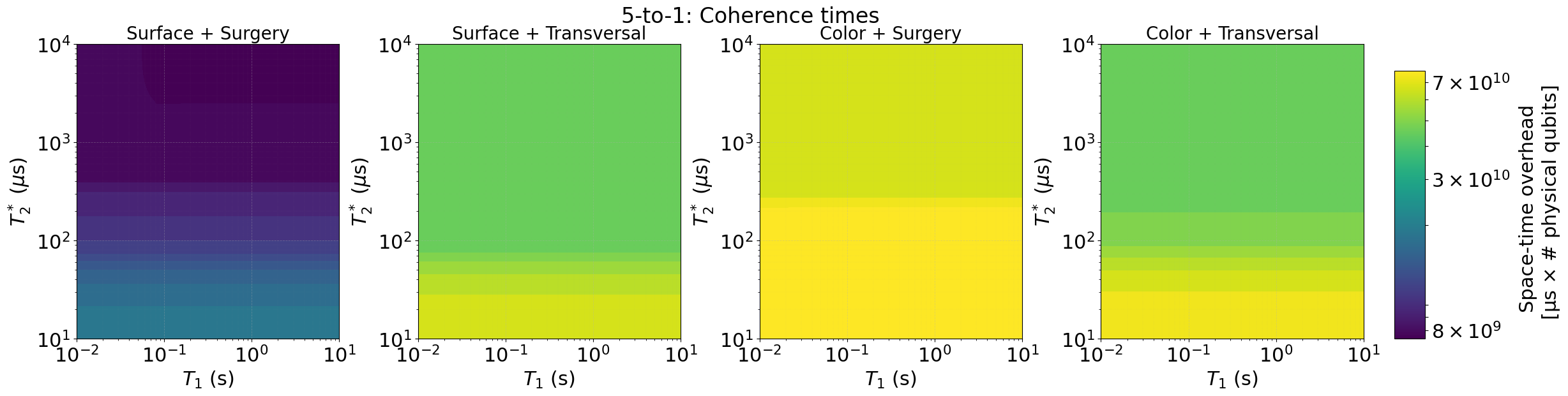}
     \includegraphics[width=\linewidth]{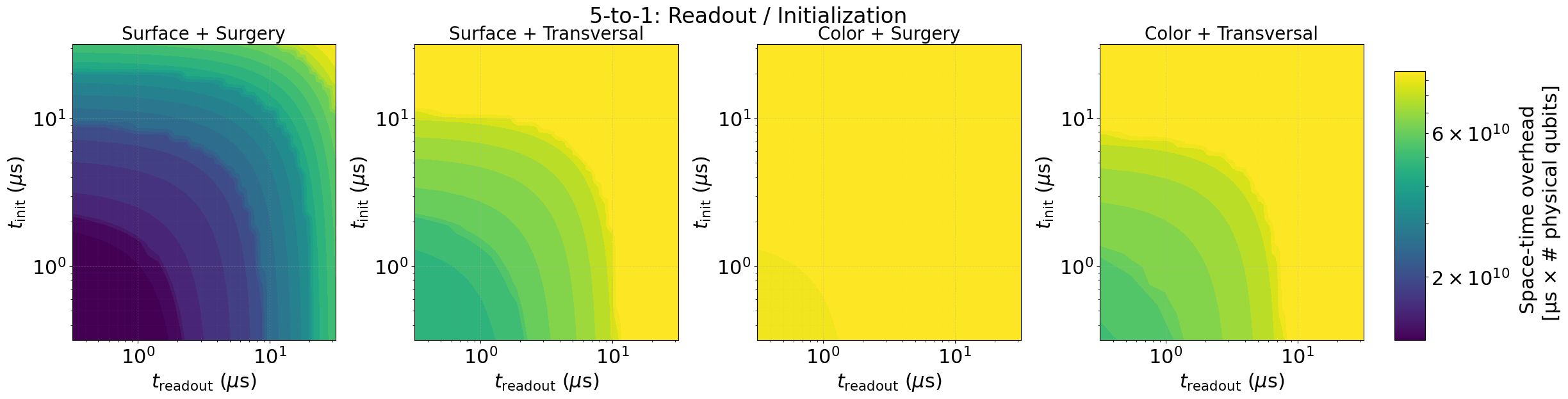}
    \caption{ Space-time overhead for the sparse architecture using $5\to1$ MSD as a function of gate times (top row), coherence times (middle row), and readout and initialization times (bottom row). The four columns from left to right correspond to the surface code with lattice surgery and transversal operations, and the color code with lattice surgery and transversal operations, respectively.}
\end{figure*}

\clearpage
\newpage

\begin{figure*}[h]
        \centering
    \includegraphics[width=\linewidth]{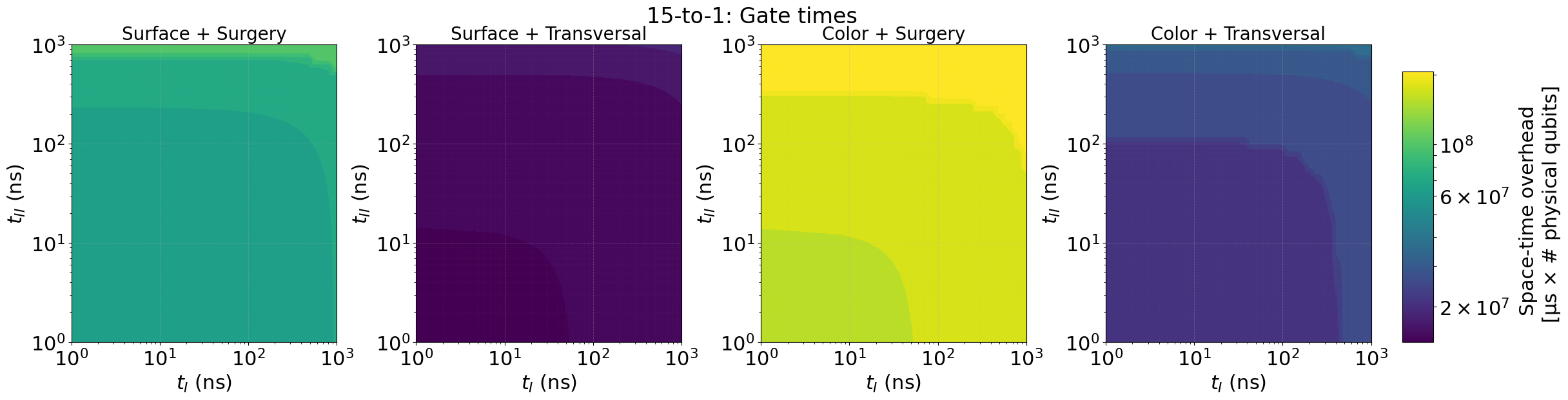}
    \includegraphics[width=\linewidth]{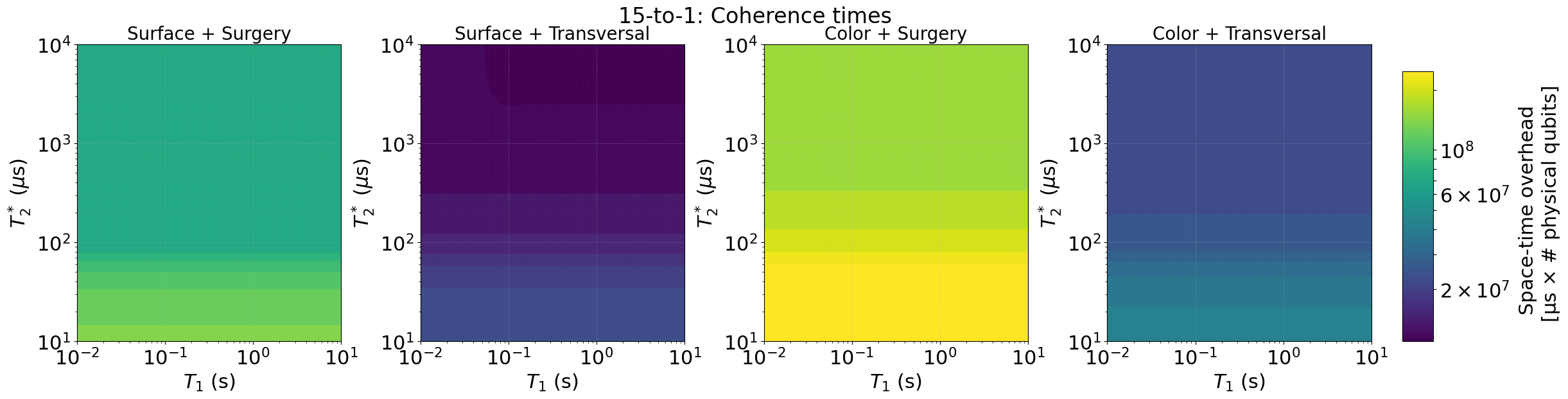}
    \includegraphics[width=\linewidth]{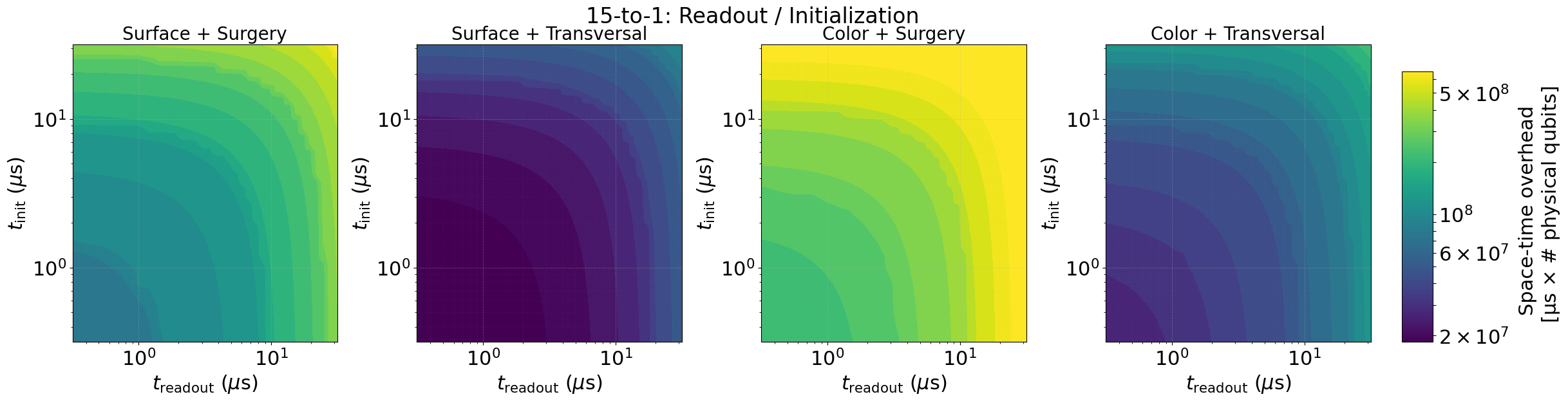}
    \caption{ Space-time overhead for the sparse architecture using $15\to1$ MSD as a function of gate times (top row), coherence times (middle row), and readout and initialization times (bottom row). The four columns from left to right correspond to the surface code with lattice surgery and transversal operations, and the color code with lattice surgery and transversal operations, respectively.}
\end{figure*}

\clearpage
\newpage

\subsection{Patched architecture data}
The figures on this page and the next show the space-time overhead for the patched layout with 2 logical qubits per patch, using the $5\to1$ and $15\to1$ MSD protocols, respectively (complementary to Fig.~\ref{fig:sweep} in the main text).
\begin{figure*}[h]
        \centering
    \includegraphics[width=\linewidth]{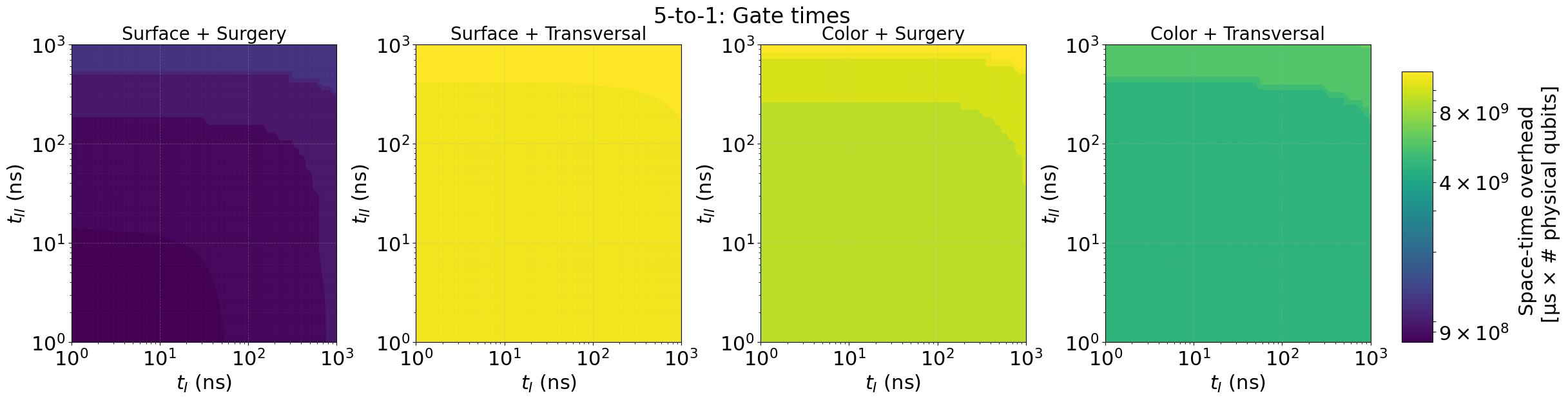}
 \includegraphics[width=\linewidth]{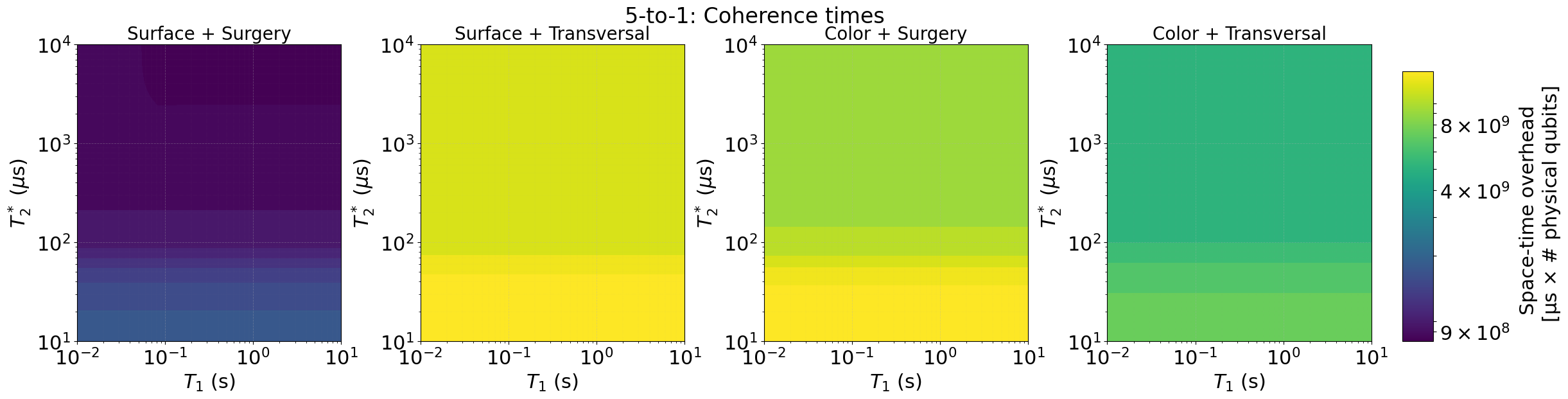}
    \includegraphics[width=\linewidth]{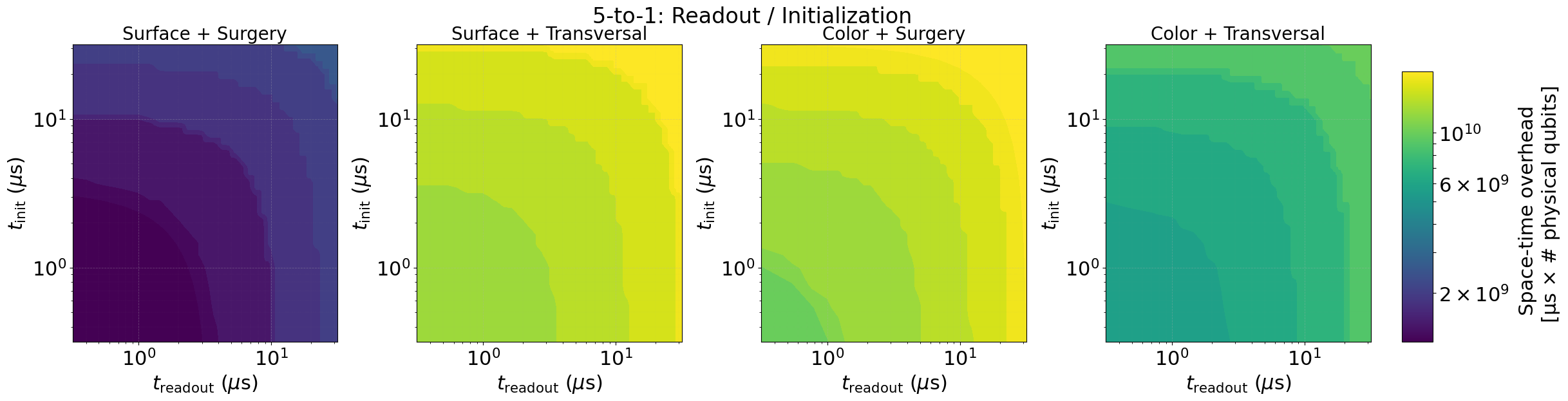}
    \caption{ Space-time overhead for the patched architecture using $5\to1$ MSD with 2 logical qubits as a function of gate times (top row), coherence times (middle row), and readout and initialization times (bottom row). The four columns from left to right correspond to the surface code with lattice surgery and transversal operations, and the color code with lattice surgery and transversal operations, respectively.}
\end{figure*}

\clearpage
\newpage

\begin{figure*}[h]
        \centering
    \includegraphics[width=\linewidth]{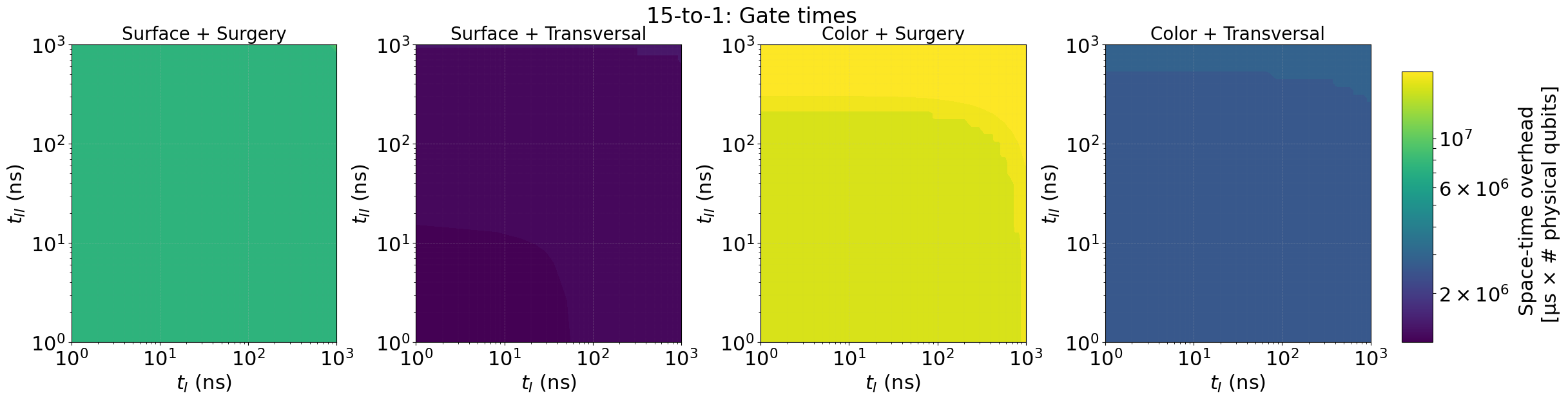}
    \includegraphics[width=\linewidth]{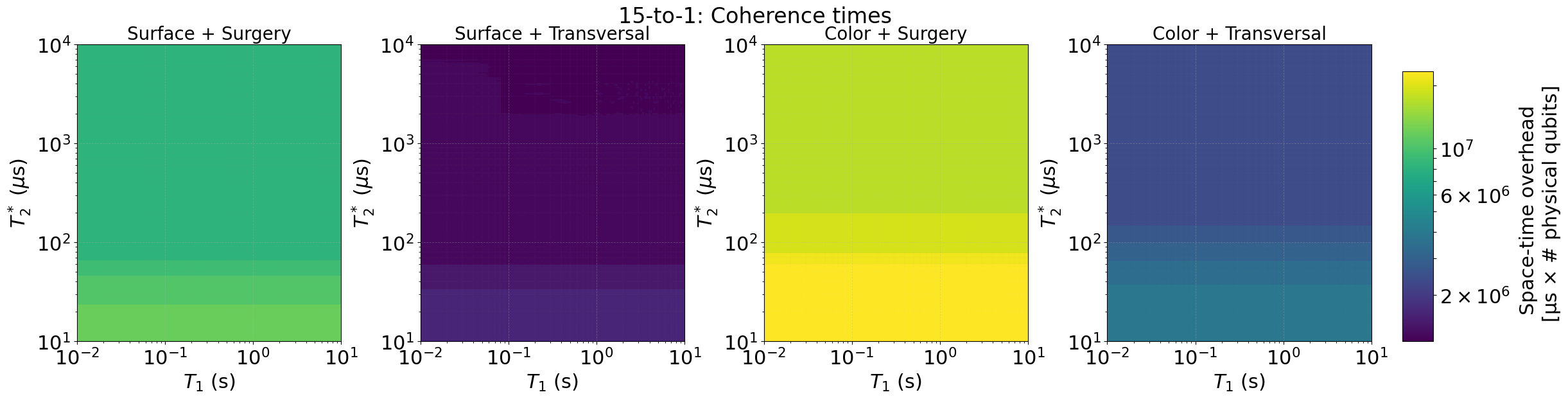}
     \includegraphics[width=\linewidth]{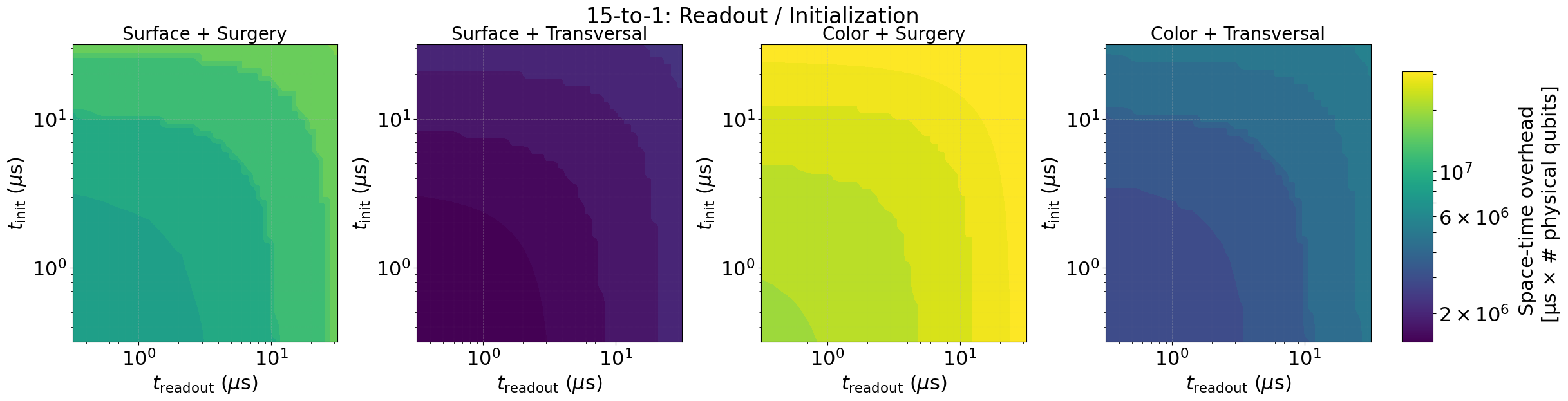}
    \caption{ Space-time overhead for the dense architecture using $15\to1$ MSD with 2 logical qubits as a function of gate times (top row), coherence times (middle row), and readout and initialization times (bottom row). The four columns from left to right correspond to the surface code with lattice surgery and transversal operations, and the color code with lattice surgery and transversal operations, respectively.}
\end{figure*}

\end{document}